\documentclass[twocolumn]{aastex631}
\usepackage{makecell}
\usepackage{multirow}
\usepackage{lineno}
\usepackage[multiple]{footmisc}

\shortauthors{Abbasi et al.}

\begin{document}

\title{IceCube search for neutrinos coincident with gravitational wave events from LIGO/Virgo run O3}
\correspondingauthor{The IceCube Collaboration}
\email{analysis@icecube.wisc.edu}
\affiliation{III. Physikalisches Institut, RWTH Aachen University, D-52056 Aachen, Germany}
\affiliation{Department of Physics, University of Adelaide, Adelaide, 5005, Australia}
\affiliation{Dept. of Physics and Astronomy, University of Alaska Anchorage, 3211 Providence Dr., Anchorage, AK 99508, USA}
\affiliation{Dept. of Physics, University of Texas at Arlington, 502 Yates St., Science Hall Rm 108, Box 19059, Arlington, TX 76019, USA}
\affiliation{CTSPS, Clark-Atlanta University, Atlanta, GA 30314, USA}
\affiliation{School of Physics and Center for Relativistic Astrophysics, Georgia Institute of Technology, Atlanta, GA 30332, USA}
\affiliation{Dept. of Physics, Southern University, Baton Rouge, LA 70813, USA}
\affiliation{Dept. of Physics, University of California, Berkeley, CA 94720, USA}
\affiliation{Lawrence Berkeley National Laboratory, Berkeley, CA 94720, USA}
\affiliation{Institut f{\"u}r Physik, Humboldt-Universit{\"a}t zu Berlin, D-12489 Berlin, Germany}
\affiliation{Fakult{\"a}t f{\"u}r Physik {\&} Astronomie, Ruhr-Universit{\"a}t Bochum, D-44780 Bochum, Germany}
\affiliation{Universit{\'e} Libre de Bruxelles, Science Faculty CP230, B-1050 Brussels, Belgium}
\affiliation{Vrije Universiteit Brussel (VUB), Dienst ELEM, B-1050 Brussels, Belgium}
\affiliation{Department of Physics and Laboratory for Particle Physics and Cosmology, Harvard University, Cambridge, MA 02138, USA}
\affiliation{Dept. of Physics, Massachusetts Institute of Technology, Cambridge, MA 02139, USA}
\affiliation{Dept. of Physics and The International Center for Hadron Astrophysics, Chiba University, Chiba 263-8522, Japan}
\affiliation{Department of Physics, Loyola University Chicago, Chicago, IL 60660, USA}
\affiliation{Dept. of Physics and Astronomy, University of Canterbury, Private Bag 4800, Christchurch, New Zealand}
\affiliation{Dept. of Physics, University of Maryland, College Park, MD 20742, USA}
\affiliation{Dept. of Astronomy, Ohio State University, Columbus, OH 43210, USA}
\affiliation{Dept. of Physics and Center for Cosmology and Astro-Particle Physics, Ohio State University, Columbus, OH 43210, USA}
\affiliation{Niels Bohr Institute, University of Copenhagen, DK-2100 Copenhagen, Denmark}
\affiliation{Dept. of Physics, TU Dortmund University, D-44221 Dortmund, Germany}
\affiliation{Dept. of Physics and Astronomy, Michigan State University, East Lansing, MI 48824, USA}
\affiliation{Dept. of Physics, University of Alberta, Edmonton, Alberta, Canada T6G 2E1}
\affiliation{Erlangen Centre for Astroparticle Physics, Friedrich-Alexander-Universit{\"a}t Erlangen-N{\"u}rnberg, D-91058 Erlangen, Germany}
\affiliation{Physik-department, Technische Universit{\"a}t M{\"u}nchen, D-85748 Garching, Germany}
\affiliation{D{\'e}partement de physique nucl{\'e}aire et corpusculaire, Universit{\'e} de Gen{\`e}ve, CH-1211 Gen{\`e}ve, Switzerland}
\affiliation{Dept. of Physics and Astronomy, University of Gent, B-9000 Gent, Belgium}
\affiliation{Dept. of Physics and Astronomy, University of California, Irvine, CA 92697, USA}
\affiliation{Karlsruhe Institute of Technology, Institute for Astroparticle Physics, D-76021 Karlsruhe, Germany }
\affiliation{Karlsruhe Institute of Technology, Institute of Experimental Particle Physics, D-76021 Karlsruhe, Germany }
\affiliation{Dept. of Physics, Engineering Physics, and Astronomy, Queen's University, Kingston, ON K7L 3N6, Canada}
\affiliation{Dept. of Physics and Astronomy, University of Kansas, Lawrence, KS 66045, USA}
\affiliation{Department of Physics and Astronomy, UCLA, Los Angeles, CA 90095, USA}
\affiliation{Centre for Cosmology, Particle Physics and Phenomenology - CP3, Universit{\'e} catholique de Louvain, Louvain-la-Neuve, Belgium}
\affiliation{Department of Physics, Mercer University, Macon, GA 31207-0001, USA}
\affiliation{Dept. of Astronomy, University of Wisconsin{\textendash}Madison, Madison, WI 53706, USA}
\affiliation{Dept. of Physics and Wisconsin IceCube Particle Astrophysics Center, University of Wisconsin{\textendash}Madison, Madison, WI 53706, USA}
\affiliation{Institute of Physics, University of Mainz, Staudinger Weg 7, D-55099 Mainz, Germany}
\affiliation{Department of Physics, Marquette University, Milwaukee, WI, 53201, USA}
\affiliation{Institut f{\"u}r Kernphysik, Westf{\"a}lische Wilhelms-Universit{\"a}t M{\"u}nster, D-48149 M{\"u}nster, Germany}
\affiliation{Bartol Research Institute and Dept. of Physics and Astronomy, University of Delaware, Newark, DE 19716, USA}
\affiliation{Dept. of Physics, Yale University, New Haven, CT 06520, USA}
\affiliation{Dept. of Astronomy, Yale University, New Haven, CT 06520, USA}
\affiliation{Dept. of Physics, Columbia University, New York, NY 10027, USA}
\affiliation{Dept. of Physics, University of Oxford, Parks Road, Oxford OX1 3PU, UK}
\affiliation{Dept. of Physics, Drexel University, 3141 Chestnut Street, Philadelphia, PA 19104, USA}
\affiliation{Physics Department, South Dakota School of Mines and Technology, Rapid City, SD 57701, USA}
\affiliation{Dept. of Physics, University of Wisconsin, River Falls, WI 54022, USA}
\affiliation{Dept. of Physics and Astronomy, University of Rochester, Rochester, NY 14627, USA}
\affiliation{Department of Physics and Astronomy, University of Utah, Salt Lake City, UT 84112, USA}
\affiliation{Oskar Klein Centre and Dept. of Physics, Stockholm University, SE-10691 Stockholm, Sweden}
\affiliation{Dept. of Physics and Astronomy, Stony Brook University, Stony Brook, NY 11794-3800, USA}
\affiliation{Dept. of Physics, Sungkyunkwan University, Suwon 16419, Korea}
\affiliation{Institute of Basic Science, Sungkyunkwan University, Suwon 16419, Korea}
\affiliation{Institute of Physics, Academia Sinica, Taipei, 11529, Taiwan}
\affiliation{Dept. of Physics and Astronomy, University of Alabama, Tuscaloosa, AL 35487, USA}
\affiliation{Dept. of Astronomy and Astrophysics, Pennsylvania State University, University Park, PA 16802, USA}
\affiliation{Dept. of Physics, Pennsylvania State University, University Park, PA 16802, USA}
\affiliation{Dept. of Physics and Astronomy, Uppsala University, Box 516, S-75120 Uppsala, Sweden}
\affiliation{Dept. of Physics, University of Wuppertal, D-42119 Wuppertal, Germany}
\affiliation{DESY, D-15738 Zeuthen, Germany}
\affiliation{Dept. of Physics, University of Florida, Gainesville, FL 32611-8440, USA}

\author[0000-0001-6141-4205]{R. Abbasi}
\affiliation{Department of Physics, Loyola University Chicago, Chicago, IL 60660, USA}

\author[0000-0001-8952-588X]{M. Ackermann}
\affiliation{DESY, D-15738 Zeuthen, Germany}

\author{J. Adams}
\affiliation{Dept. of Physics and Astronomy, University of Canterbury, Private Bag 4800, Christchurch, New Zealand}

\author{N. Aggarwal}
\affiliation{Dept. of Physics, University of Alberta, Edmonton, Alberta, Canada T6G 2E1}

\author[0000-0003-2252-9514]{J. A. Aguilar}
\affiliation{Universit{\'e} Libre de Bruxelles, Science Faculty CP230, B-1050 Brussels, Belgium}

\author[0000-0003-0709-5631]{M. Ahlers}
\affiliation{Niels Bohr Institute, University of Copenhagen, DK-2100 Copenhagen, Denmark}

\author{M. Ahrens}
\affiliation{Oskar Klein Centre and Dept. of Physics, Stockholm University, SE-10691 Stockholm, Sweden}

\author[0000-0002-9534-9189]{J.M. Alameddine}
\affiliation{Dept. of Physics, TU Dortmund University, D-44221 Dortmund, Germany}

\author{A. A. Alves Jr.}
\affiliation{Karlsruhe Institute of Technology, Institute for Astroparticle Physics, D-76021 Karlsruhe, Germany }

\author{N. M. Amin}
\affiliation{Bartol Research Institute and Dept. of Physics and Astronomy, University of Delaware, Newark, DE 19716, USA}

% ...
\date{\today}
\collaboration{1000}{The IceCube Collaboration}
\noaffiliation

\begin{abstract}
Using data from the IceCube Neutrino Observatory, we searched for high-energy neutrino emission from the gravitational-wave events detected by advanced LIGO and Virgo detectors during their third observing run. We did a low-latency follow-up on the public candidate events released during the detectors' third observing run and an archival search on the 80 confident events reported in GWTC-2.1 and GWTC-3 catalogs. An extended search was also conducted for neutrino emission on longer timescales from neutron star containing mergers. Follow-up searches on the candidate optical counterpart of GW190521 were also conducted. We used two methods; an unbinned maximum likelihood analysis and a Bayesian analysis using astrophysical priors, both of which were previously used to search for high-energy neutrino emission from gravitational-wave events. No significant neutrino emission was observed by any analysis and upper limits were placed on the time-integrated neutrino flux as well as the total isotropic equivalent energy emitted in high-energy neutrinos. 
\end{abstract}


\section{Introduction} \label{sec:intro}

Since the initial discoveries of astrophysical high-energy neutrinos in 2013 \citep{IceCube:2013low, PhysRevLett.113.101101} and gravitational waves (GWs) in 2015 \citep{LIGOScientific:2016aoc}, we have entered an exciting era of multi-messenger astronomy. We now have over 10~years of IceCube neutrino data from the full detector configuration \citep{IceCube:2016zyt} and 90 reported GW events with high astrophysical probability by the LIGO Scientific, Virgo and KAGRA Collaborations (LVK) \citep{Abbott_2019, LIGOScientific:2021usb, LIGOScientific:2021djp}. This abundance of multi-messenger data allows for statistically robust searches for common sources of GWs and high-energy neutrinos. Searches dating back before the individual confident discoveries of astrophysical GWs and high-energy neutrinos have not found a significant joint emission \citep{ 2008CQGra..25k4039A, 2009IJMPD..18.1655V, 2011PhRvL.107y1101B, 2012PhRvD..85j3004B, 2013JCAP...06..008A, 2014PhRvD..90j2002A, 2016PhRvD..93l2010A}. Following the first confident GW observation \citep{PhysRevLett.116.061102}, several attempts from IceCube and ANTARES have not found significant emission of coincident high energy neutrinos \citep{, 2017ApJ...850L..35A,2017PhRvD..96b2005A,Albert_2020, Aartsen:2020mla,Veske:2021Q6}. Searches for neutrinos in the low-energy regime have also been conducted by IceCube \citep{IceCube:2021ddq}, Super-Kamiokande \citep{2021ApJ...918...78A}, KamLAND \citep{2021ApJ...909..116A}, and Borexino \citep{2017ApJ...850...21A}.

The discovery of such a joint emission would provide important information about physics of the source and improve our understanding of the sources of the individual messengers. Currently, the emission of high-energy neutrinos from a GW source is expected to come from formed jets during the GW emission, which accelerates charged particles \citep{RevModPhys.85.1401}. These charged particles would produce mesons. From their decays and the decays of their secondaries, high-energy neutrino emission is expected \citep{Fang:2017tla,PhysRevD.98.043020}. Moreover, the inclusion of neutrino information to the gravitational-wave observation would help in constraining the location of the source more precisely in the sky, enabling more explorations to be done on it via the telescopes with narrow field of views. These motivations keep the search efforts vibrant despite the estimated low chance for joint detections with the current detectors \citep{2011PhRvL.107y1101B, 2012PhRvD..85j3004B, Fang:2017tla}.

In this article, we present our low-latency follow-up searches and archival searches for high-energy neutrino emission from the GW events detected during the complete third observing run of advanced LIGO and Virgo detectors (O3). In Section \ref{sec:detectors}, we describe the IceCube detector and its neutrino data used for this analysis, and the GW detector runs followed up in this paper. In Section \ref{sec:Methods}, we provide relevant details about the searches done by two main analysis methods; unbinned maximum likelihood (UML) and Low Latency Algorithm for Multi-messenger Astrophysics (LLAMA). More detailed discussions on the methods can also be found in our previous publication \citep{Aartsen:2020mla}. Section \ref{sec:pipelines} describes the low-latency operation of the pipelines for following-up the candidate GW event alerts reported during the O3 run at the Gravitational-wave Candidate Event Database (GraceDB) \footnote{\url{https://gracedb.ligo.org/}}, and summarizes the results. In Section \ref{sec:results}, we present the results of our archival searches using both LLAMA and UML methods. These archival searches were performed on the 44 confident GW events from GWTC-2.1 \citep{LIGOScientific:2021usb}  \footnote{Most of the events in GWTC-2.1 were already reported in the catalog GWTC-2 \citep{LIGOScientific:2020ibl}. These were previously analyzed by both LLAMA and UML searches \citep{Veske:2021Q6}. The LLAMA pipeline reanalyzed them with a refined background distribution, while the UML results remained the same.} and 36 GW events from GWTC-3 \citep{LIGOScientific:2021djp}. These analyses include a search within a time window of $\pm$500~s around the GW events, a dedicated follow-up on the candidate optical flare from GW190521 \citep{PhysRevLett.125.101102,Graham:2020gwr}, and an extended two-week search on the neutron star containing events by the UML pipeline.
\section{The Neutrino and Gravitational Wave Observations}
\label{sec:detectors}
\subsection{The IceCube Detector}
\label{sec:icecube}
The IceCube Neutrino Observatory is a cubic-kilometer detector array located at the geographic South Pole \citep{IceCube:2016zyt}.
The detector consists of 86 strings drilled deep into ice. These strings hold 60 Digital Optical Modules (DOMs) between depths
of 1.5 km and 2.5 km in the Antarctic ice. The main component of the DOMs are photomultiplier tubes used to detect the
Cherenkov light emitted by charged particles produced when neutrinos interact in ice.

There are two main event topologies seen within IceCube data: tracks and cascades. Tracks are produced when muon neutrinos undergo charged-current interactions and produce muons that travel along a straight line and deposit Cherenkov light along its path. Cascades, which mainly consist of electromagnetic showers, are generated via charged-current interactions of electron neutrinos and neutral-current interactions of neutrinos of all flavors within the ice. Tracks are excellent for pointing towards various astrophysical sources since they have an angular resolution of $\lesssim \, 1^\circ$, which is much better than the pointing resolution of cascades ($\gtrsim \,10^\circ$)\citep{2014PhRvD..89j2004A, 2014JInst...9P3009A}.

The analyses presented here use neutrino data from a low-latency data stream known as the Gamma-ray Follow-Up (GFU) Online event stream. The GFU Online event selection is able to rapidly reconstruct neutrino events observed in the IceCube detector and the data is made available within roughly 30~s, allowing for rapid neutrino follow-ups. The GFU dataset uses track events detected with IceCube, since their pointing resolutions are well suited for follow-up analyses. The details of the selection can be found in \cite{Aartsen:2016qbu} and the online version of the dataset, which we use in this article, is described further in \cite{Kintscher:2016uqh}.

The dataset consists of through-going muon tracks originating primarily from cosmic-ray backgrounds from the atmosphere. In the southern sky, the sample is dominated by the atmospheric muons while in the northern sky, the sample is dominated by atmospheric neutrinos. Atmospheric muons do not contribute to the rate in the northern sky due to Earth absorption. The all-sky neutrino event rate ranges from 6-7~mHz depending on seasonal variation of atmospheric neutrinos \citep{Heix:2019jib}. Overall the rate of astrophysical neutrinos is roughly three orders of magnitude lower than that of the atmospheric backgrounds \citep{IceCube:2016xci}.

\subsection{The third observing run of ground-based gravitational-wave detectors}
\label{sec:lvc}
On April 1$^{\rm st}$ 2019 at 15:00 UTC the LIGO and Virgo detectors network \citep{TheLIGOScientific:2014jea,TheVirgo:2014hva} started their third observing run with an increased sensitivity enabling the detection of gravitational waves from compact binary coalescences at a rate of greater than 1 merger per week \citep{LIGOScientific:2021usb, LIGOScientific:2021djp}. During the period of October 1$^{\rm st}$ 15:00~UTC to November 1$^{\rm st}$ 15:00~UTC the detectors were not collecting data, thus separating the observation run to two segments, O3a followed by O3b, which ended on March 27$^{\rm th}$ 2020 at 17:00~UTC. The near-realtime analysis of LIGO-Virgo data by the LIGO Scientific and Virgo  Collaborations (LVC) allows for the broadcasting of open public alerts. On the other hand, an in-depth offline analysis provides an update to the catalog of GW events.

\begin{figure}
    \centering
    \includegraphics[width=0.48\textwidth]{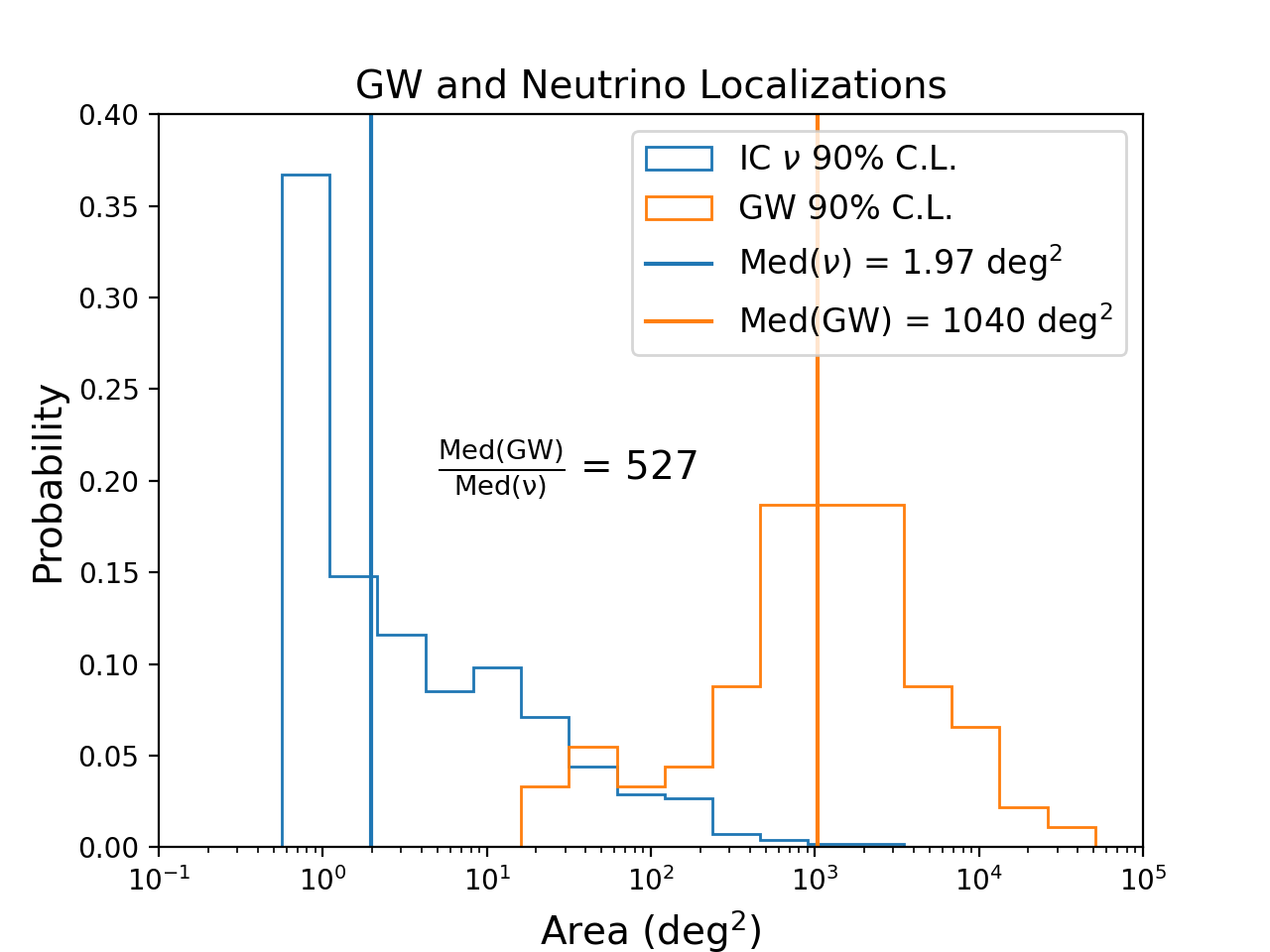}
    \caption{A comparison of the sky localizations of the  90~\% probability regions of the GW skymaps (orange) and the 90~\% contours of the neutrino localizations (blue). The skymaps shown here include all 91 GW events from O1, O2 and O3. It can be seen that we are mainly limited by the large areas of the GW skymaps, which reduces if three detectors from LVC detect the event simultaneously.}
    \label{fig:localizations}
\end{figure}

In this paper, since a combination of the events from IceCube and the GW events from O3 are used, the analyses becomes dependent on the localizations of both the neutrino and the GW events. Figure \ref{fig:localizations} compares the sky localizations of the skymaps of the candidate GW events published in the GW catalogs (O1 to O3) and the neutrino events detected by IceCube, within the GFU dataset. The 90\% localizations of both are used to make the comparison. It is seen that we are mainly limited by the localization uncertainties of the GW skymaps. These uncertainties are expected to reduce within the future runs of the ground-based gravitational-wave detectors \citep{KAGRA:2013rdx}.

\section{Methods} \label{sec:Methods}
There are two main searches that we employed: the UML and LLAMA searches. Both the UML and LLAMA analyses performed short time scale follow-ups for each reported GW event. The analyses searched for neutrino emission within a $\pm$500~s time window centered around the GW merger time. This time window was used both in the realtime and archival searches. The time window is a conservative empirical estimate of the delay between the GW and neutrino emission for a model based on gamma-ray bursts \citep{Baret:2011tk}.

Additionally, the UML analysis performed a long time scale analysis on all binary neutron star (BNS) and neutron star-black hole (NSBH) candidates. This search, called the 2-week follow-up,  is motivated by models which predict neutrino emission on longer time scales from binaries with at least one neutron star \citep{Fang:2017tla,Decoene:2019eux}. We searched within a time period of [-0.1,+14]~days around the GW merger times.

Both analyses also performed a neutrino follow-up search on the candidate optical counterpart to the binary black hole (BBH) merger GW190521 observed by Zwicky Transient Facility (ZTF) \citep{Graham:2020gwr}. ZTF observed a flaring active galactic nuclei (AGN), J124942.3+344929, which coincided with the 90\% credible region of the GW event's sky localization. This flare can be explained by the accretion of the gas in the AGN disk to the kicked final black hole of the merger \citep{McKernan_2019}. The motivation for the neutrino follow-up was the expected formation of a jet accelerating particles due to the chaotic accretion dynamics around the kicked black hole travelling through the AGN disk.

\subsection{Unbinned Maximum Likelihood} \label{sec:UML}
The unbinned maximum likelihood (UML) method tests for a point-like neutrino source coincident with the GW localization region. The likelihood takes into account the direction, angular error, and reconstructed energy of each neutrino on the sky. 
The sky is divided into equal area bins using the Healpix pixelization scheme \citep{2005ApJ...622..759G}. We then perform a likelihood ratio test where the test statistic (TS) is the log-likelihood ratio. The TS is computed at each pixel in the sky by maximizing the log-likelihood ratio and weighting the result by the GW localization probability in the given pixel. The pixel with the largest TS value is taken to be the best-fit location for a joint GW-neutrino source and the associated TS is considered the final observed TS for the analysis. For a full detailed description of the likelihood and TS used here, see \cite{Hussain:2019xzb}.

To compute the significance for each GW follow-up, we perform 30,000 pseudo-experiments with scrambled neutrino data to generate a background TS distribution. Then scrambling is carried out by randomly assigning a time for the neutrinos, which is equivalent to a scramble in right ascension, while maintaining the declination dependence of the data. The final observed TS for a given GW event is then compared to its background distribution to compute a $p$-value.

In the case where the observed TS is consistent with background, we place 90\% confidence level (CL) upper limits (ULs) on the time-integrated neutrino flux, $E^2F$, assuming an $E^{-2}$ spectrum, where $F\,=\, dN/dE\,dA$. The limits are computed by injecting simulated signal neutrinos into the sky according to the GW localization probability. We then follow the all-sky scan procedure described above to compute a TS for a given value of injected neutrino flux. We run 500 trials for a given injected neutrino flux with a random injection location chosen for each trial. The 90\% UL on the neutrino flux is then defined as the flux for which 90\% of trials produce a TS value greater than the observed TS value for the GW event.

Upper limits to the isotropical equivalent energy ($E_{\rm iso}$ ULs) are computed in a similar manner. Once again we assume an $E^{-2}$ spectrum and convert our injected $E_{\rm iso}$ into a flux at Earth by sampling a location on the sky as well as a distance to the GW source according the the 3D localization probability provided by LIGO/Virgo. The flux is then converted to an expected number of events observed at IceCube using the dataset's declination dependence and effective area.

Note that all reported ULs are only valid within a certain range of energies. The energy range of our data sample depends strongly on declination. The central 68\% energy range in the southern hemisphere is roughly 5 $\times$ 10$^5$~GeV~-~10$^7$~GeV and in the northern hemisphere ranges from roughly 5 $\times$ 10$^3$~GeV to 10$^5$~GeV.

For the follow-up of the potential optical counterpart of GW190521, AGN J124942.3+344929, we do not include any of the GW spatial information because we are testing for neutrino emission from the precise location of the AGN rather than the full GW contour. We search for neutrinos correlated with the location of the AGN in a 112 day time window after the merger, which is a conservative estimate based on the time profile of the optical flare. This is done in a model independent manner, with no assumptions on the emission profile in the entire time window. This method is equivalent to the full all-sky scan method described above except the localization skymap is a delta function at the single pixel containing the AGN.

\subsection{Low Latency Algorithm for Multi-messenger Astrophysics (LLAMA)} \label{sec:LLAMA}
The LLAMA analysis is based on the calculation of Bayesian probabilities of the observed coincidences of GWs and high-energy neutrinos \citep{PhysRevD.100.083017}. The odds ratio of the coincidence arising from a joint astrophysical emission of GWs and neutrinos being unrelated, considering any of them being not astrophysical as well, is used as a test statistic. For the analysis of confirmed GW detections followed up in this study, the GW events are assumed to be certainly astrophysical. The origins of the neutrinos are quantified for astrophysical or background scenarios. This requires the effective area of IceCube, past triggers of the GFU stream (which are predominantly of atmospheric origin), and the reconstructed energies of the neutrinos and their sky localizations. In addition to this, an $E^{-2}$ astrophysical spectrum is assumed. The relation between the GW and neutrinos are quantified via the difference between their detection times, their respective sky localizations, and the mean distance reconstruction of the GW event. Together with the astrophysical emission energy $E_{\mathrm{iso}}$, which is log-uniform between $10^{46}-10^{51}$ erg, the distance reconstruction of the GW event accounts for the propagation of the neutrinos in space.

Precomputed background distributions are used for calculating the $p$-values. In order to include the distance information of the GW events appropriately, different background distributions are constructed for different source types (BNS, NSBH, BBH coalescences). For this purpose, GW events are simulated for each source category and they are randomly matched with scrambled past GFU detections. The number of neutrinos matched with each GW event is drawn according to a Poisson distribution with a mean corresponding to the average GFU trigger count in 1000 s. The 90\% CL upper limits (frequentist limits) on the time-integrated neutrino flux are calculated as described in the appendix of \cite{Aartsen:2020mla}. 

The neutrino follow-up on the candidate optical counterpart of GW190521 in the LLAMA analysis follows the assumed emission model described in \cite{Graham:2020gwr}. The model assumes a linearly decreasing gas density around the active galactic nuclei. The kicked black hole from the merger travels through and accretes gas from the AGN disk. For our neutrino search, we hypothesized a neutrino emission from the particle accelerating jets which are expected to form due to the accretion. Hence, the intensity of the expected neutrino emission is assumed to be proportional to the accretion rate, which is assumed to be a linearly decreasing emission intensity over time in this model. The start and end times of the accretion were found from the observed light curves by following the same model, which also includes an optical diffusion delay obeying a Maxwell-Boltzmann distribution. The least-squares estimations for the start and end times of the accretion were found to be 23 and 80 days after the merger respectively, the same as that found in \cite{Graham:2020gwr}. So, we searched for a neutrino emission from a point source located at the AGN's position, free of any diffusion effect, which starts 23 days after the BBH merger and linearly decreases for the following 57 days until it ends.

\section{Low-Latency Operation} \label{sec:pipelines}
\label{sec:low-latency}
Both UML \citep{Aartsen:2020mla} and LLAMA \citep{2019arXiv190105486C,PhysRevD.100.083017} analyses deployed low-latency pipelines designed to perform automated neutrino follow-up searches after receiving notices from LVC through the Gamma-ray Coordinates Network (GCN) \citep{llama:2019icrc,Hussain:2019xzb}.

\begin{figure}
    \centering
    \includegraphics[width=0.48\textwidth]{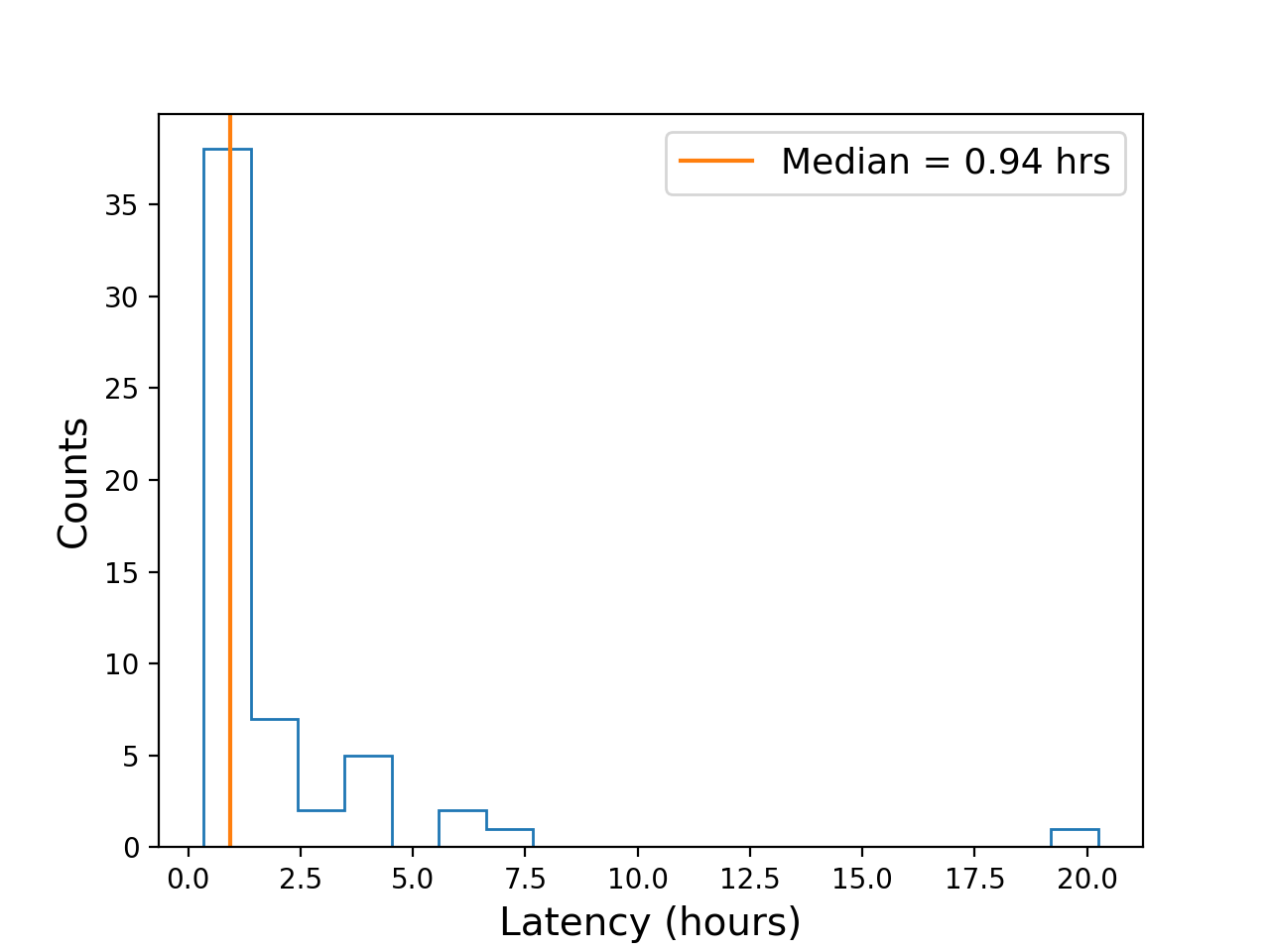}
    \caption{Latency of IceCube GCN circulars relative to GW merger times for all 56 events reported during the O3 observing run\footnote{All of the GCN notices of these GW events can be found at \url{https://gcn.gsfc.nasa.gov/lvc_events.html}.}. The outlier near 20 hours is S190421ar, where the LVC GCN notice was not received until roughly 19 hours after the GW merger. This plot only shows the follow-ups which were triggered automatically via GCN notices sent by LVC.}
    \label{fig:latency}
\end{figure}

These pipelines allow for rapid neutrino follow-ups and the dissemination of results to the astronomical community via GCN circulars. Low-latency neutrino information can help inform the observing strategies of electromagnetic observatories searching for electromagnetic counterparts to GW events. For example, observatories such as \textit{Swift}-XRT were able to use IceCube's neutrino follow-up results to narrow the search region for several GW events  \citep{Keivani:2020utg}. While no electromagnetic counterparts were found during the O3 observing run, these pipelines show the discovery potential of low-latency multi-messenger astronomy in identifying joint sources of photons, GWs, and neutrinos. 

Both analyses take advantage of the GCN notices to receive information about a given GW event and trigger a dedicated neutrino follow-up search. The pipelines use a python package, PyGCN \citep{pygcn}, to continuously monitor the GCN system for GCN notices sent by LVC. Due to the low-latency of the GFU Online stream ($\sim$30~s) and the speed of the follow-up analyses ($\sim$56~min), IceCube was able to rapidly circulate results from neutrino follow-ups to the astronomical community by using subsequent GCN circulars. Figure \ref{fig:latency} shows the distribution of response times between the IceCube GCN circulars and the GW merger time. The latency shown in the figure takes into account the time taken by LVC to send the initial GCN notice. Also included in the latency is the final vetting of the IceCube results by the collaboration's Realtime Oversight Committee (ROC) before sending the IceCube follow-up results via GCN circulars. Follow-ups with observed $p$-value $\leq$ 1\% in either pipeline or any follow-ups that were deemed interesting to the astronomical community by the ROC, resulted in releasing the directional information of the potentially significant neutrino candidate via GCN circulars.

During O3, there were a total of 56 non-retracted candidate GW events that were publicly shared. We ran follow-ups on these events and 4 of them resulted in the release of the directional information of a neutrino to the astronomical community. These released coincidences were further followed-up by different telescopes and observatories, e.g. \textit{Swift}-XRT~\citep{10.1093/mnras/staa3032,Keivani:2020utg} and Karl G. Jansky Very Large Array~\citep{Bhakta_2021}. For each of these events, the LVC GCN notices and the GCN circular archives are linked. The archives show all follow-ups performed by each observatory, including the follow-ups that use IceCube information. These events were the following:

\begin{figure*}
    \centering
    \includegraphics[width=0.9\textwidth]{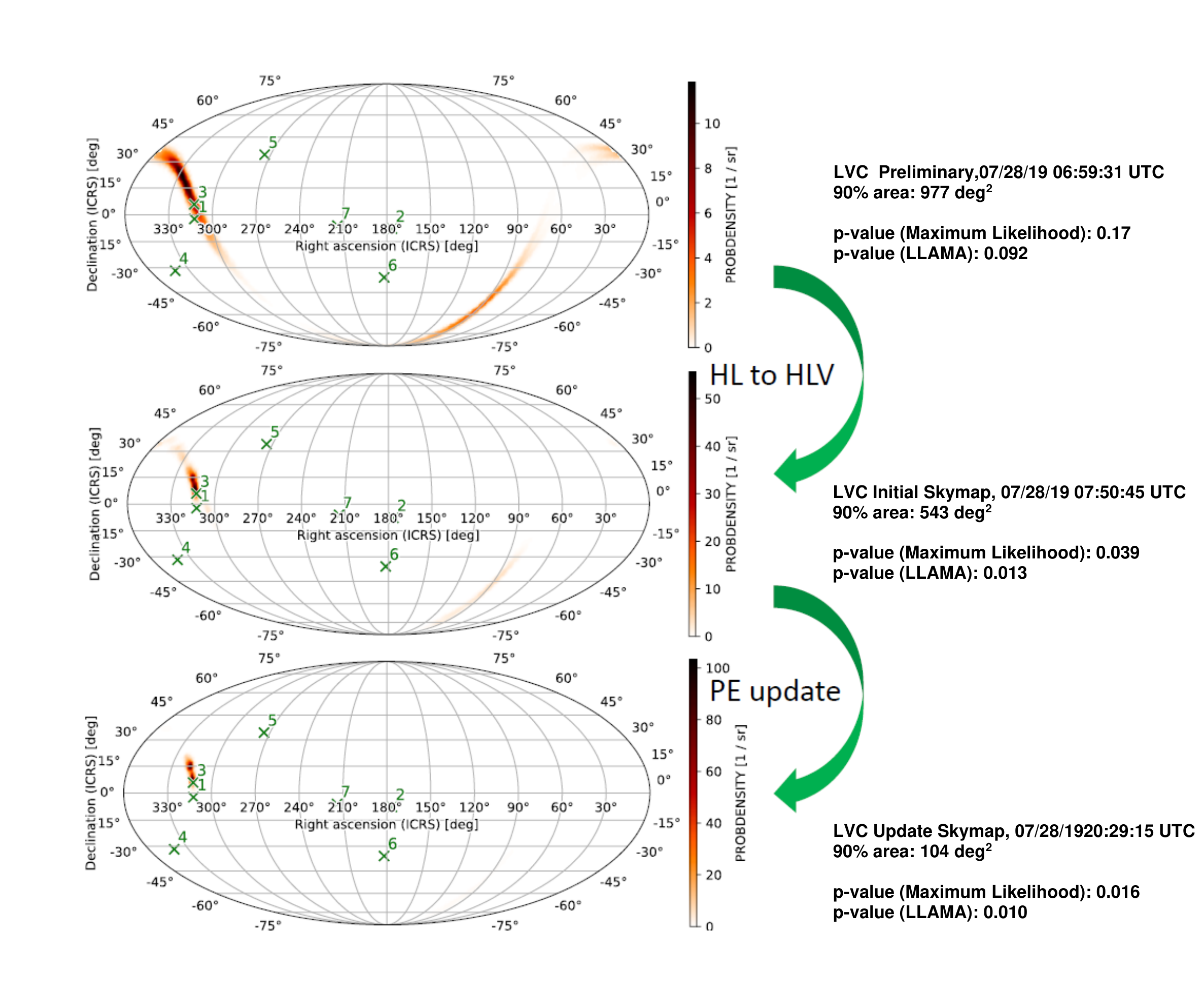}
    \caption{Evolution of the localization skymap for S190728q \footnote{\url{https://gcn.gsfc.nasa.gov/notices_l/S190728q.lvc}} and the associated follow-up results from each pipeline which were sent via GCN circulars. As the localization is refined, the $p$-values from both pipeline become more significant. The colormap in the figure represents the probability per pixel in the skymap and the green crosses show the neutrino observations.}
    \label{fig:S190728q_updates}
\end{figure*}

\begin{itemize}
    \item S190517h\footnote{GW event GCN notice \url{https://gcn.gsfc.nasa.gov/notices_l/S190517h.lvc}}\textsuperscript{,\,}\footnote{GCN circular archive \url{https://gcn.gsfc.nasa.gov/other/GW190517h.gcn3}}: This candidate BBH merger event had one neutrino located in the 90\% credible sky region of the GW localization. Due to this spatial coincidence the neutrino's localization was shared with the community \footnote{\url{https://gcn.gsfc.nasa.gov/gcn/gcn3/24573.gcn3}}, despite its low statistical significance.
    
    \item S190728q\footnote{GW event GCN notice \url{https://gcn.gsfc.nasa.gov/notices_l/S190728q.lvc}}\textsuperscript{,\,}\footnote{GCN circular archive \url{https://gcn.gsfc.nasa.gov/other/GW190728q.gcn3}}: This candidate BBH merger event originally had a two-detector localization which did not yield any significant neutrino coincidence. The localization was later improved by the incorporation of the Virgo data, which increased the significance of one of the neutrinos. With the final online skymap the coincidence had the $p$-values 1.0\% and 1.6\% for the LLAMA and UML searches, respectively\footnote{\url{https://gcn.gsfc.nasa.gov/gcn/gcn3/25210.gcn3}}. Figure \ref{fig:S190728q_updates} shows the various localization skymaps sent by LVC and the associated results from each pipeline, which were reported in low-latency via GCN circulars. The skymaps were refined over a period of roughly 14 hours following the initial GCN notice sent by LVC. It is seen that the $p$-values from both pipelines become more significant as the localization is refined, since the neutrino candidate 3 remains within the high probability region of the skymap as the GW localization shrinks. Figure \ref{fig:S190728q} shows the zoomed in updated skymap of GW190728\_064510 with the coincident neutrino overlaid. 
    \begin{figure}[ht]
    \centering
    \includegraphics[width=0.48\textwidth]{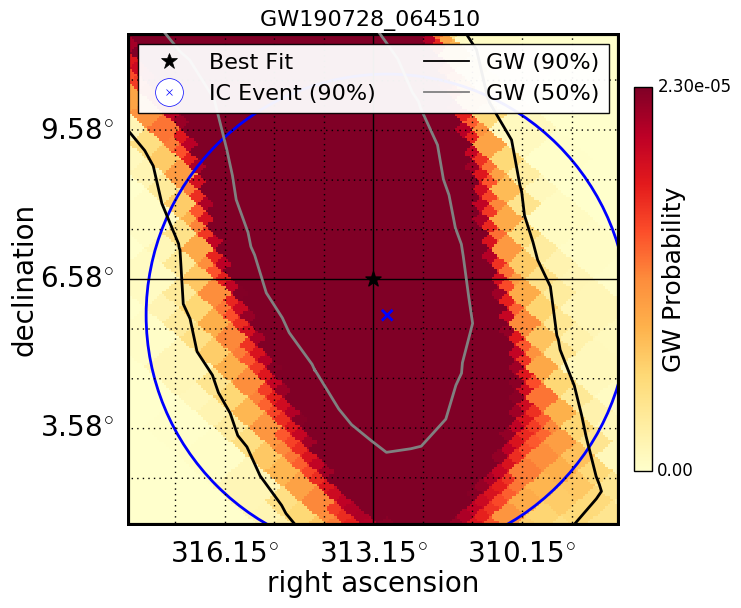}
    \caption{Skymap of GW190728\_064510 overlaid with the coincident neutrino. The red region represents the GW localization probability per pixel. The blue cross shows the best-fit neutrino direction with the circle showing the 90\% containment angular error region. The neutrino arrived 360~s before the GW merger. The final pre-trial $p$-values for this event are $p=0.013$ and $p=0.04$ with the LLAMA and UML analyses, respectively. The GCN circular describing this event was also sent in realtime (\url{https://gcn.gsfc.nasa.gov/gcn3/25210.gcn3}).}
    \label{fig:S190728q}
\end{figure}
    
    \item S191216ap\footnote{GW event GCN notice\\ \url{https://gcn.gsfc.nasa.gov/notices_l/S191216ap.lvc}}\textsuperscript{,\,}
    \footnote{GCN circular archive\\ \url{https://gcn.gsfc.nasa.gov/other/GW191216ap.gcn3}}: This candidate BBH merger event was one of the events for which the results of the two analyses disagreed. It was located relatively close, at $\sim400$ Mpc. Due to this atypically close distance for a BBH merger, the neutrino-GW coincidence was favored by the LLAMA search which assigned a $p$-value of 0.6\%, whereas the UML search obtained a $p$-value of 22\%
    \footnote{\url{https://gcn.gsfc.nasa.gov/gcn/gcn3/26460.gcn3}}. The most interesting response to our GCN notices came after the release of the neutrino coinciding with this event. HAWC observatory sent out another notice saying their most significant \emph{subthreshold} gamma-ray trigger coincides both with the neutrino and GW's localizations \footnote{\url{https://gcn.gsfc.nasa.gov/gcn/gcn3/26472.gcn3}}. No further counterpart was found from the region and due to the uncertain nature of the gamma-ray trigger the state of the triple coincidence remained inconclusive.
    
    \item S200213t\footnote{GW event GCN notice\\ \url{https://gcn.gsfc.nasa.gov/notices_l/S200213t.lvc}}\textsuperscript{,\,}\footnote{GCN circular archive\\ \url{https://gcn.gsfc.nasa.gov/other/GW200213t.gcn3}}: This event was the only candidate BNS merger for which a coincident neutrino was released. However, it was excluded in the published GW catalogs, since it did not meet the threshold requirements in the offline analysis from LVC \citep{LIGOScientific:2021djp}. The UML and LLAMA searches obtained $p$-values of 0.3\% and 1.7\% respectively for the neutrino coincidence\footnote{\url{https://gcn.gsfc.nasa.gov/gcn/gcn3/27043.gcn3}}.
\end{itemize}



Both of these low-latency pipelines are being prepared to continue neutrino follow-ups during the fourth observing run of LIGO, Virgo and KAGRA detectors, planned to start in 2023.

\section{Archival Searches on Catalogs} \label{sec:results}

Once the catalogs containing the confident GW detections were published by LVC, we performed archival searches on these events. There were several GW events added or subtracted in the catalog when compared to the the candidate events shared with the community by LVC during the O3 run. Initially, LVC released the catalog GWTC-2 \citep{LIGOScientific:2020ibl}, which contained 39 events from the first half of O3. These events were analyzed using both UML and LLAMA methods and no significant neutrino emission was found \citep{Veske:2021Q6}. Later, this catalog was renewed by LVC resulting in the publication of GWTC-2.1\citep{LIGOScientific:2021usb}, which has an updated statistic used for the classification of the events as confident detections. This updated catalog has 44 GW events out of which 8 were new when compared to GWTC-2. Three events from GWTC-2 were retracted in the updated catalog. Here, we present the results of the 44 confident events in GWTC-2.1. The 36 common events were reanalyzed by the LLAMA search with a renewed background distribution, which was generated with the latest population estimates for the binary black holes. No appreciable change was found with the previous analysis. The results of the UML analysis for the common events stayed the same. Finally, LVC also published GWTC-3 \citep{LIGOScientific:2021djp}, a catalog containing the confident GW events observed during the second half of the O3 run \citep{LIGOScientific:2021djp}. These events were also analyzed as a part of the archival search.

\begin{figure}
    \centering
    \includegraphics[width=0.48\textwidth]{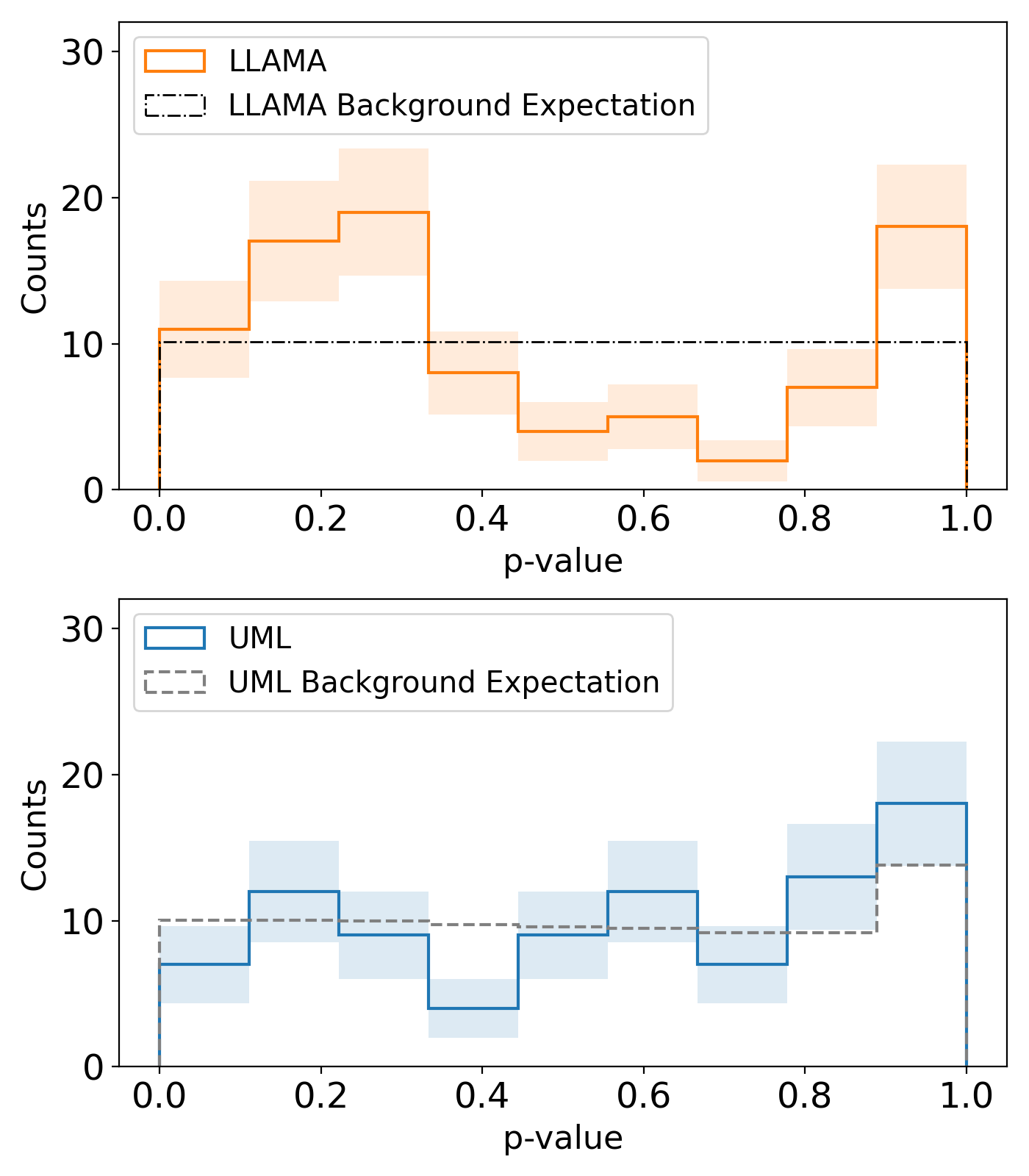}
    \caption{$p$-value distribution for the LLAMA (top panel) and UML (bottom panel) analyses of the 11 events in GWTC-1 \citep{Abbott_2019}, 44 GW events in GWTC-2.1 \citep{LIGOScientific:2021usb}, and the 36 GW events in GWTC-3 \citep{LIGOScientific:2021djp}. The distributions are consistent with background expectations. The $p$-value distributions obtained for the events in GWTC-1 were already published in \cite{Aartsen:2020mla}. The LLAMA background expectations shown here is taken from that of one representative GW, and scaled to 91 GW events.  The orange and blue bands represent the Poisson errors on the observed distribution of LLAMA and UML $p$-values, respectively.}
    \label{fig:pvals}
\end{figure}

First, we present the results of the searches for neutrino emission within a time window of $\pm$500~s around the 80 mergers in GWTC-2.1 and GWTC-3. We did not observe a significant neutrino emission from any GW event by any analysis. ULs were placed on the time-integrated, energy scaled neutrino flux, $E^2F$, as well as the $E_{\rm iso}$, emitted in high-energy muon neutrinos. Table \ref{tab:results} summarizes the results for each follow-up of GW events in GWTC-2.1 performed by both analyses. Similarly, Table \ref{tab:results2} shows the results for the GW events in GWTC-3. Figure \ref{fig:pvals} shows the histogram of the $p$-values for the collection of GW events from GWTC-1 \citep{Abbott_2019}, GWTC-2.1 \citep{LIGOScientific:2021usb} and GWTC-3 \citep{LIGOScientific:2021djp} from both analyses and the background expectations. The set of events did not show any significant sign of emission. The shown background expectation for the UML analysis was derived from the background TS distributions of each GW. The LLAMA analysis' background $p$-value distribution is seen to be uniform for all kinds of events. The different results for the LLAMA and the UML analyses arise from the inherent differences in the statistical approaches of the two --- one being a Bayesian approach including priors of the GW source and the other being a purely frequentist approach. This is also true for the $p$-values obtained in the low-latency search described in section \ref{sec:low-latency}.

\begin{figure}
    \centering
    \includegraphics[width=0.48\textwidth]{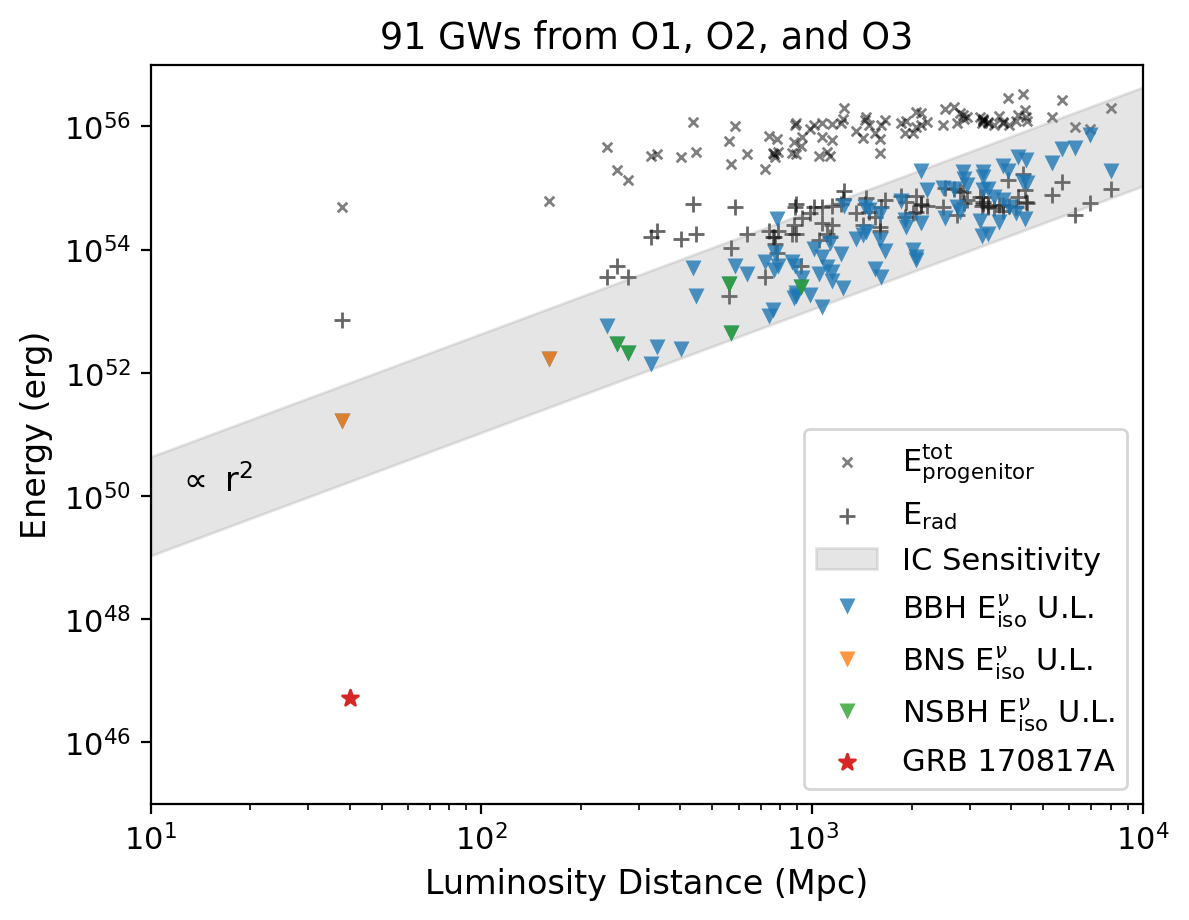}
    \caption{$E_{\rm iso}$ ULs for 91 GW candidates in GWTC-1, GWTC 2.1 and GWTC-3. The blue, green, and orange triangles show the BBH, BNS and NSBH events, respectively. Note that the 36 GW candidates present in both GWTC-2 and GWTC-2.1 do not have updated skymaps available so the results shown here are using the GWTC-2 skymaps. The black crosses represent the total rest mass energy of the progenitors of the binary merger. The grey plusses represent the total energy radiated by the binary system. Also shown for reference is the observed $E_{\rm iso}$ in photons by $Fermi$ GBM for GW170817 (red star) \citep{Monitor:2017mdv}. The grey band represents the best and worst median UL that IceCube can place based on a point source hypothesis. The $E_{\rm iso}$ ULs for the 11 GW events from GWTC-1 remain the same as those published in \cite{Aartsen:2020mla}.}
    \label{fig:eiso}
\end{figure}

Figure \ref{fig:eiso} shows the $E_{\rm iso}$ ULs for all GW events in GWTC-1, GWTC-2.1 and GWTC-3 along with the total rest mass energy of the initial compact objects and the total energy radiated by the system post-merger. The total radiated energy is computed by taking the difference of the total rest mass energy of the two progenitors and the final remnant object. 

No significant neutrino emission was observed in the second archival search presented here, which is the 2-week follow-up.
There are only 3 GW events in GWTC-2.1 which may have at least one neutron star in the binary system: GW190425, GW190814, and GW190917\_114630. Also 4 NSBH events were published in the GWTC-3 catalog: GW191219\_163120, GW200105\_162426 (marginal event), GW200115\_042309, GW200210\_092254. All of these events have at least one progenitor object with a mass estimate lower than 3~M$_{\odot}$ \citep{LIGOScientific:2021usb,LIGOScientific:2021djp}. The 2-week follow-up is performed on these 7 GW events. Once again, we place 90\% ULs on the time-integrated neutrino emission from each of the 7 GWs tested here. Table \ref{tab:2week_results} shows the $p$-values and ULs for these events and Fig. \ref{fig:2week_ts_maps_gwtc2} shows the final test statistic maps for these events. There was no difference between the neutron star containing events in GWTC-2 and GWTC-2.1.

Finally, no significant neutrino emission was found for the follow-ups on the candidate optical counterpart of GW190521 by both analysis methods. The modelled search of LLAMA yielded a $p$-value of 0.79, 90\% CL upper limit on the $E^2F$ of 0.05 GeV cm$^{-2}$ and 90\% CL upper limit on $E_{\rm iso}$ of 8$\times10^{53}$ erg. The UML analysis found a $p$-value of 0.25, with a 90\% CL upper limit on the time-integrated flux of $E^2F$=0.081 GeV cm$^{-2}$.

\section{Conclusion} \label{sec:conclusion}
Finding joint sources of GWs and high-energy neutrinos can help shed light on the sources of the highest energy neutrinos and cosmic rays \citep{doi:10.1146/annurev-nucl-101918-023510}. Studying these joint sources will also further expand our understanding of energetic outflows from the mergers of compact objects. The completion of the O3 realtime observing run and the release of the update to the second GW catalog, GWTC-2.1, followed by the release of GWTC-3 have provided a substantial increase in the number of reported GW candidates available for follow-up searches.

We developed low-latency pipelines which ran automated neutrino follow-ups for all GW events reported by LVC during the O3 observing run. Two different analyses, UML and LLAMA, both ran in low-latency and followed up each of the 56 candidate events reported during the O3 run. Four of the follow-up searches resulted in the release of the neutrino candidate's direction to the public via GCN circulars. This information prompted follow-up searches in electromagnetic observatories such as \textit{Swift}-XRT, demonstrating the power of low-latency multi-messenger observations in informing the observing strategies of other observatories. The unresolved triple coincidence for GW191216, involving a subthreshold gamma-ray trigger from HAWC observatory, triggered the development of general multi-messenger search methods for many messengers \citep{Veske_2021}.

In addition to the low-latency follow-ups, we performed three offline analyses of the GW events reported in GWTC-2.1 and GWTC-3. The first analysis searched for neutrino emission within a $\pm$500~s time window centered around the GW merger time. Both the UML and LLAMA methods performed this search and no significant neutrino emission was observed in either search.

The second analysis was a 2-week follow-up of all BNS and NSBH candidate events with the UML search. All the GW events followed up in this analysis had at least one progenitor object with a mass estimate of $<$3~M$_{\odot}$. No significant neutrino emission was observed and 90\% ULs were placed on the time-integrated neutrino flux from each source. 

The third analysis searched for neutrino emission from the potential optical counterpart of the BBH merger GW190521 reported by ZTF. The UML analysis tested a time window of 112~days following the GW merger time which covers the entire flare in the optical light curve. The UML analysis assumed a  uniform neutrino emission within the time window. The LLAMA analysis assumed linearly decreasing neutrino emission in a 57 day time window according to the contemplated emission scenario for the optical flare. No significant neutrino emission was observed in both analysis methods and we derived 90\% ULs on the time-integrated flux and the $E_{\mathrm{iso}}$ from the AGN J124942.3+344929.

Apart from the analyses presented here, there also exists a gravitational wave follow-up analysis with neutrinos of a few 10 -100s of GeV energies detected by IceCube \citep{2022icrc.confE.939B}. This upcoming analysis will provide additional information, complimentary to the analyses with high-energy neutrinos presented here. Additionally, a search for extremely low energy neutrinos, with 0.5-5 GeV energies, from IceCube was conducted, and found no significant emission of neutrinos \citep{IceCube:2021ddq}.

The low-latency and archival searches will continue to function during the upcoming O4 run of LVK. It is expected that the O4 operational run will demonstrate enhanced performance, thereby increasing the rate of expected mergers. This would provide more opportunities to conduct multi-messenger studies which may lead to a potential discovery of neutrino and gravitational wave correlations. Additionally, the inclusion of more detectors from LVK will reduce the area of the sky localizations of the GW skymaps. This is also expected to contribute towards higher significances in case of coincident detections \citep{KAGRA:2013rdx}.

Future GW detectors like the Einstein Telescope \citep{et_science_team_2011} and Cosmic Explorer \citep{2019BAAS...51g..35R} aim at achieving improved sensitivities and lowering their frequency regime of operation. These improved detectors are expected to enhance the rate of observed merger events with better precision, which will in turn boost the capabilities of multi-messenger observations of these sources \citep{2019BAAS...51c.239K}. The next generation of the IceCube Neutrino Observatory, IceCube-Gen2, is planned to be an 8-fold extension to the instrumented volume of the current detector array. It is expected to extend the current energy range of IceCube to several 100s of PeV \citep{2021JPhG...48f0501A}. IceCube-Gen2 can potentially help in addressing the question of joint emission of neutrinos and GWs, when used in tandem with the future GW detectors.

\begin{table*}
    
    \begin{tabular}{|ccc|cc|ccc|}
    \hline
    \multicolumn{3}{|c}{GWTC-2.1}&\multicolumn{2}{|c}{ LLAMA } &\multicolumn{3}{|c|}{UML} \\ \hline
      &  & Area &  & $E{^2F}$ UL &  & $E{^2F}$ UL &  \\
    \multirow{-2}{*}{Event} & \multirow{-2}{*}{Type} & {[deg$^2$]} & \multirow{-2}{*}{$p$-value} & [GeVcm$^{-2}$] & \multirow{-2}{*}{$p$-value} & [GeVcm$^{-2}$] & \multirow{-2}{*}{$E_{\rm iso}$ UL [erg]}\\ \hline
    GW190403\_051519  & BBH  & 5589.4& 0.51  & 0.14   & 0.46  & 0.101  & 1.86 $\times$ 10$^{55}$ \\ \hline
    GW190408\_181802  & BBH  & 148.8 & 0.22  & 0.048 & 0.17 & 0.0512 &  4.85 $\times$ 10$^{53}$ \\ \hline
    GW190412          & BBH  & 20.9 & 0.27  & 0.041 & 0.13 & 0.0459 &  8.31 $\times$ 10$^{52}$ \\ \hline
    GW190413\_052954  & BBH  & 1484.5 & 0.30  & 0.087 & 0.28 & 0.133  &  7.01 $\times$ 10$^{54}$ \\ \hline
    GW190413\_134308  & BBH  &730.6 & 0.27  & 0.34  & 0.34 & 0.270  &  2.84 $\times$ 10$^{55}$ \\ \hline
    GW190421\_213856  & BBH  & 1211.5& 0.81  & 0.46  & 0.56 & 0.393  &  1.40 $\times$ 10$^{55}$ \\ \hline
    GW190425          & BNS  &9958.2 & 0.16  & 0.22  & 0.94 & 0.176  &  1.66 $\times$ 10$^{52}$ \\ \hline
    GW190426\_190642  & BBH  & 8214.5& 0.42  & 0.17  & 0.18 & 0.282  &  1.25 $\times$ 10$^{55}$ \\ \hline
    GW190503\_185404  & BBH  & 94.4& 0.94  & 0.54  & 0.34 & 0.584  &  4.99 $\times$ 10$^{54}$ \\ \hline
    GW190512\_180714  & BBH  & 218.0& 0.81  & 0.23  & 0.85 & 0.199  &  1.74 $\times$ 10$^{54}$ \\ \hline
    GW190513\_205428  & BBH  &518.4 & 0.99  & 0.043 & 0.94 & 0.0514 &  6.73 $\times$ 10$^{53}$ \\ \hline
    GW190514\_065416  & BBH  & 3009.7& 0.25  & 0.089 & 0.44 & 0.0453 &  3.96 $\times$ 10$^{54}$ \\ \hline
    GW190517\_055101  & BBH  & 473.3& 0.21  & 0.48  & 0.26 & 0.366  &  6.05 $\times$ 10$^{54}$ \\ \hline
    GW190519\_153544  & BBH  & 857.1& 0.067 & 0.15  & 0.21 & 0.0914 &  3.20 $\times$ 10$^{54}$ \\ \hline
    GW190521          & BBH  & 1008.2& 0.62  & 0.37  & 0.63 & 0.359  &  1.90 $\times$ 10$^{55}$ \\ \hline
    GW190521\_074359  & BBH  &546.5 & 0.11  & 0.049 & 0.15 & 0.0451 &  2.36 $\times$ 10$^{53}$ \\ \hline
    GW190527\_092055  & BBH  &3662.4 & 0.65  & 0.41  & 0.88 & 0.326  &  1.01 $\times$ 10$^{55}$ \\ \hline
    GW190602\_175927  & BBH  & 694.5& 0.31  & 0.34  & 0.17 & 0.370  &  9.73 $\times$ 10$^{54}$ \\ \hline
    GW190620\_030421  & BBH  &7202.1 & 0.20  & 0.36  & 0.23 & 0.121  &  4.13 $\times$ 10$^{54}$ \\ \hline
    GW190630\_185205  & BBH  & 1216.9& 0.64  & 0.15  & 0.81 & 0.427  &  5.31 $\times$ 10$^{53}$ \\ \hline
    GW190701\_203306  & BBH  & 46.1& 1.0   & 0.039 & 0.87 & 0.0385 &  7.65 $\times$ 10$^{53}$ \\ \hline
    GW190706\_222641  & BBH  & 653.8& 0.99  & 0.036 & 0.92 & 0.0356 &  3.17 $\times$ 10$^{54}$ \\ \hline
    GW190707\_093326  & BBH  & 1346.& 0.43  & 0.24  & 0.63 & 0.202  &  4.74 $\times$ 10$^{53}$ \\ \hline
    GW190708\_232457  & BBH  & 13675.4& 0.11  & 0.11  & 0.56 & 0.0720 &  1.62 $\times$ 10$^{53}$ \\ \hline
    GW190719\_215514  & BBH  &2890.1 & 0.83  & 0.054 & 0.91 & 0.0512 &  4.90 $\times$ 10$^{54}$ \\ \hline
    GW190720\_000836  & BBH  &463.4 & 0.99  & 0.13  & 0.94 & 0.0872 &  5.34 $\times$ 10$^{53}$ \\ \hline
    GW190725\_174728  & BBH  & 2292.5&  0.048  & 0.19  & 0.59 & 0.0918 &  4.04 $\times$ 10$^{53}$ \\ \hline
    GW190727\_060333  & BBH  &833.8 & 0.89  & 0.38  & 0.74 & 0.324  &  1.53 $\times$ 10$^{55}$ \\ \hline
    GW190728\_064510  & BBH  & 395.5& 0.0084 & 0.89  & 0.04 & 0.315  &  6.36 $\times$ 10$^{53}$ \\ \hline
    GW190731\_140936  & BBH  & 3387.3& 0.25  & 0.93  & 0.61 & 0.385  &  1.81 $\times$ 10$^{55}$ \\ \hline
    GW190803\_022701  & BBH  & 1519.5& 0.31  & 0.037 & 0.64 & 0.0354 &  1.69 $\times$ 10$^{54}$ \\ \hline
    GW190805\_211137  & BBH  & 3949.1& 0.74  & 0.20  & 0.93 & 0.180  &  2.56 $\times$ 10$^{55}$ \\ \hline
    GW190814          & BBH*  & 19.3& 1.0   & 0.24  & 1.0  & 0.259  &  5.68 $\times$ 10$^{52}$ \\ \hline
    GW190828\_063405  & BBH  &520.1 & 0.93  & 0.21  & 0.98 & 0.178  &  2.74 $\times$ 10$^{54}$ \\ \hline
    GW190828\_065509  & BBH  & 664.0& 0.84  & 0.38  & 0.84 & 0.368  &  3.73 $\times$ 10$^{54}$ \\ \hline
    GW190910\_112807  & BBH  & 10880.3& 0.22  & 0.45  & 0.77 & 0.177  &  1.90 $\times$ 10$^{54}$ \\ \hline
    GW190915\_235702  & BBH  & 396.9& 0.56  & 0.036 & 0.44 & 0.0354 &  3.61 $\times$ 10$^{53}$ \\ \hline
    GW190916\_200658  & BBH  & 4499.2& 0.52  & 0.16  & 0.85 & 0.108  &  1.22 $\times$ 10$^{55}$ \\ \hline
    GW190917\_114630  & NSBH* & 2050.6& 0.20  & 0.19  & 0.72 & 0.203  &  6.37 $\times$ 10$^{53}$ \\ \hline
    GW190924\_021846  & BBH  & 357.9& 0.031 & 0.037 & 0.23 & 0.0346 &  4.46 $\times$ 10$^{52}$ \\ \hline
    GW190925\_232845  & BBH  & 1233.5& 0.39   & 0.11  & 0.59 & 0.0908 &  3.41 $\times$ 10$^{53}$ \\ \hline
    GW190926\_050336  & BBH  & 2505.9&  0.13  & 0.78  & 0.33 & 0.280  &  2.30 $\times$ 10$^{55}$ \\ \hline
    GW190929\_012149  & BBH  & 2219.3& 0.11 & 0.34  & 0.22 & 0.276  &  1.85 $\times$ 10$^{55}$ \\ \hline
    GW190930\_133541  & BBH  & 1679.6& 0.14  & 0.038 & 0.31 & 0.0427 &  1.05 $\times$ 10$^{53}$ \\ \hline
\end{tabular}
\caption{Results for the events in GWTC-2.1 \citep{LIGOScientific:2021usb} for the 1000~s follow-up. GW190814 is labelled as a BBH merger here although the type of the lighter object with $\sim2.6$ M$_\odot$ is unknown  \citep{Abbott:2020khf}. GW190917\_114630 is labelled as NSBH since its estimated source properties are more like that of an NSBH event although the event was found to be significant by a BBH template. The table also shows the area on the sky containing 90\% probabilities from the GW skymap.}
\label{tab:results}
\end{table*}

\begin{table*}
    
    \begin{tabular}{|ccc|cc|ccc|}
    \hline
    \multicolumn{3}{|c}{GWTC-3}&\multicolumn{2}{|c}{ LLAMA } &\multicolumn{3}{|c|}{UML} \\ \hline
     &  & Area &  & $E{^2F}$ UL &  & $E{^2F}$ UL &  \\
    \multirow{-2}{*}{Event} & \multirow{-2}{*}{Type} & {[deg$^2$]} & \multirow{-2}{*}{$p$-value} & [GeVcm$^{-2}$] & \multirow{-2}{*}{$p$-value} & [GeVcm$^{-2}$] & \multirow{-2}{*}{$E_{\rm iso}$ UL [erg]}\\ \hline
    GW191103\_012549  & BBH  & 2519.6 & 0.53  & 0.049  &  0.71 &0.049  &  $1.96\,\times \,10^{53}$\\ \hline
    GW191105\_143521  &   BBH & 728.7 & 0.27  & 0.28  & 0.54 & 0.267 &  $1.28\,\times  \,10^{54}$\\ \hline
    GW191109\_010717&  BBH & 1784.3 & 0.14  &  0.48 & 0.05 & 0.508 &  $5.03\,\times\, 10^{54}$\\ \hline
    GW191113\_071753&  BBH & 2993.3 & 0.076  & 0.52  & 0.19 & 0.441 &  $3.12\,\times \,10^{54}$\\ \hline
    GW191126\_115259& BBH  & 1514.5 & 0.77  &  0.13 & 1.00 & 0.138 &  $1.42 \, \times \, 10^{54}$\\ \hline
    GW191127\_050227& BBH  & 1499.2 & 0.38 &  0.078 & 0.83 & 0.081 &  $2.96\,\times\,10^{54}$\\ \hline
    GW191129\_134029& BBH  & 848.3 & 0.25  &  0.35 & 0.30 & 0.425 &  $8.95\,\times\,10^{53}$\\ \hline
    GW191204\_110529& BBH  & 4747.7 & 0.16  &  0.36 & 0.49 & 0.085 &  $1.46\,\times\,10^{54}$\\ \hline
    GW191204\_171526& BBH  & 344.9 &  0.97 &  0.26 & 1.00 & 0.280 &  $3.96\,\times\,10^{53}$\\ \hline
    GW191215\_223052& BBH  & 595.8 &  0.98 &  0.26 & 1.00 & 0.211 &  $2.98\,\times\,10^{54}$\\ \hline
    GW191216\_213338& BBH  & 480.1 &  0.0049 & 0.093  & 0.10 & 0.071 & $2.57\,\times\,10^{52}$ \\ \hline
    GW191219\_163120& NSBH  & 2232.1 &  0.09 & 0.26  & 0.71 & 0.219 &  $2.80\,\times\,10^{53}$\\ \hline
    GW191222\_033537& BBH  & 2299.2 &  0.95  & 0.36  & 1.00 & 0.375 & $1.1\,\times\,10^{55}$ \\ \hline
    GW191230\_180458& BBH  & 1012.2 & 0.37 &  0.36 & 0.28 & 0.488 & $3.18\,\times\,10^{55}$ \\ \hline
    GW200105\_162426& NSBH  & 7881.8 & 0.20  & 0.13  & 0.81 & 0.095 & $2.98\,\times\,10^{52}$ \\ \hline
   GW200112\_155838&  BBH &  4250.4 & 0.58 &  0.18 & 0.79 & 0.133 & $8.43\,\times\,10^{53}$ \\ \hline
    GW200115\_042309&NSBH   &  511.9 & 0.34 &  0.038 & 0.45 & 0.045 & $2.12\,\times\,10^{52}$ \\ \hline
    GW200128\_022011& BBH  & 2677.5 & 0.46 &  0.25 & 0.47 & 0.243 &  $9.31\,\times\,10^{54}$\\ \hline
    GW200129\_065458& BBH  & 81.8 & 0.033  &  0.041 & 0.05 & 0.406 & $1.73\,\times\,10^{53}$ \\ \hline
   GW200202\_154313& BBH  & 159.3 &  0.0057 &  0.039 & 0.06 & 0.038 & $2.43\,\times\,10^{52}$ \\ \hline
    GW200208\_130117& BBH  & 38.0 &  0.94 &  0.33 & 1.00 & 0.518 & $9.25\,\times\,10^{54}$ \\ \hline
    GW200208\_222617& BBH  & 1889.2 &  0.41 &  0.045 & 0.90 & 0.043 & $4.98\,\times\,10^{54}$ \\ \hline
    GW200209\_085452& BBH  & 924.5 &  0.84 &  0.50 & 1.00 & 0.041 &  $1.81\,\times\,10^{54}$\\ \hline
    GW200210\_092254&BBH   & 1830.7 &  0.28 &  0.071 & 0.79 & 0.081 &  $2.51\,\times\,10^{53}$\\ \hline
    GW200216\_220804&BBH   & 3009.5 &  0.065 & 0.066  & 0.46 & 0.236 &  $2.82\,\times\,10^{54}$\\ \hline
    GW200219\_094415&BBH   & 702.1 &  0.98 &  0.23 & 1.00 & 0.035 &  $9.57\,\times\,10^{54}$\\ \hline
    GW200220\_061928&BBH   & 3484.7 &  0.23 &  0.22 & 0.05 & 0.357 & $4.23\,\times\,10^{55}$ \\ \hline
    GW200220\_124850& BBH  & 3168.9 &  0.42 &  0.13 & 0.53 & 0.118 & $6.31\,\times\,10^{54}$ \\ \hline
    GW200224\_222234&   BBH& 49.9 &  0.90 &  0.068 & 1.00 & 0.079 & $9.33\,\times\,10^{53}$ \\ \hline
    GW200225\_060421&   BBH& 509.0 & 0.0048 &  0.10 & 0.20 & 0.055 & $3.03\,\times\,10^{53}$ \\ \hline
    GW200302\_015811&   BBH& 7010.8 &  0.16 &  0.67 & 0.21 & 0.531 & $4.34\,\times\,10^{54}$ \\ \hline
   GW200306\_093714 &  BBH & 4371.2 &  0.15 &  0.074 & 0.57 & 0.046 & $9.99\,\times\,10^{53}$  \\ \hline
    GW200308\_173609  &  BBH & 18705.7 &  0.24 &  0.38 & 0.29 & 0.326 & $7.18\,\times\,10^{55}$ \\ \hline
    GW200311\_115853&  BBH & 35.6 & 1.0  &  0.047 & 1.00 & 0.076 & $4.38\,\times\,10^{53}$ \\ \hline
    GW200316\_215756&  BBH & 410.4 & 0.17 &   0.066& 0.04 & 0.110 & $5.19\,\times\,10^{53}$ \\ \hline
    GW200322\_091133  &  BBH & 31571.1 &  0.23 &  0.18 & 0.87 & 0.148 & $4.39\,\times\,10^{55}$ \\ \hline
\end{tabular}
\caption{Results for the events in GWTC-3 \citep{LIGOScientific:2021djp} for the 1000~s follow-up. The central 68\% energy range of the events contributing to the limits shown here ranges from $5\,\times$ 10$^5$~GeV~-~10$^7$~GeV in the southern hemisphere and $5\,\times$ 10$^3$~GeV~-~10$^5$~GeV in the northern hemisphere.}
\label{tab:results2}
\end{table*}

\begin{table*}
    \centering
    \begin{tabular}{|c|c|c|c|}
    \hline
  Event & Type &  $p$-value &  $E{^2F}$ UL [GeVcm$^{-2}$] \\ \hline
  GW190425         & BNS   &   0.43 & 0.661  \\ \hline
  GW190917\_114630  & NSBH  &   0.84 & 0.442  \\ \hline
  GW190814         & BBH   &   0.59 & 0.309  \\ \hline
  GW191219\_163120  & NSBH   &   0.67 &  0.347  \\ \hline
  GW200105\_162426  & NSBH   &   0.47 &  0.382  \\ \hline
  GW200115\_042309  & NSBH   &   0.68 &  0.078  \\ \hline
  GW200210\_092254  & NSBH   &   0.13 &  0.303  \\ \hline
    \end{tabular}
    \caption{Results for the 2 week follow-up analysis using the UML method. 3 events from GWTC-2.1 \citep{LIGOScientific:2021usb} and 4 events from GWTC-3 \citep{LIGOScientific:2021djp} were followed up as they were the only potential BNS/NSBH candidates.}
    \label{tab:2week_results}
\end{table*}

\section*{Acknowledgements}
The IceCube collaboration acknowledges the significant contributions to this manuscript from Aswathi Balagopal V., Raamis Hussain and Do\u{g}a Veske.
The authors gratefully acknowledge the support from the following agencies and institutions: USA – U.S. National Science Foundation-Office of Polar Programs, U.S. National Science Foundation-Physics Division, U.S. National Science Foundation-EPSCoR, Wisconsin Alumni Research Foundation, Center for High Throughput Computing (CHTC) at the University of Wisconsin–Madison, Open Science Grid (OSG), Extreme Science and Engineering Discovery Environment (XSEDE), Frontera computing project at the Texas Advanced Computing Center, U.S. Department of Energy-National Energy Research Scientific Computing Center, Particle astrophysics research computing center at the University of Maryland, Institute for Cyber-Enabled Research at Michigan State University, and Astroparticle physics computational facility at Marquette University; Belgium – Funds for Scientific Research (FRS-FNRS and FWO), FWO Odysseus and Big Science programmes, and Belgian Federal Science Policy Office (Belspo); Germany – Bundesministerium für Bildung und Forschung (BMBF), Deutsche Forschungsgemeinschaft (DFG), Helmholtz Alliance for Astroparticle Physics (HAP), Initiative and Networking Fund of the Helmholtz Association, Deutsches Elektronen Synchrotron (DESY), and High Performance Computing cluster of the RWTH Aachen; Sweden – Swedish Research Council, Swedish Polar Research Secretariat, Swedish National Infrastructure for Computing (SNIC), and Knut and Alice Wallenberg Foundation; Australia – Australian Research Council; Canada – Natural Sciences and Engineering Research Council of Canada, Calcul Québec, Compute Ontario, Canada Foundation for Innovation, WestGrid, and Compute Canada; Denmark – Villum Fonden and Carlsberg Foundation; New Zealand – Marsden Fund; Japan – Japan Society for Promotion of Science (JSPS) and Institute for Global Prominent Research (IGPR) of Chiba University; Korea – National Research Foundation of Korea (NRF); Switzerland – Swiss National Science Foundation (SNSF); United Kingdom – Department of Physics, University of Oxford. 

This research has made use of data or software obtained from the Gravitational Wave Open Science Center (gw-openscience.org) \citep{RICHABBOTT2021100658}, a service of LIGO Laboratory, the LIGO Scientific Collaboration, the Virgo Collaboration, and KAGRA. LIGO Laboratory and Advanced LIGO are funded by the United States National Science Foundation (NSF) as well as the Science and Technology Facilities Council (STFC) of the United Kingdom, the Max-Planck-Society (MPS), and the State of Niedersachsen/Germany for support of the construction of Advanced LIGO and construction and operation of the GEO600 detector. Additional support for Advanced LIGO was provided by the Australian Research Council. Virgo is funded, through the European Gravitational Observatory (EGO), by the French Centre National de Recherche Scientifique (CNRS), the Italian Istituto Nazionale di Fisica Nucleare (INFN) and the Dutch Nikhef, with contributions by institutions from Belgium, Germany, Greece, Hungary, Ireland, Japan, Monaco, Poland, Portugal, Spain. The construction and operation of KAGRA are funded by Ministry of Education, Culture, Sports, Science and Technology (MEXT), and Japan Society for the Promotion of Science (JSPS), National Research Foundation (NRF) and Ministry of Science and ICT (MSIT) in Korea, Academia Sinica (AS) and the Ministry of Science and Technology (MoST) in Taiwan.

\bibliography{main}{}
\bibliographystyle{aasjournal}

\appendix
\vspace{-0.36cm}
\section{Skymaps}
\begin{figure}[hb!]
\vspace{-0.2cm}
\gridline{\fig{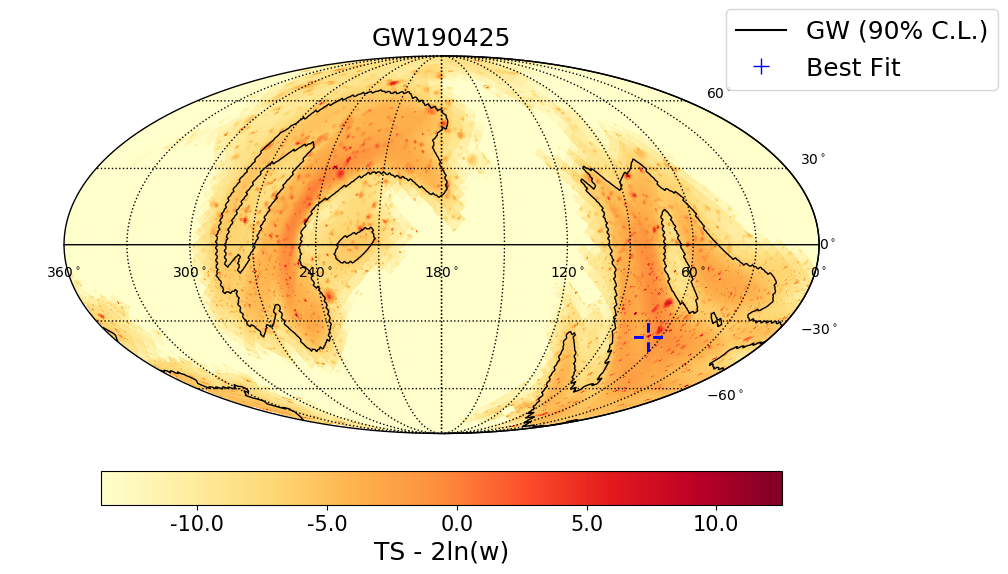}{0.34\textwidth}{(1)}
    \fig{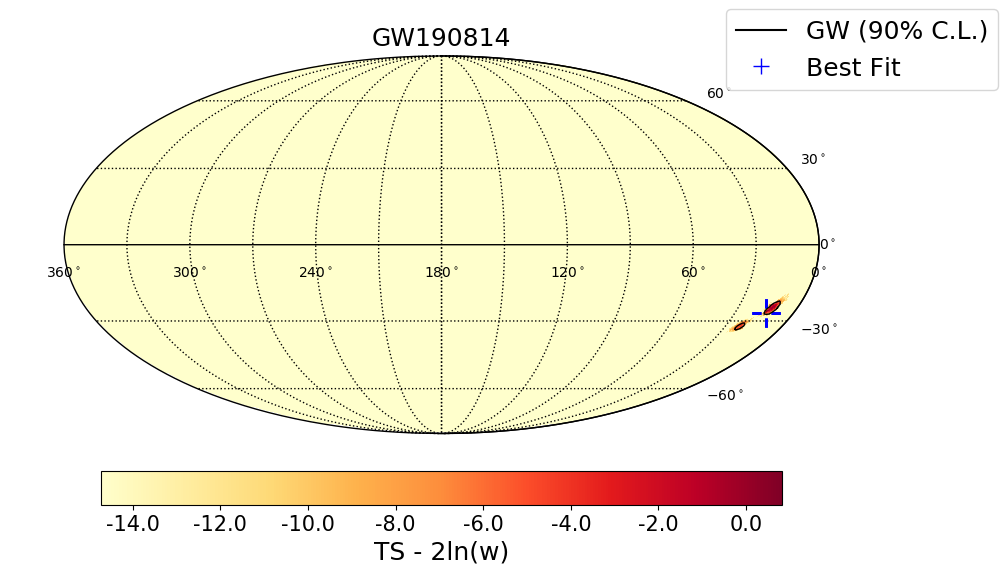}{0.34\textwidth}{(2)}}
    \vspace{-0.2cm}
\gridline{\fig{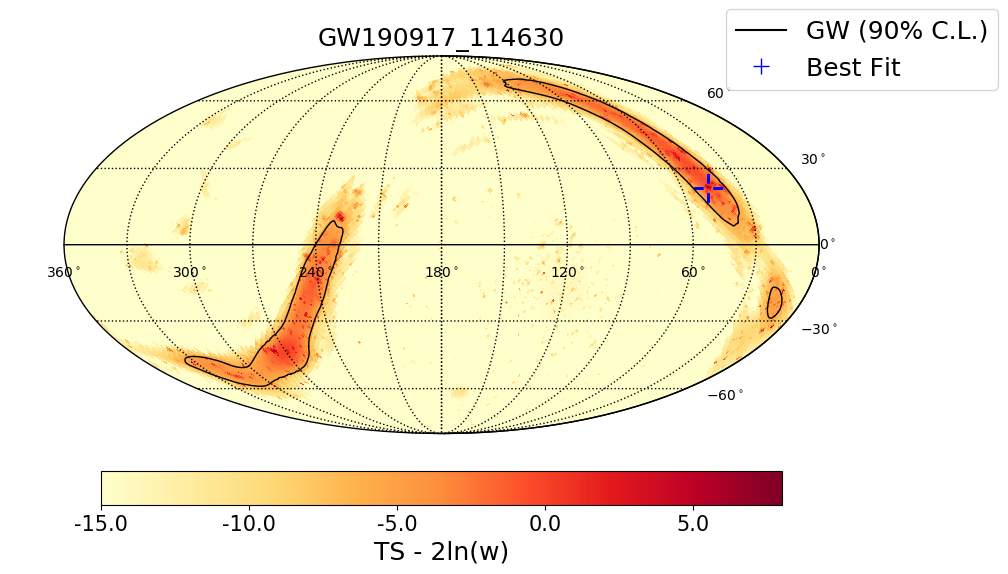}{0.34\textwidth}{(3)}}
\vspace{-0.15cm}
\gridline{\fig{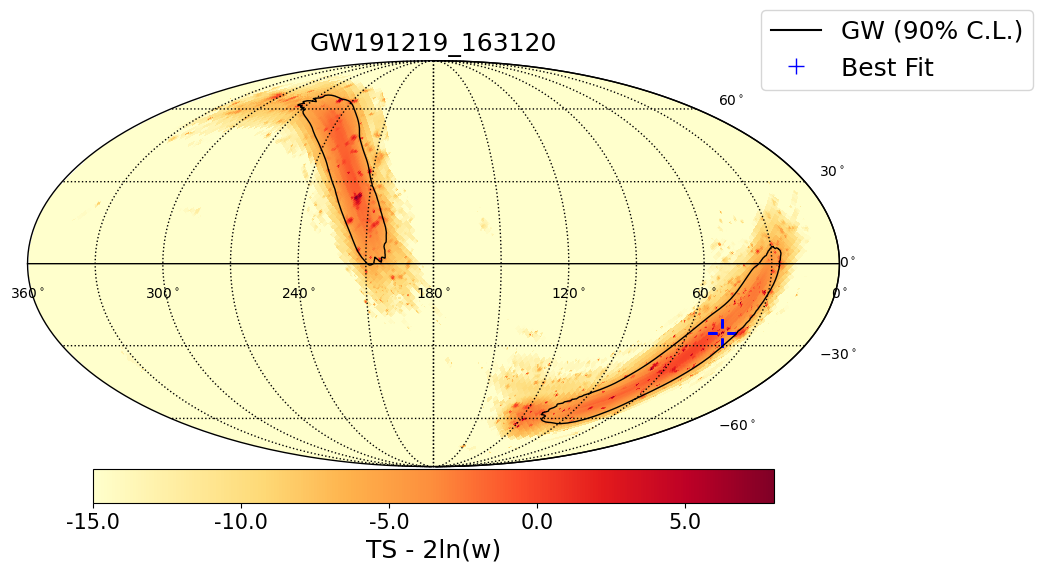}{0.34\textwidth}{(4)}
    \fig{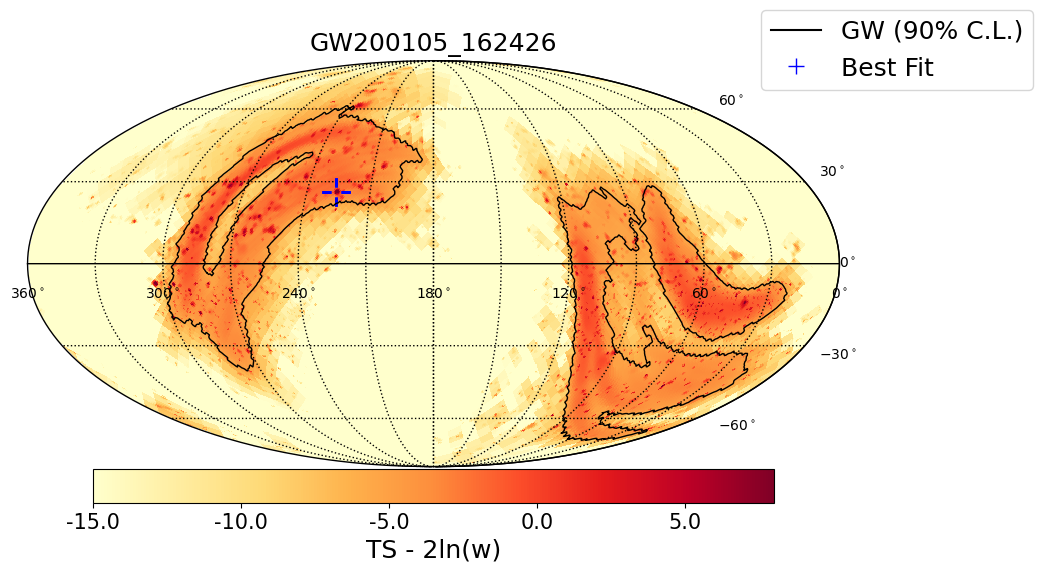}{0.34\textwidth}{(5)}}
    \vspace{-0.15cm}
\gridline{\fig{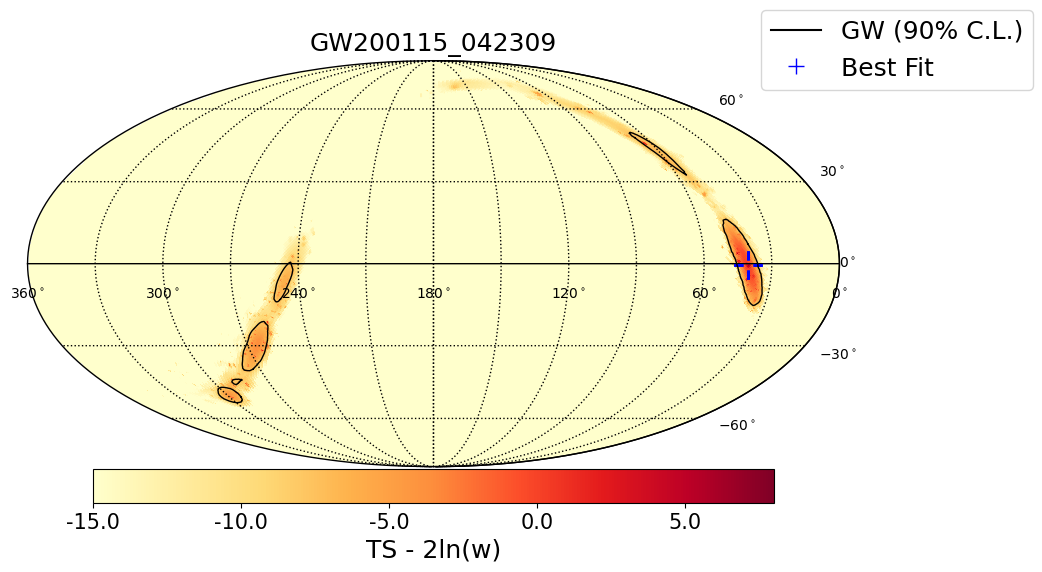}{0.34\textwidth}{(6)}
    \fig{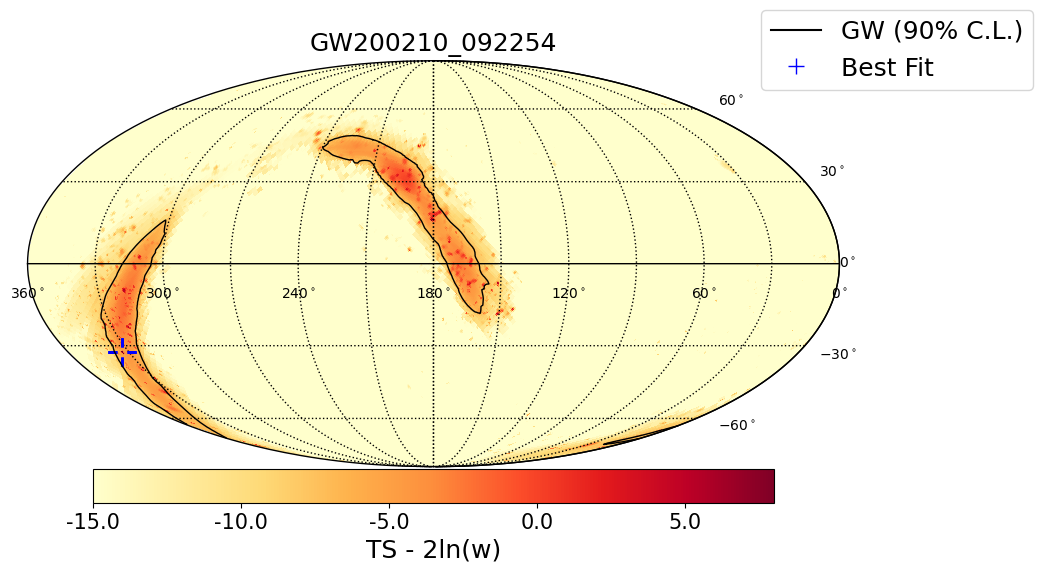}{0.34\textwidth}{(7)}}
\caption{Final test statistic maps for the 3 BNS and NSBH candidates in GWTC-2.1 \citep{LIGOScientific:2021usb}, and the 4 NSBH candidates in GWTC-3 \citep{LIGOScientific:2021djp}. The pixel with highest test statistic in the sky is shown in the blue crosshairs. The color scale shows the test statistic weighted by the GW localization information. Here $w$ = $P_{\mathrm{\rm GW}}(\Omega)/A_{\rm pix}$ where $P_{\rm GW}(\Omega)$ is the probability of the GW source being in a given pixel and $A_{\rm pix}$ is the pixel area.}
\label{fig:2week_ts_maps_gwtc2}
\end{figure}

This appendix includes the skymaps obtained in the context of this analysis. Figure \ref{fig:2week_ts_maps_gwtc2} shows the TS maps for the two-week follow-up analysis. Figures \ref{fig:1000s_skymaps} and \ref{fig:1000s_skymaps_gwtc3} show the skymaps with the GW probabilities and the observed neutrinos within the 1000~s time window in the archival search.
\vspace{-0.3cm}
\vspace{-1.26cm}
\begin{figure}[h!]
\gridline{\fig{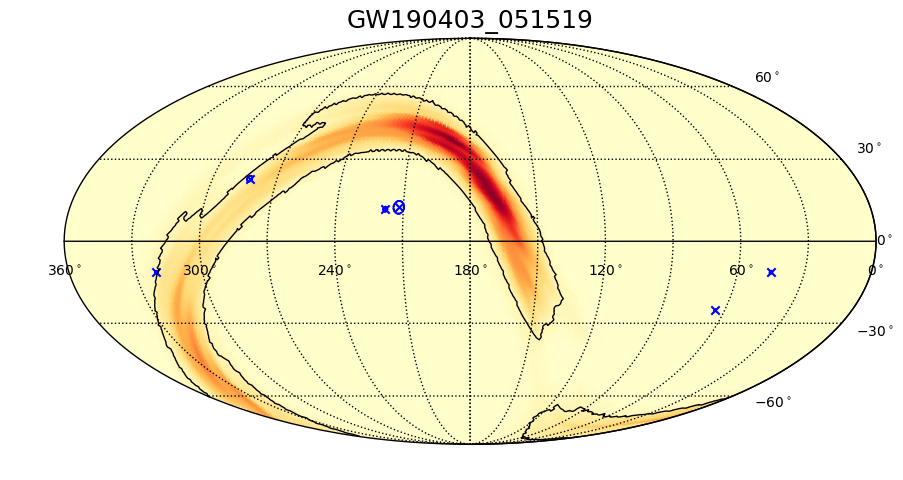}{0.31\textwidth}{(1)}
    \fig{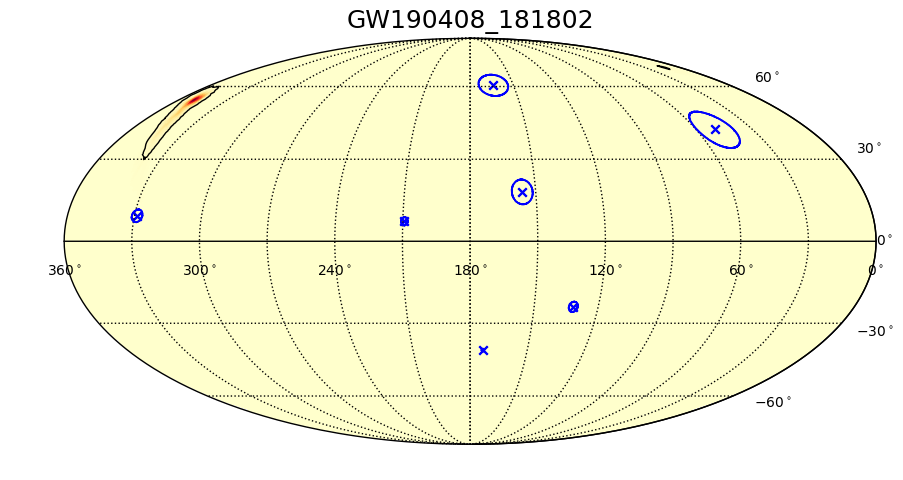}{0.31\textwidth}{(2)}
    \fig{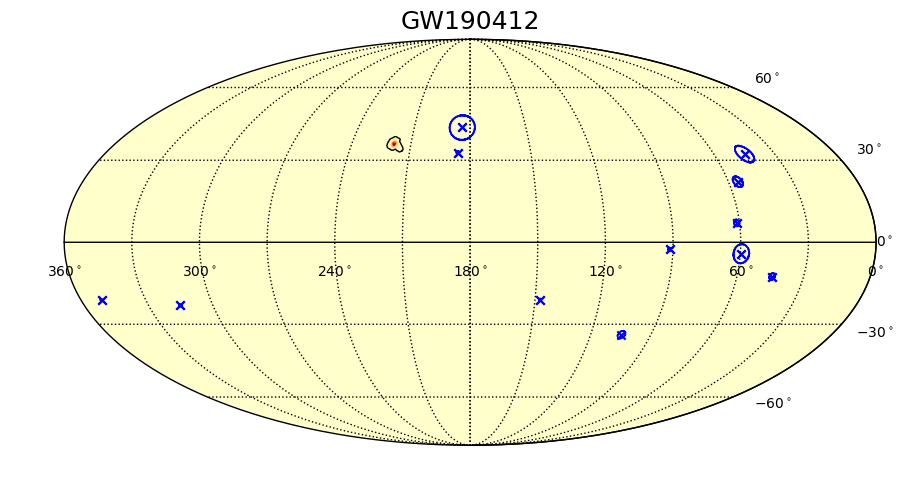}{0.31\textwidth}{(3)}}
\gridline{\fig{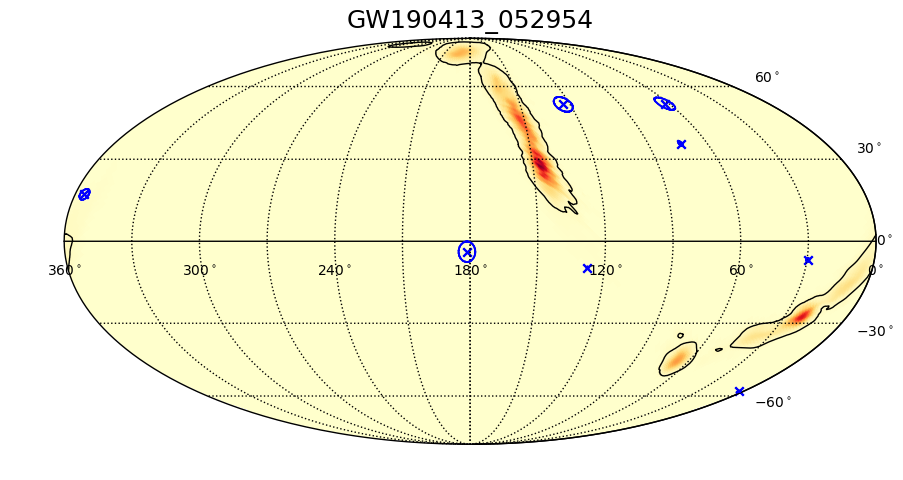}{0.31\textwidth}{(4)}
    \fig{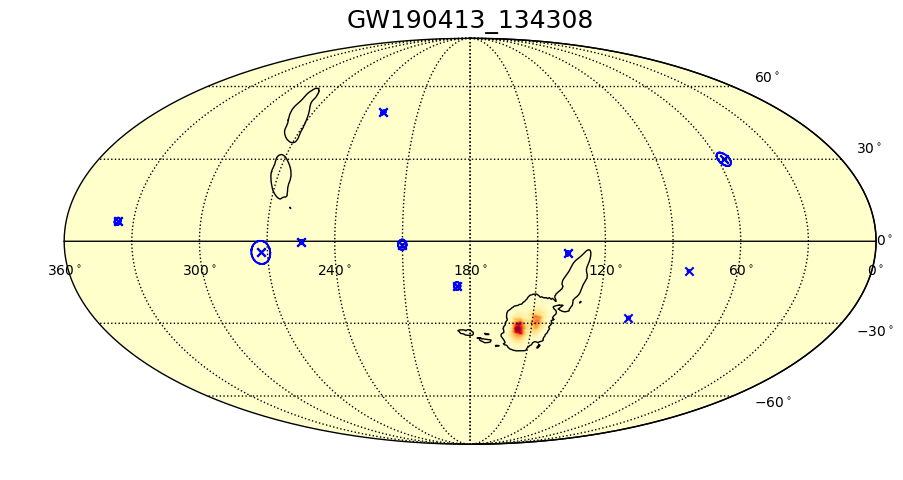}{0.31\textwidth}{(5)}
    \fig{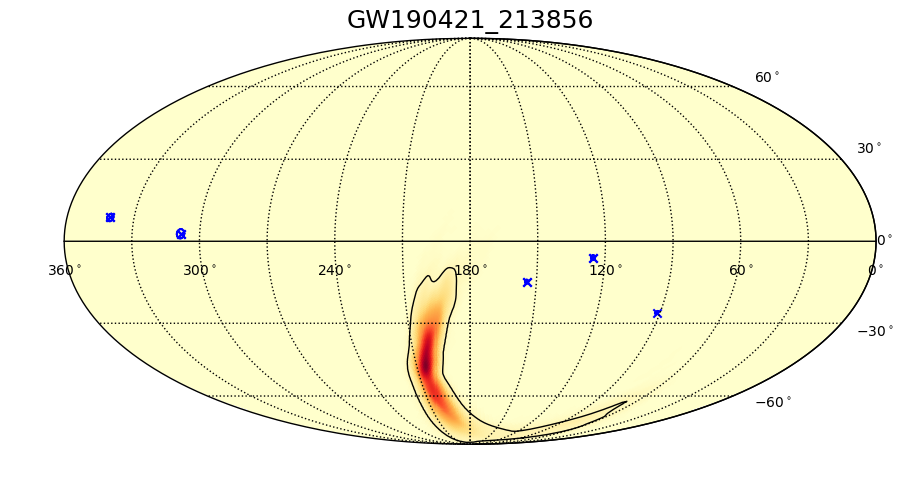}{0.31\textwidth}{(6)}}
\gridline{\fig{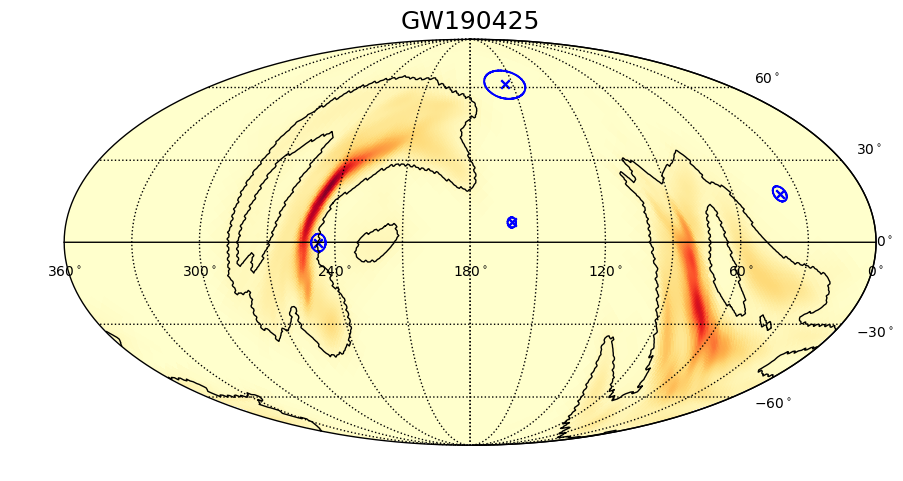}{0.31\textwidth}{(7)}
    \fig{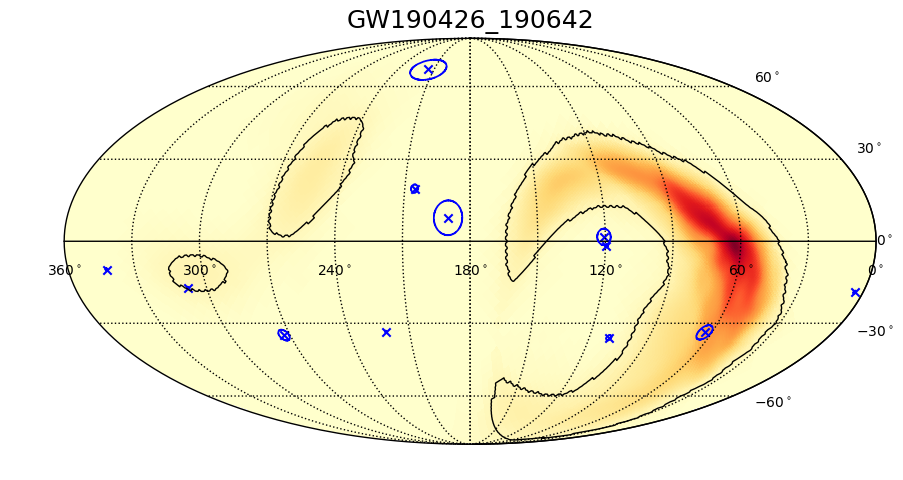}{0.31\textwidth}{(8)}
    \fig{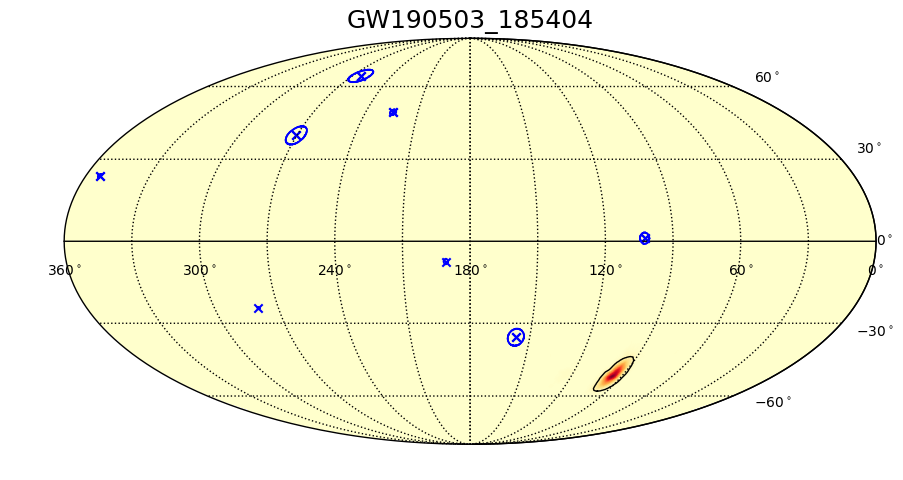}{0.31\textwidth}{(9)}}
\gridline{\fig{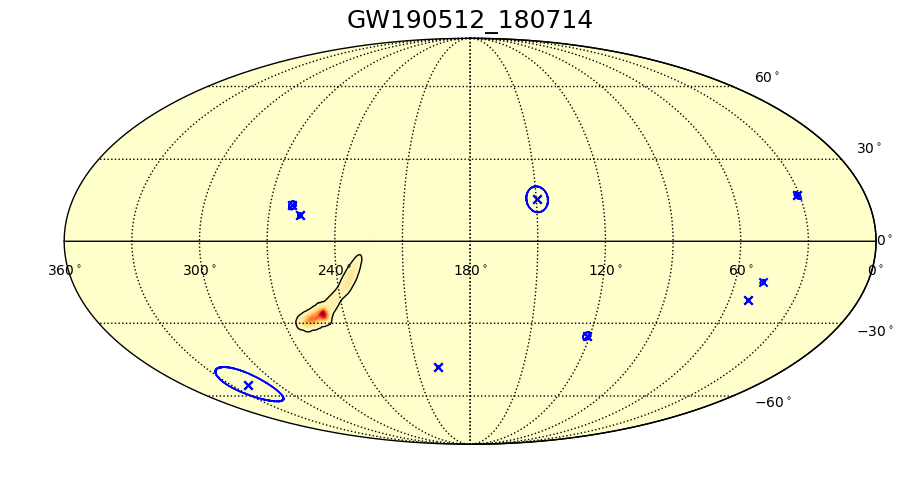}{0.31\textwidth}{(10)}
    \fig{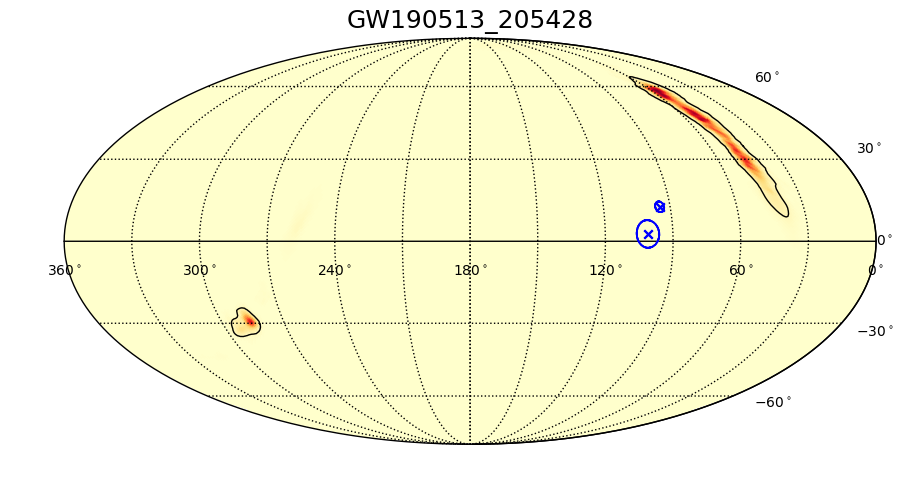}{0.31\textwidth}{(11)}
    \fig{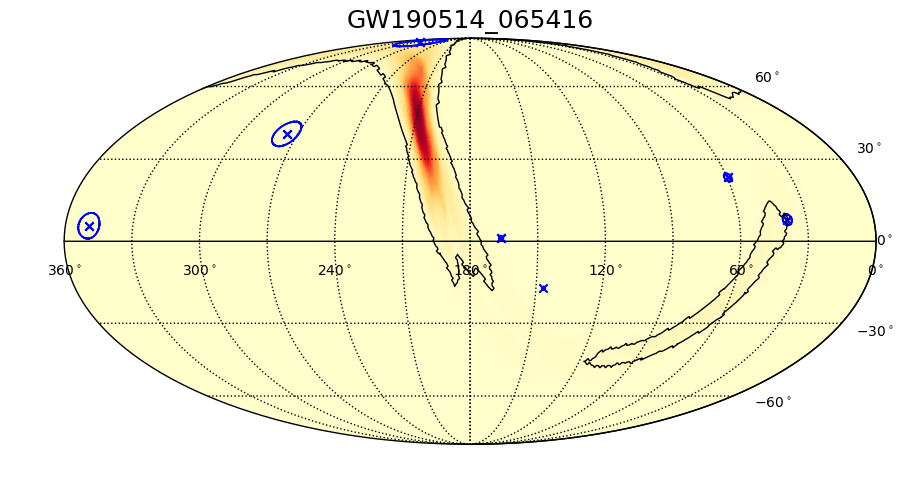}{0.31\textwidth}{(12)}}
\gridline{\fig{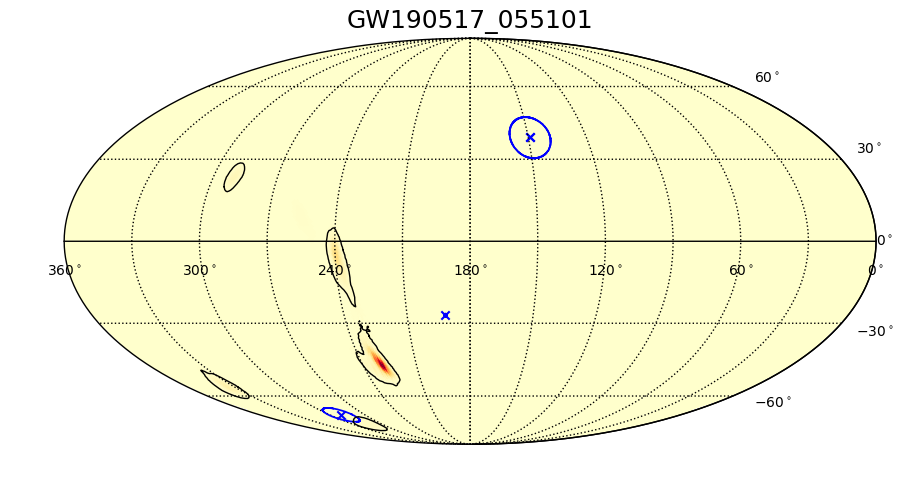}{0.31\textwidth}{(13)}
    \fig{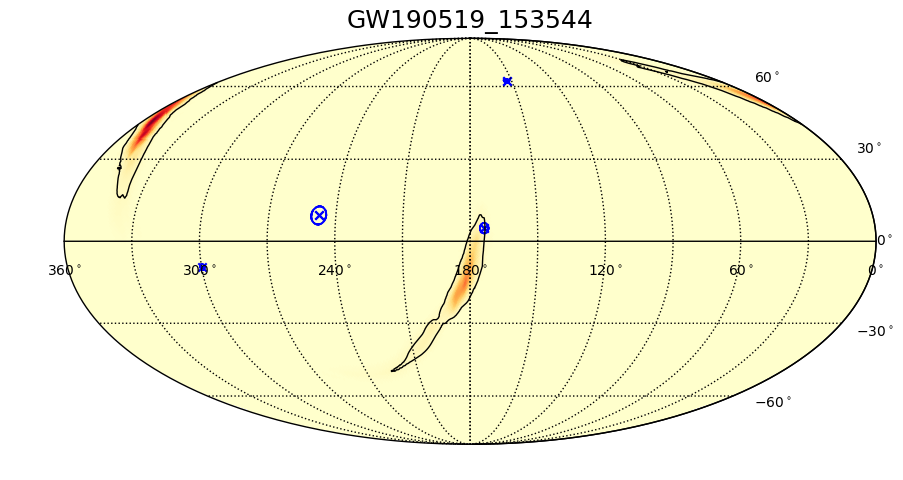}{0.31\textwidth}{(14)}
    \fig{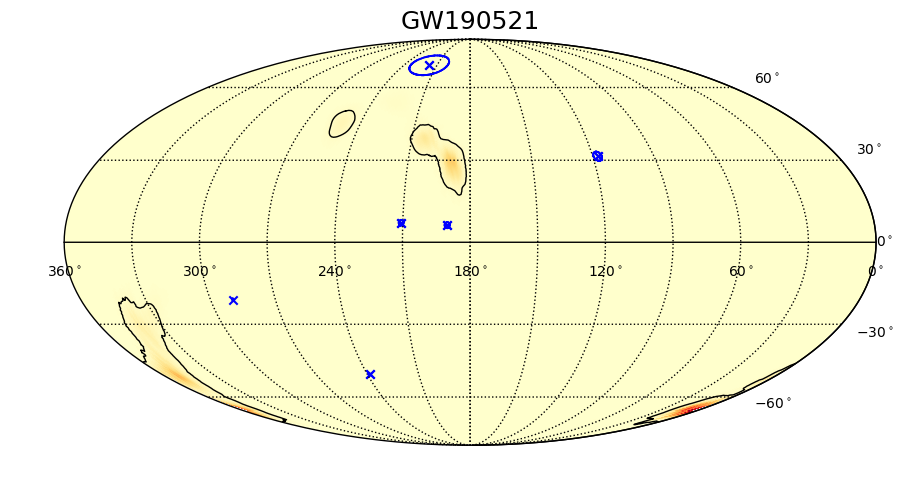}{0.31\textwidth}{(15)}}
\gridline{\fig{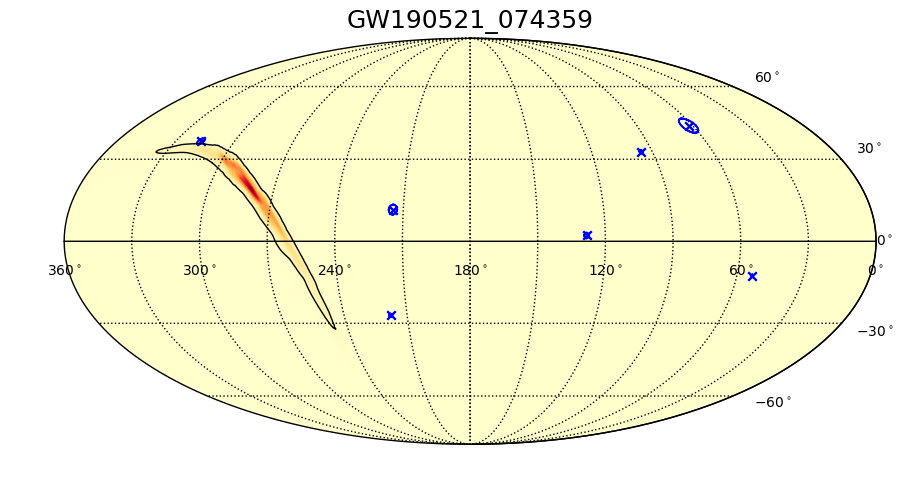}{0.31\textwidth}{(16)}
    \fig{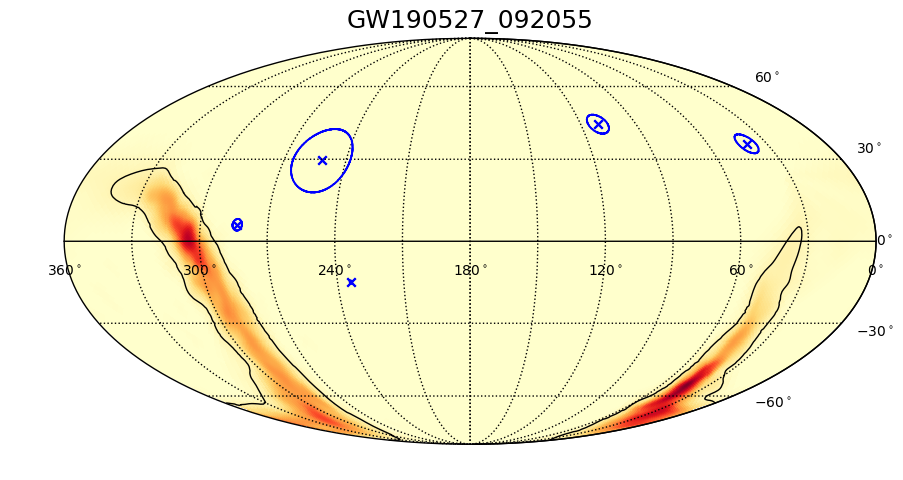}{0.31\textwidth}{(17)}
    \fig{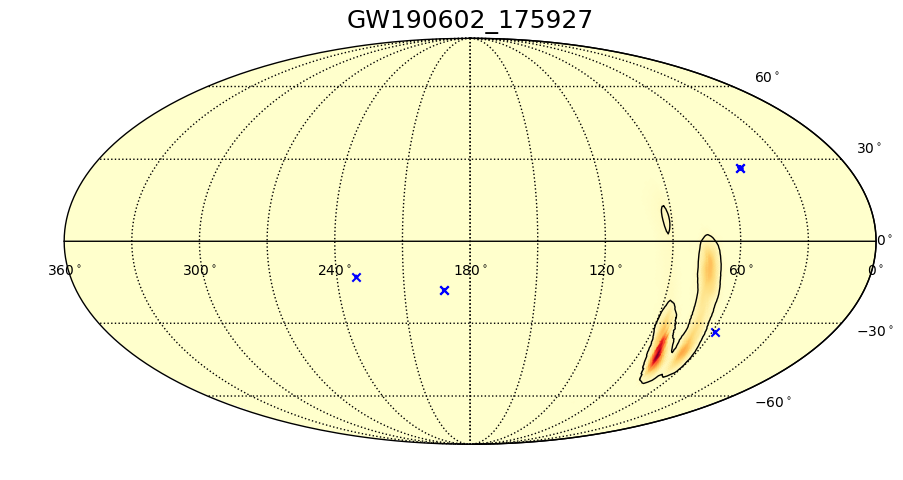}{0.31\textwidth}{(18)}}
\end{figure}
\begin{figure} 
\gridline{\fig{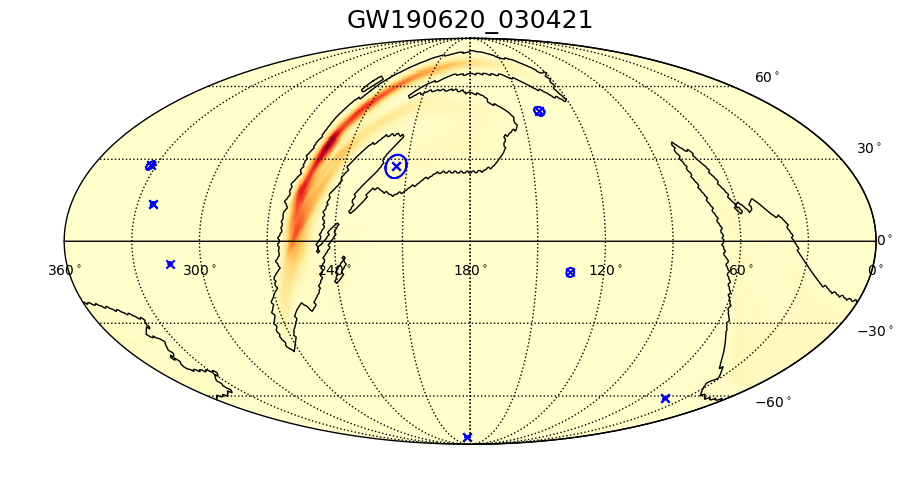}{0.31\textwidth}{(19)}
    \fig{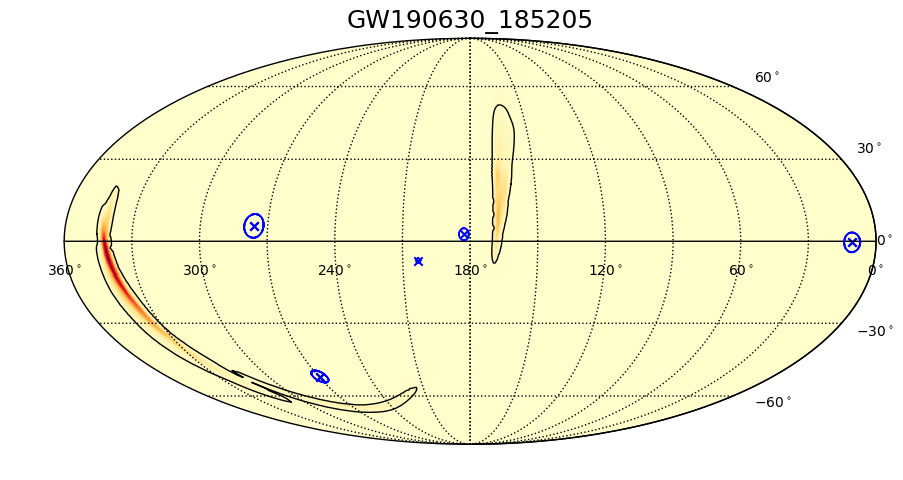}{0.31\textwidth}{(20)}
    \fig{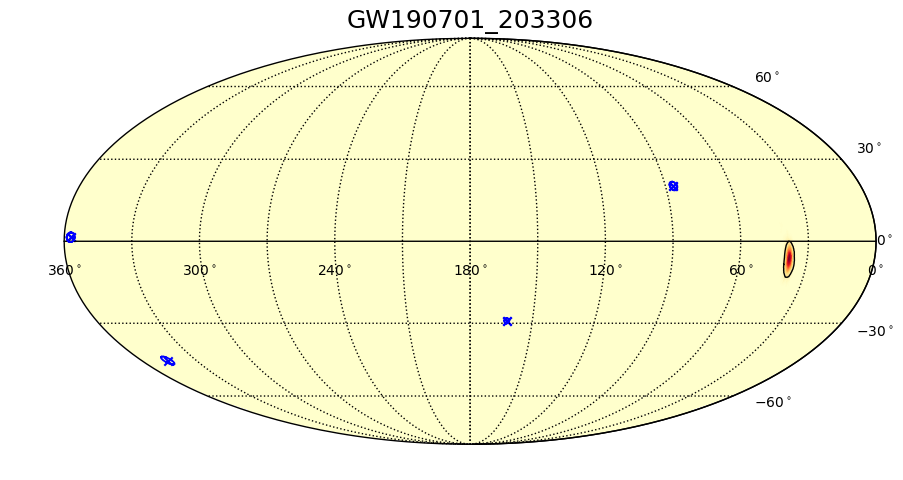}{0.31\textwidth}{(21)}}
\gridline{\fig{figures/skymaps/skymap_GW190620_030421.png}{0.31\textwidth}{(22)}
    \fig{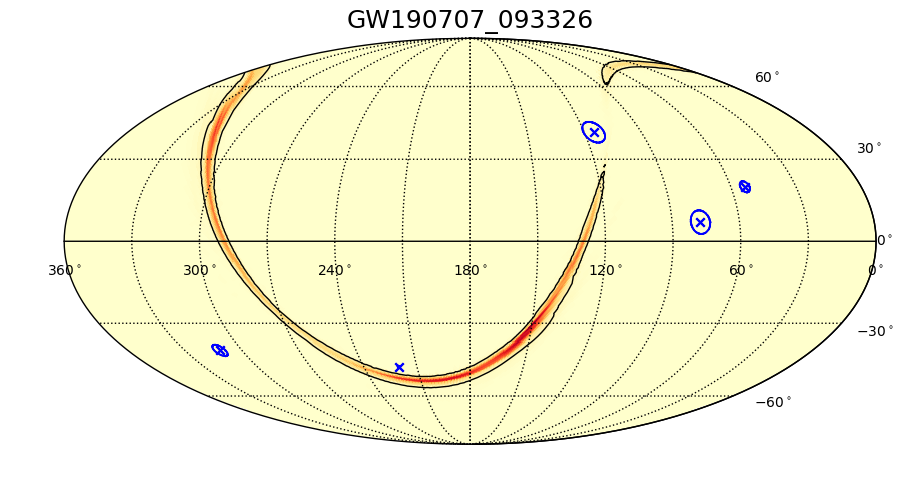}{0.31\textwidth}{(23)}
    \fig{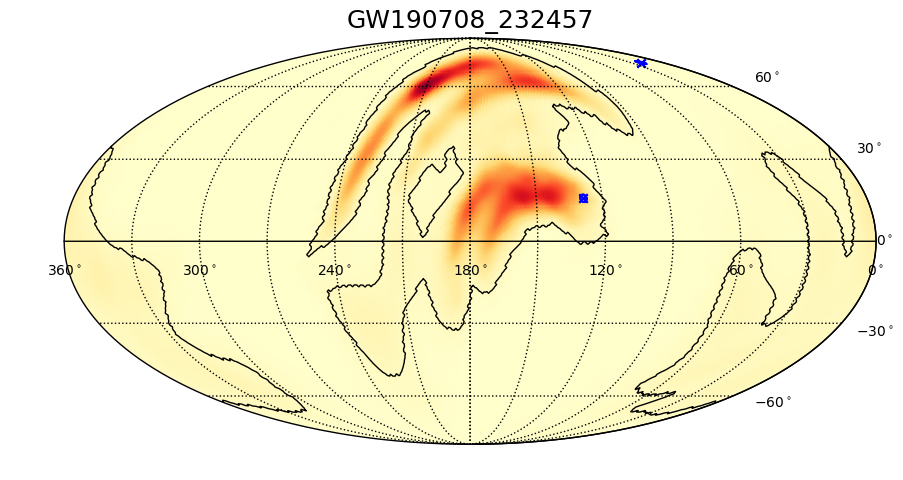}{0.31\textwidth}{(24)}}
\gridline{\fig{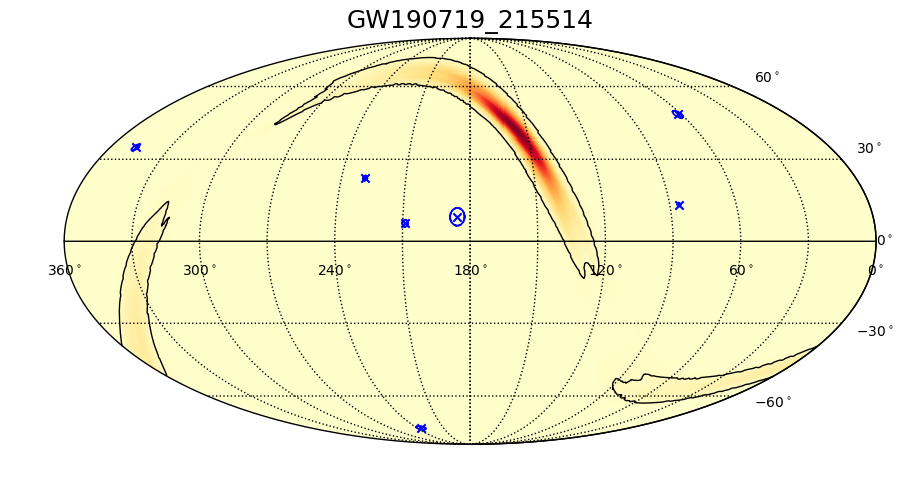}{0.31\textwidth}{(25)}
    \fig{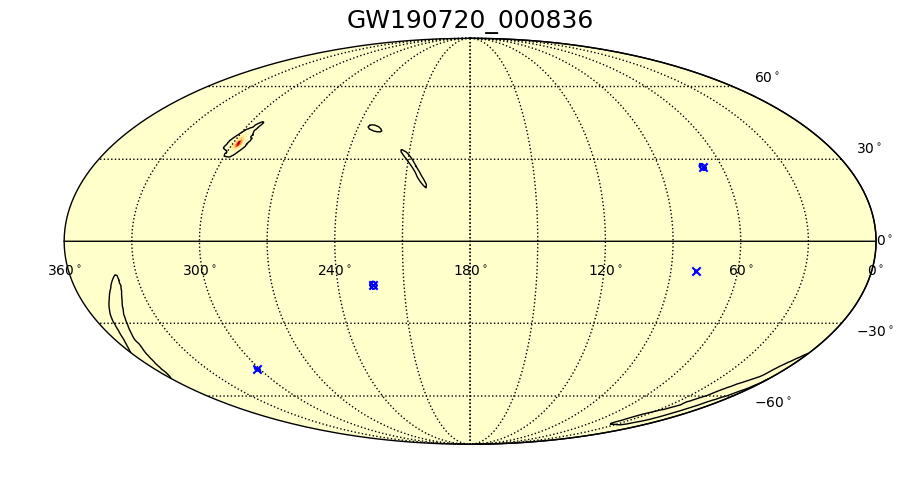}{0.31\textwidth}{(26)}
    \fig{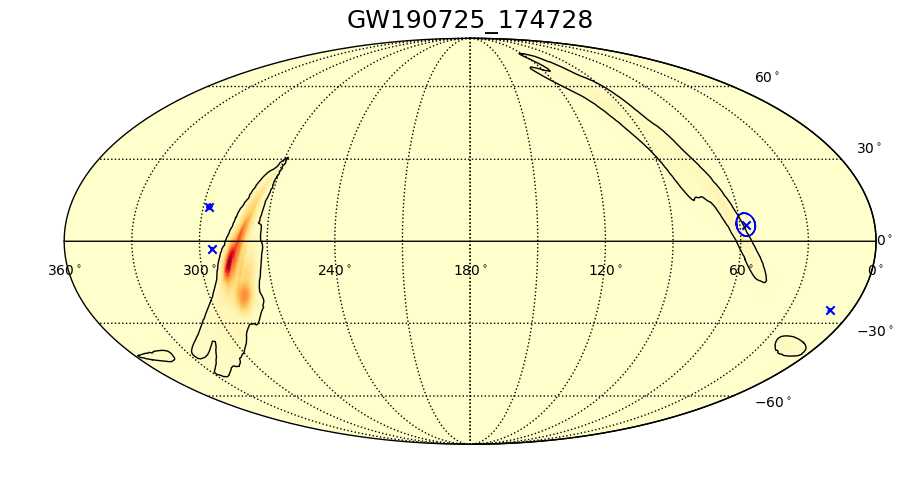}{0.31\textwidth}{(27)}}
\gridline{\fig{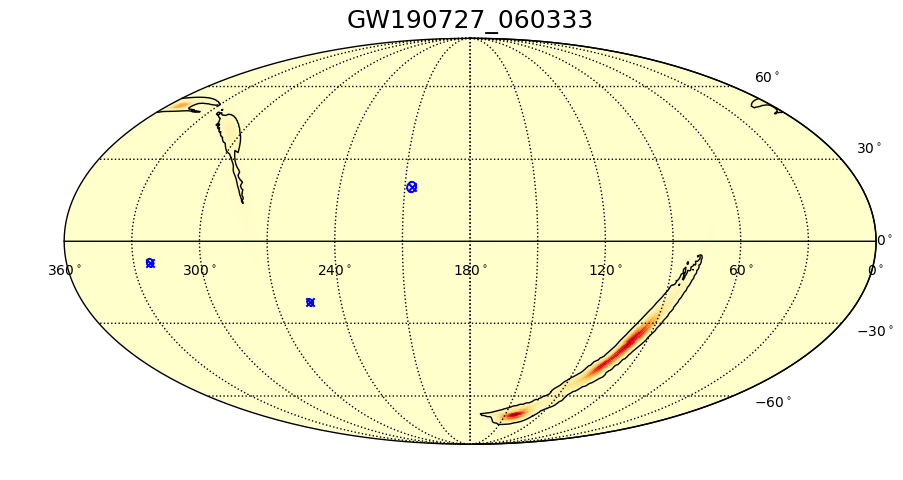}{0.31\textwidth}{(28)}
    \fig{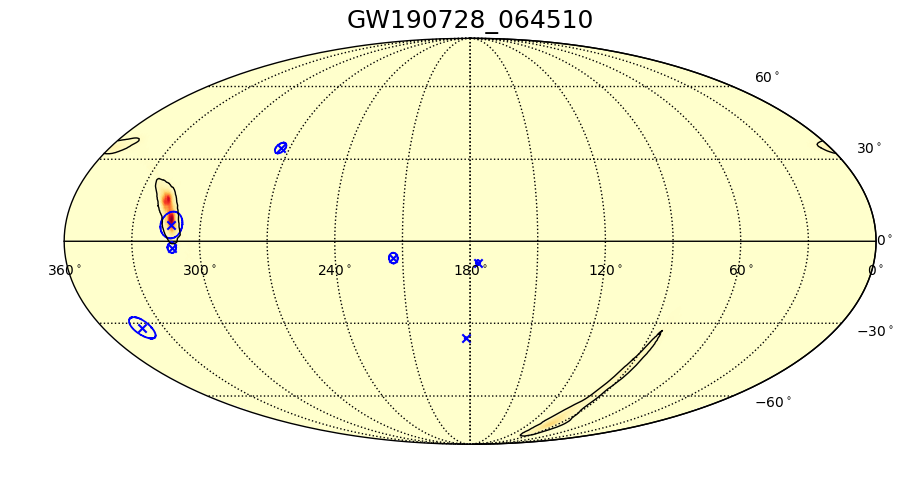}{0.31\textwidth}{(29)}
    \fig{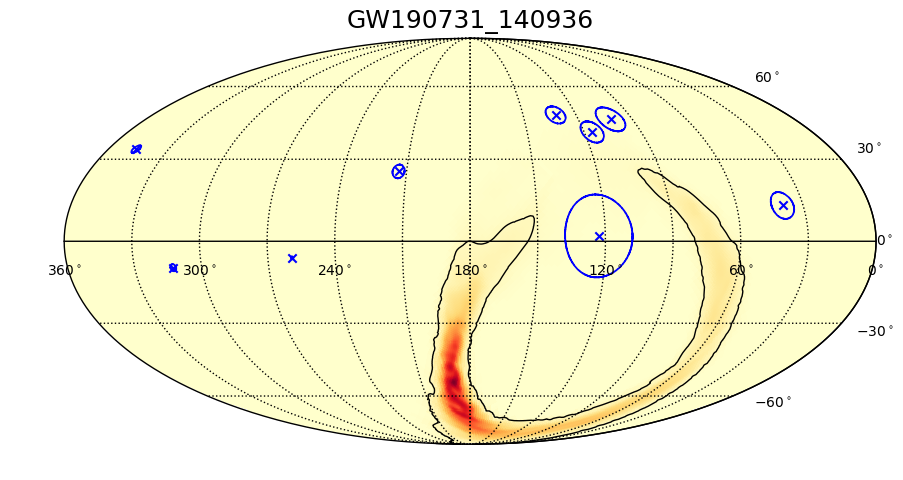}{0.31\textwidth}{(30)}}
\gridline{\fig{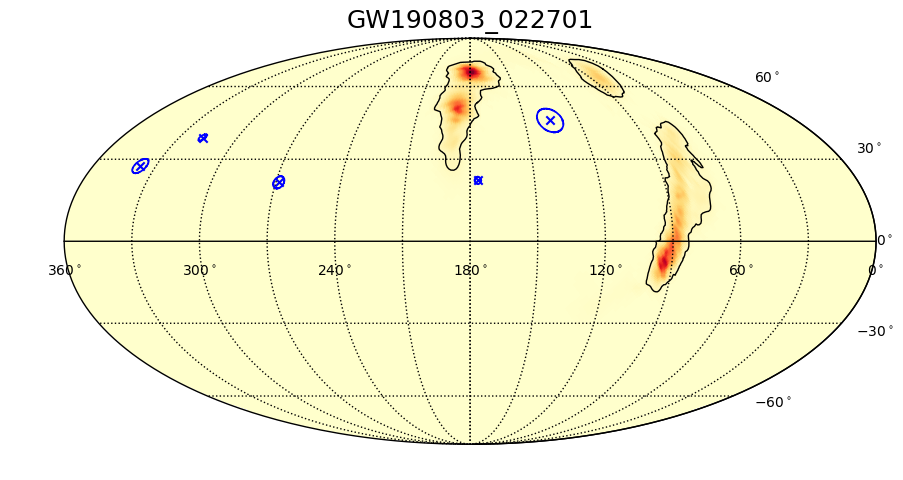}{0.31\textwidth}{(31)}
    \fig{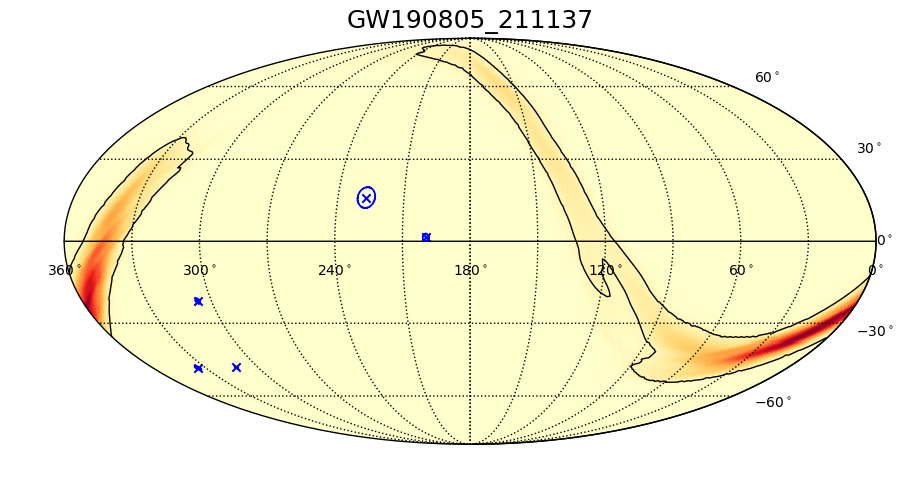}{0.31\textwidth}{(32)}
    \fig{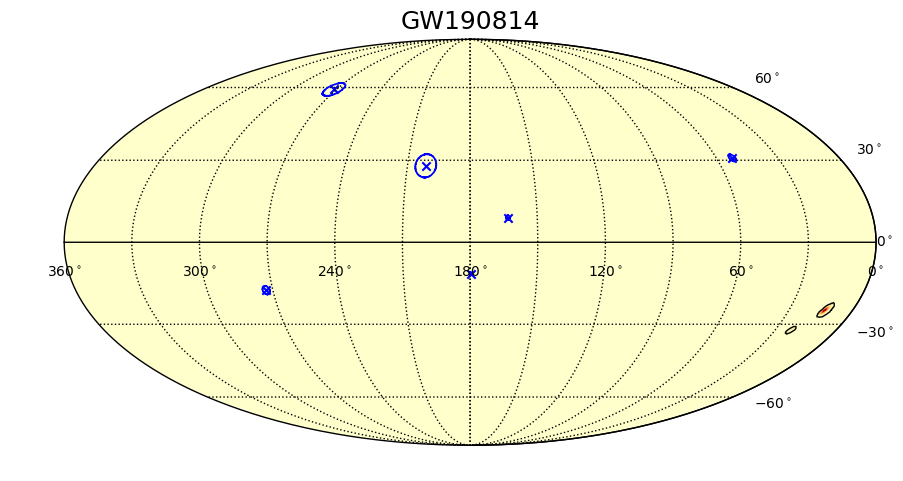}{0.31\textwidth}{(33)}}
\gridline{\fig{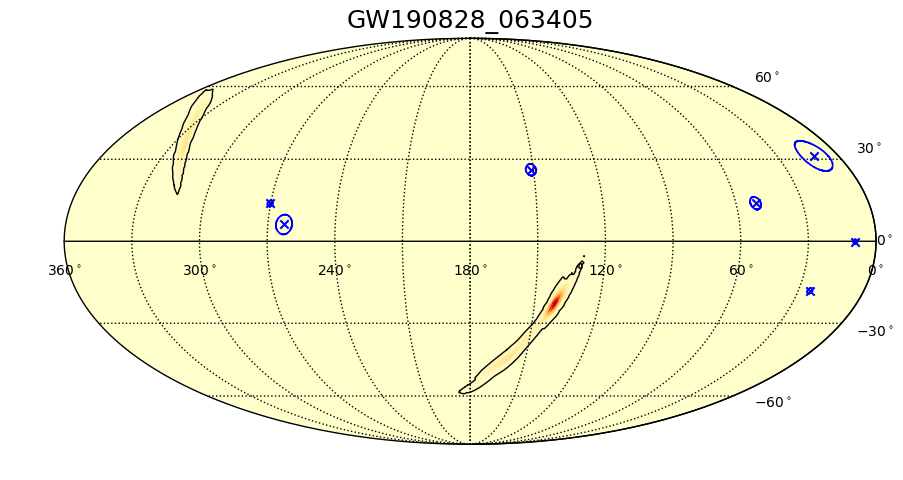}{0.31\textwidth}{(34)}
    \fig{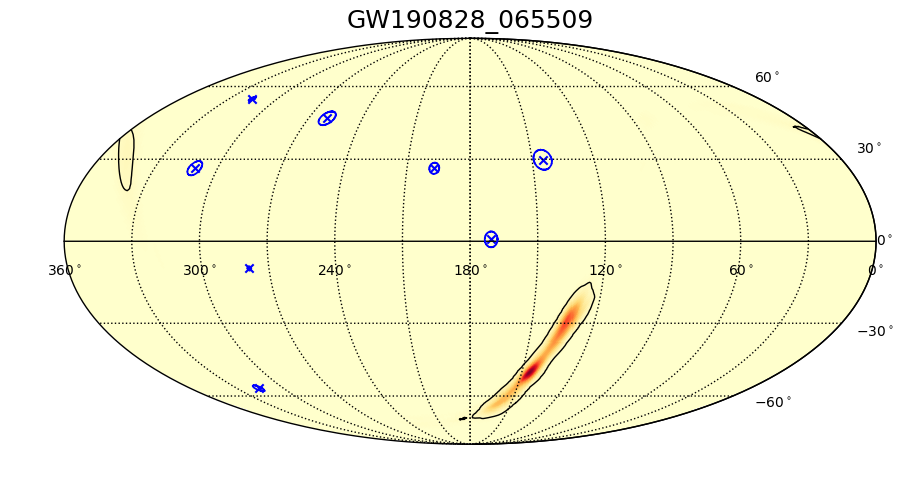}{0.31\textwidth}{(35)}
    \fig{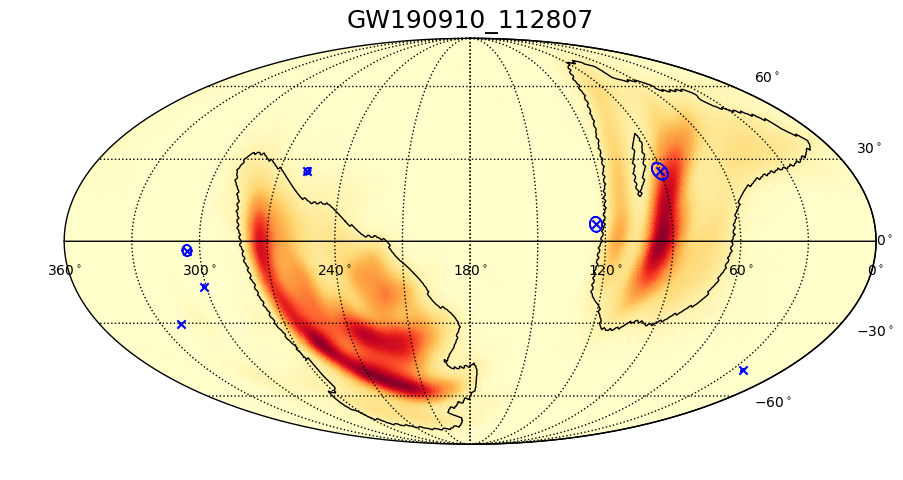}{0.31\textwidth}{(36)}}
\end{figure}
\setcounter{figure}{7}
\begin{figure} 
\gridline{\fig{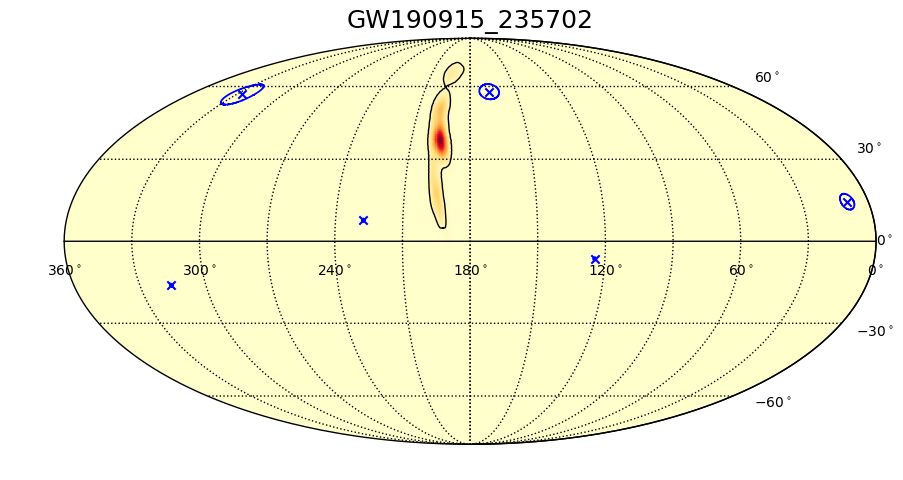}{0.31\textwidth}{(37)}
    \fig{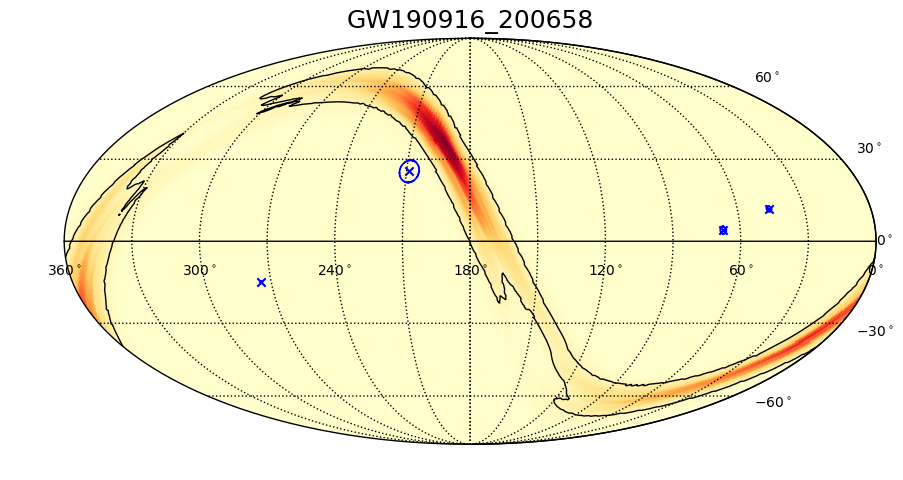}{0.31\textwidth}{(38)}
    \fig{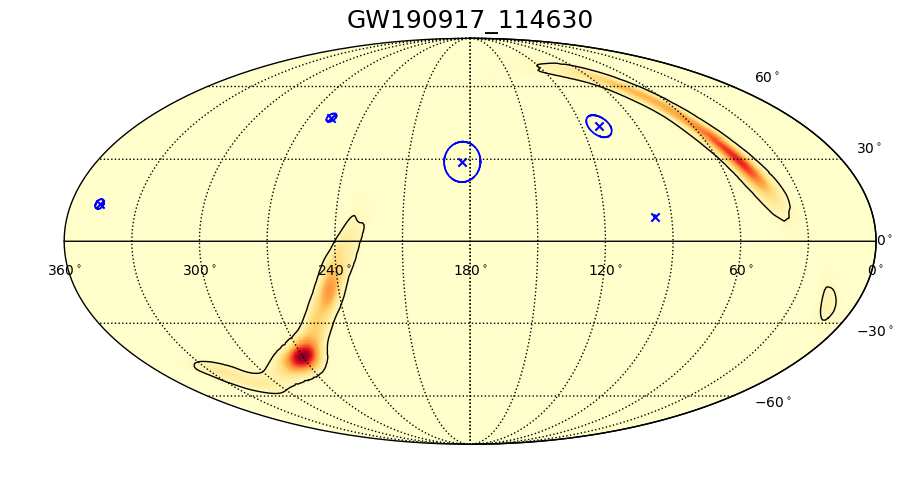}{0.31\textwidth}{(39)}}
\gridline{\fig{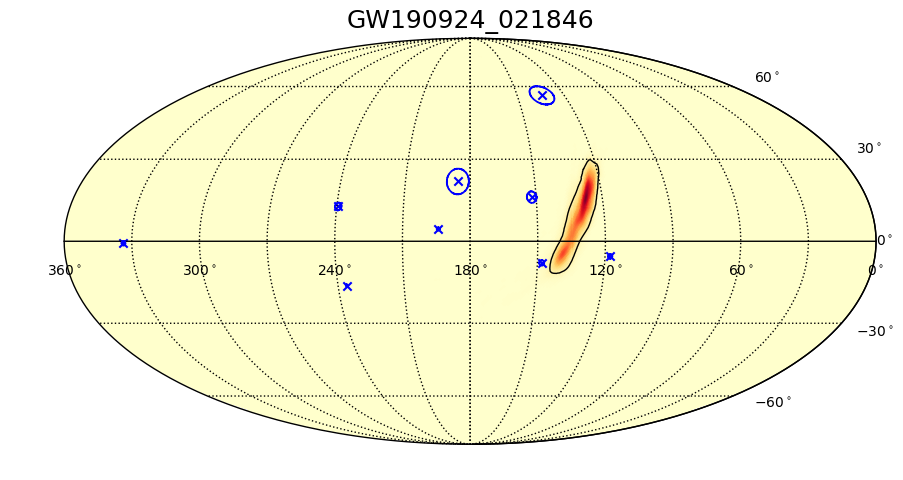}{0.31\textwidth}{(40)}
    \fig{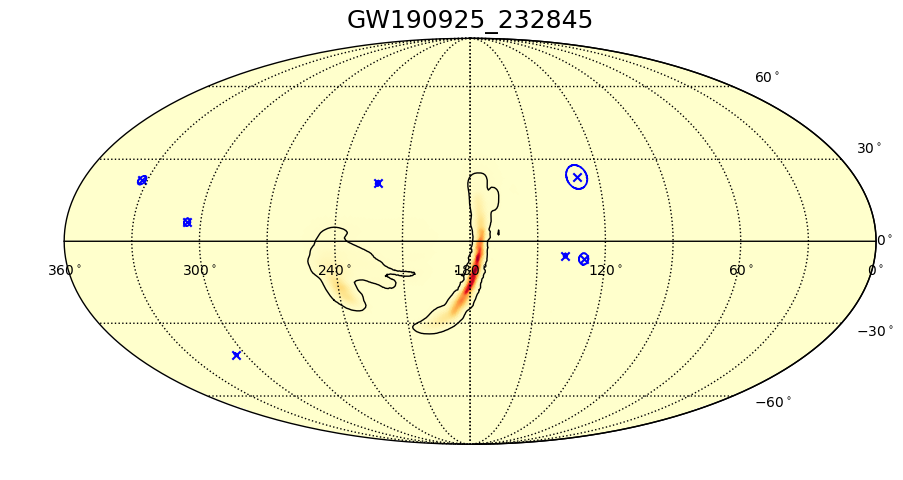}{0.31\textwidth}{(41)}
    \fig{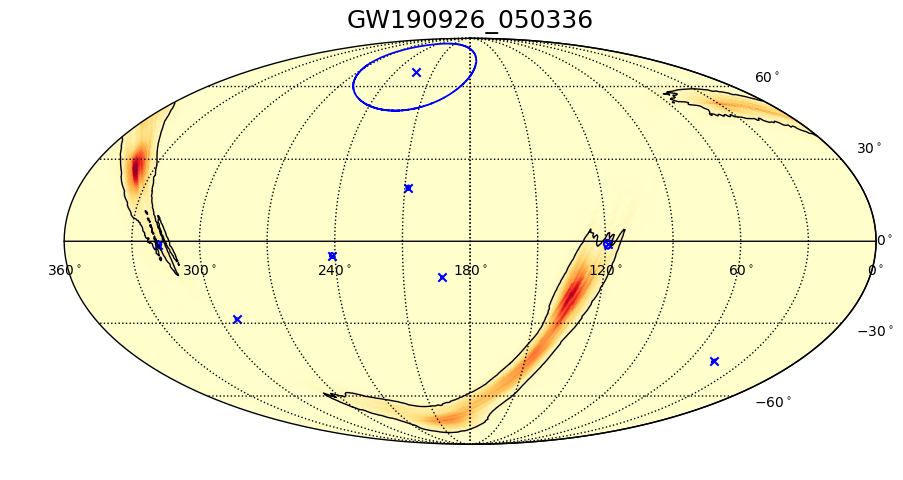}{0.31\textwidth}{(42)}}
\gridline{\fig{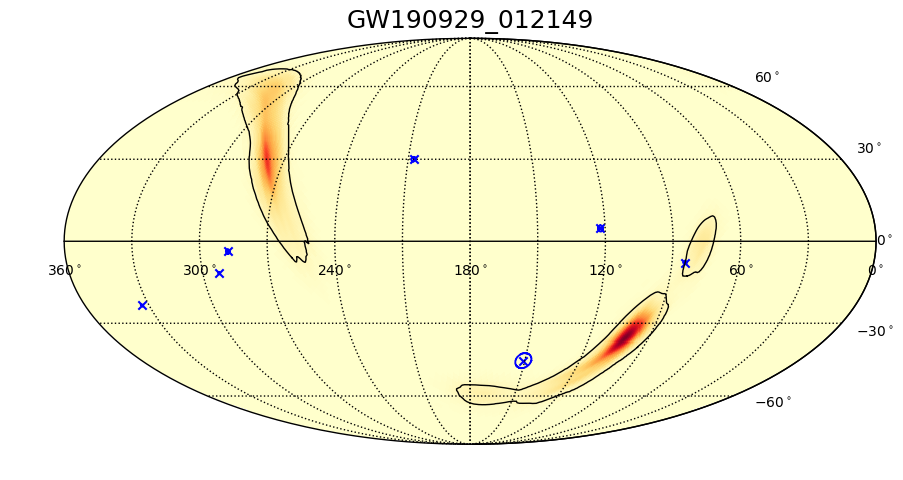}{0.31\textwidth}{(43)}
    \fig{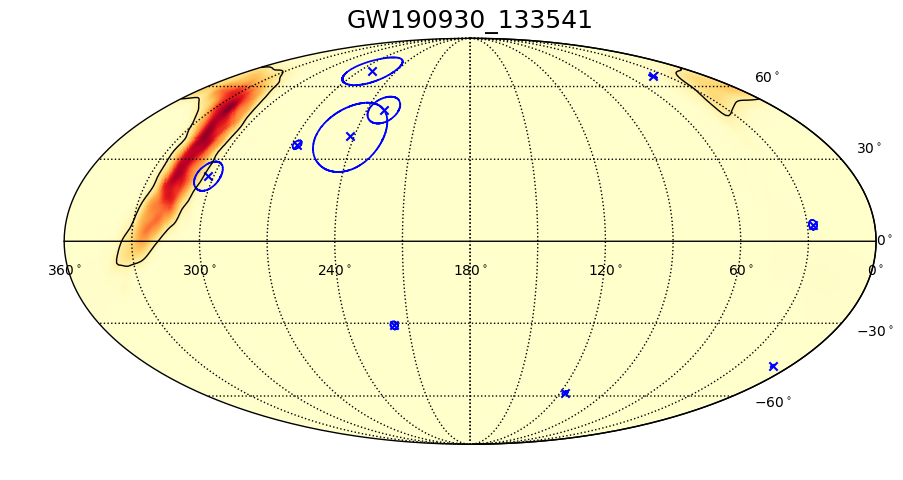}{0.31\textwidth}{(44)}}
\caption{Skymaps for the 1000~s follow-up of all events from the GWTC-2.1 \citep{LIGOScientific:2021usb} catalog. Shown in red is the localization probability of the GW event with the black contour representing the 90\% containment region of the GW localization. The blue crosses show the best fit neutrino candidate directions with the blue circles representing the 90\% angular error region of the neutrino candidates.}
\label{fig:1000s_skymaps}
\end{figure}

\begin{figure}
\gridline{\fig{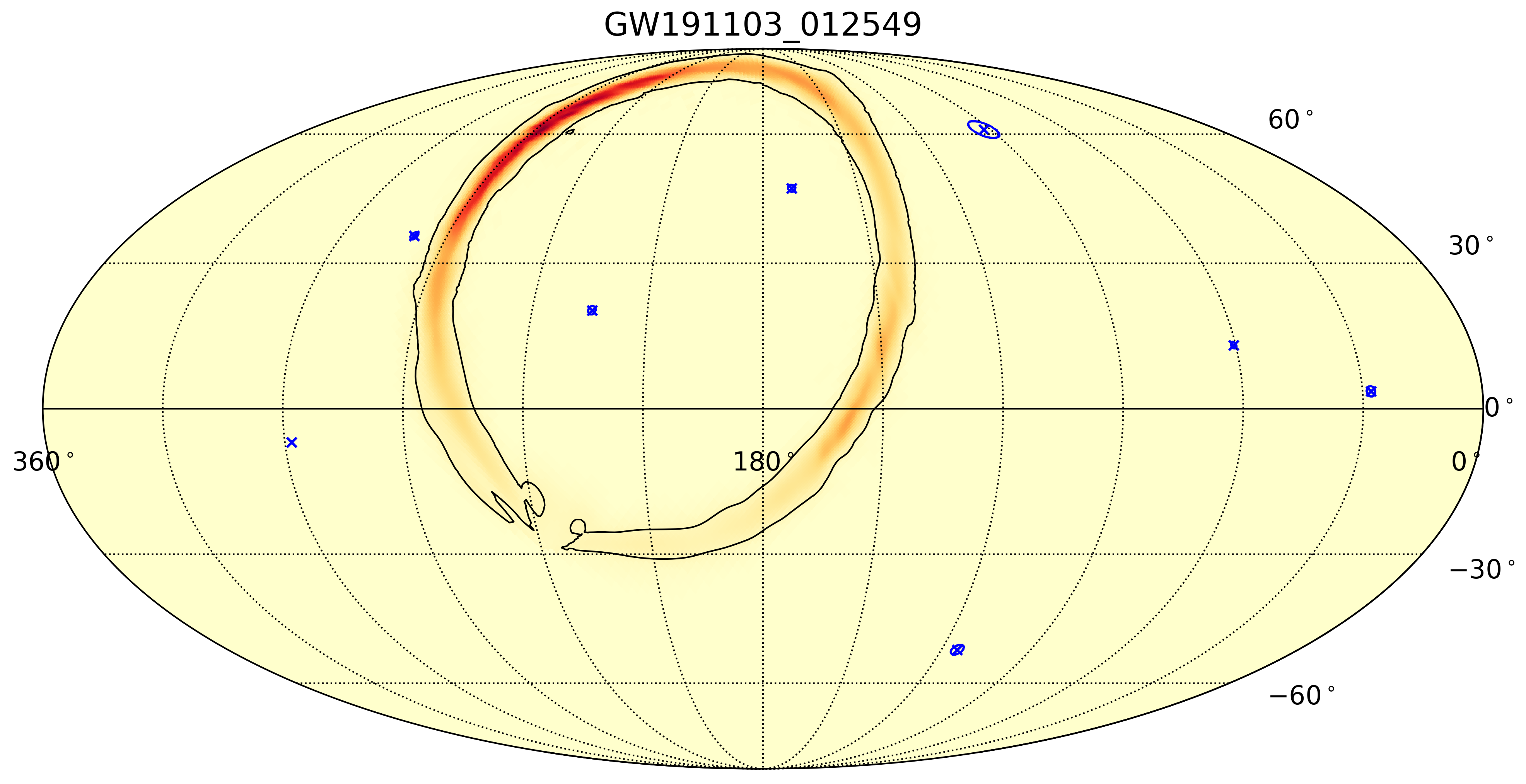}{0.31\textwidth}{(1)}
    \fig{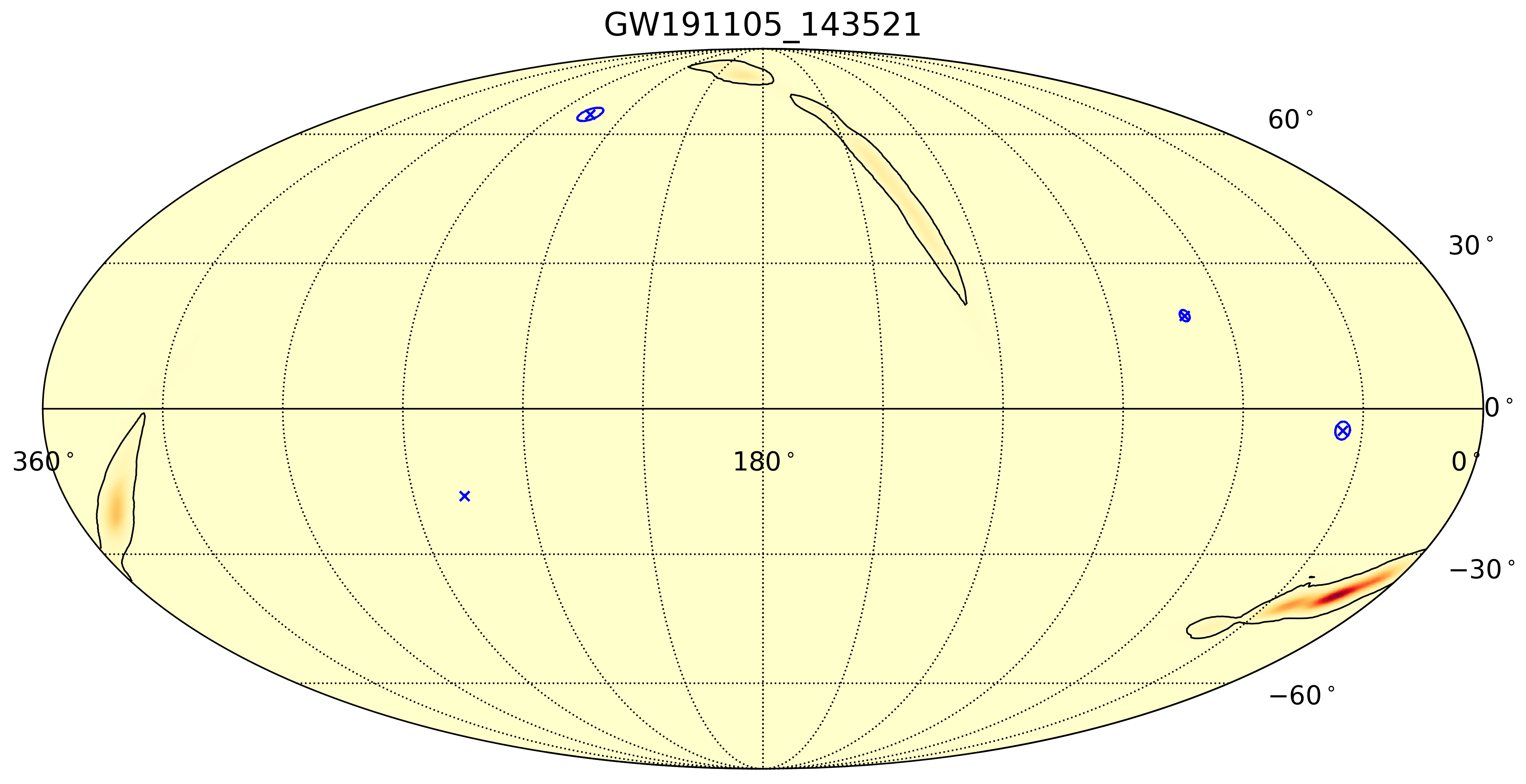}{0.31\textwidth}{(2)}
    \fig{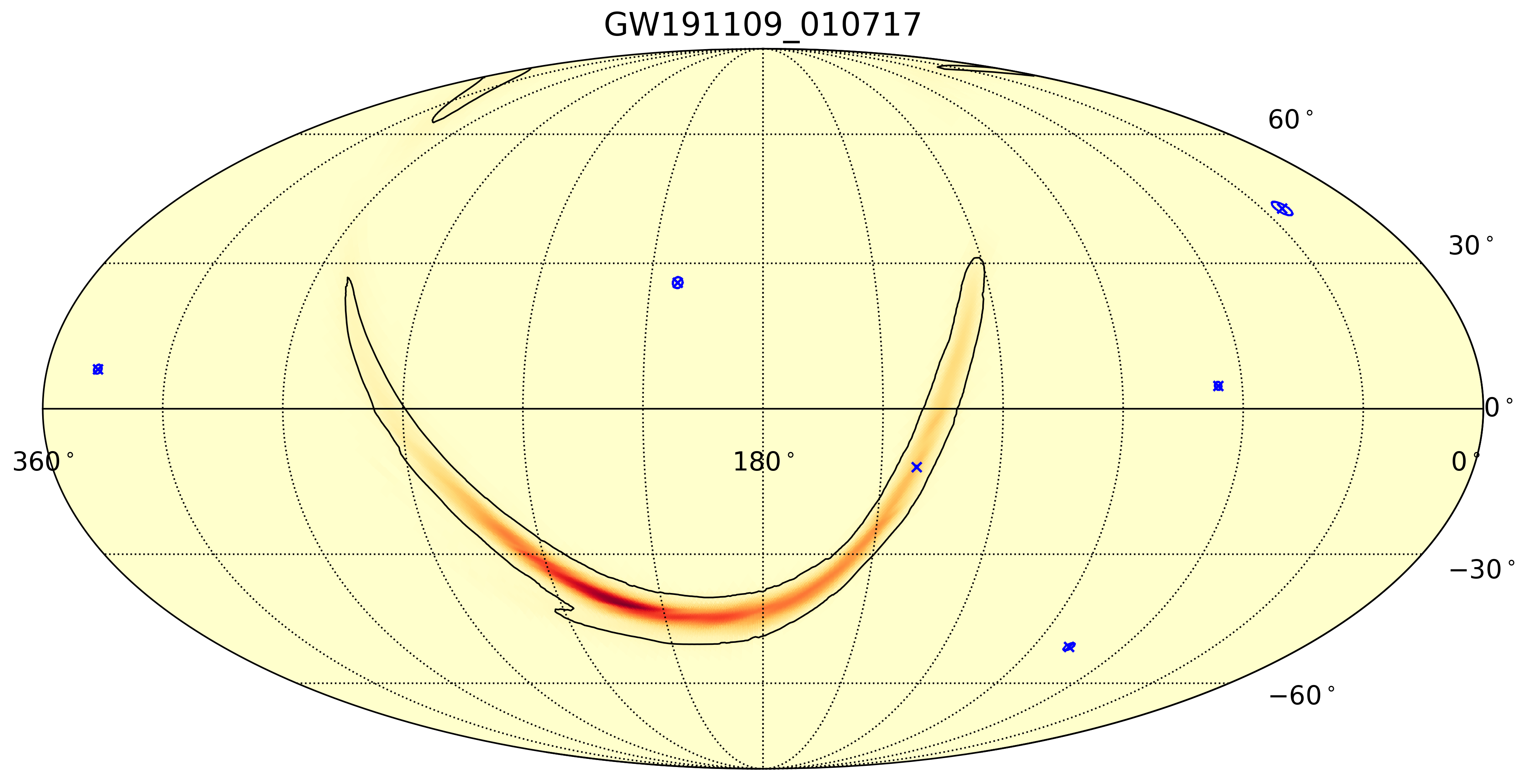}{0.31\textwidth}{(3)}}
\gridline{\fig{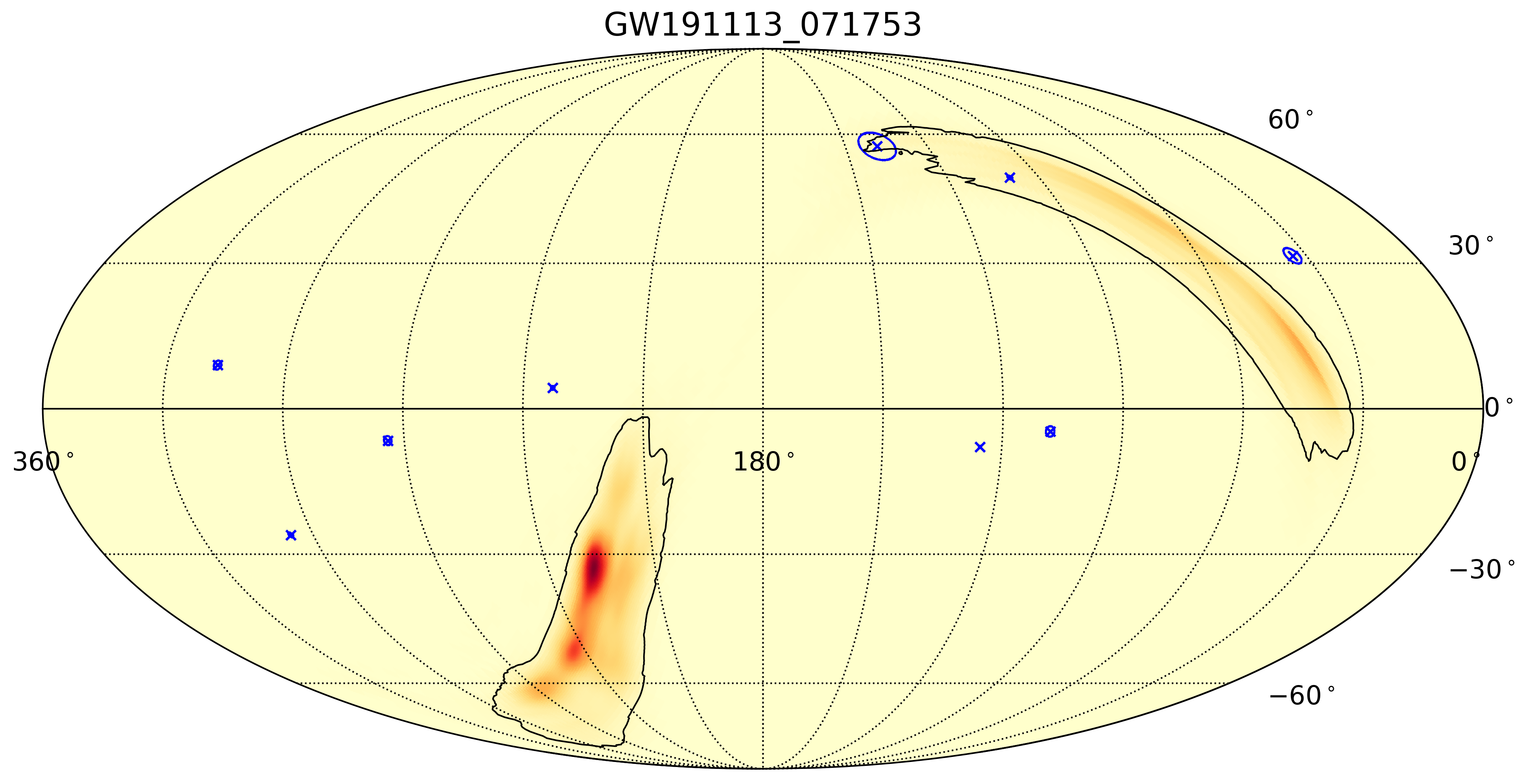}{0.31\textwidth}{(4)}
    \fig{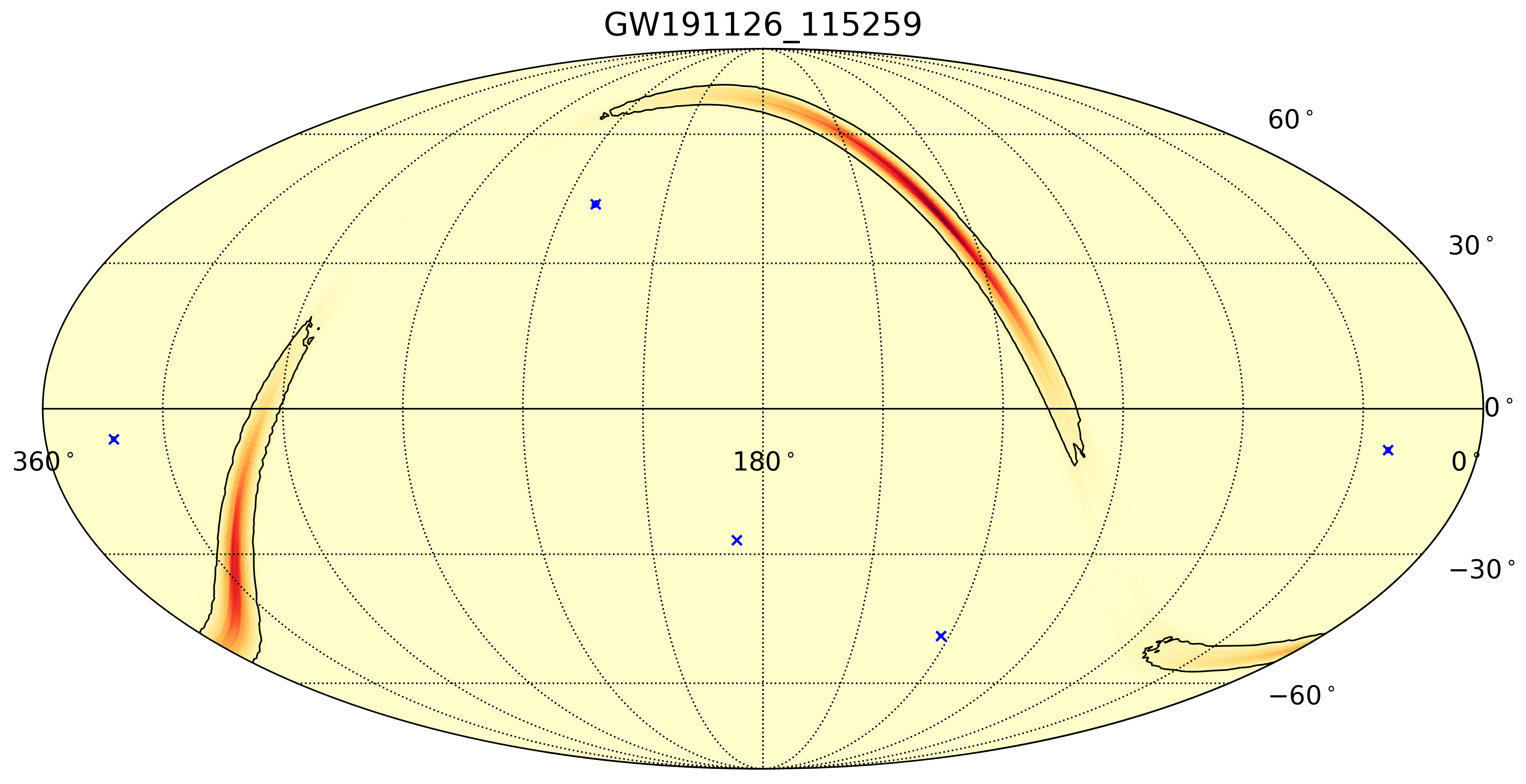}{0.31\textwidth}{(5)}
    \fig{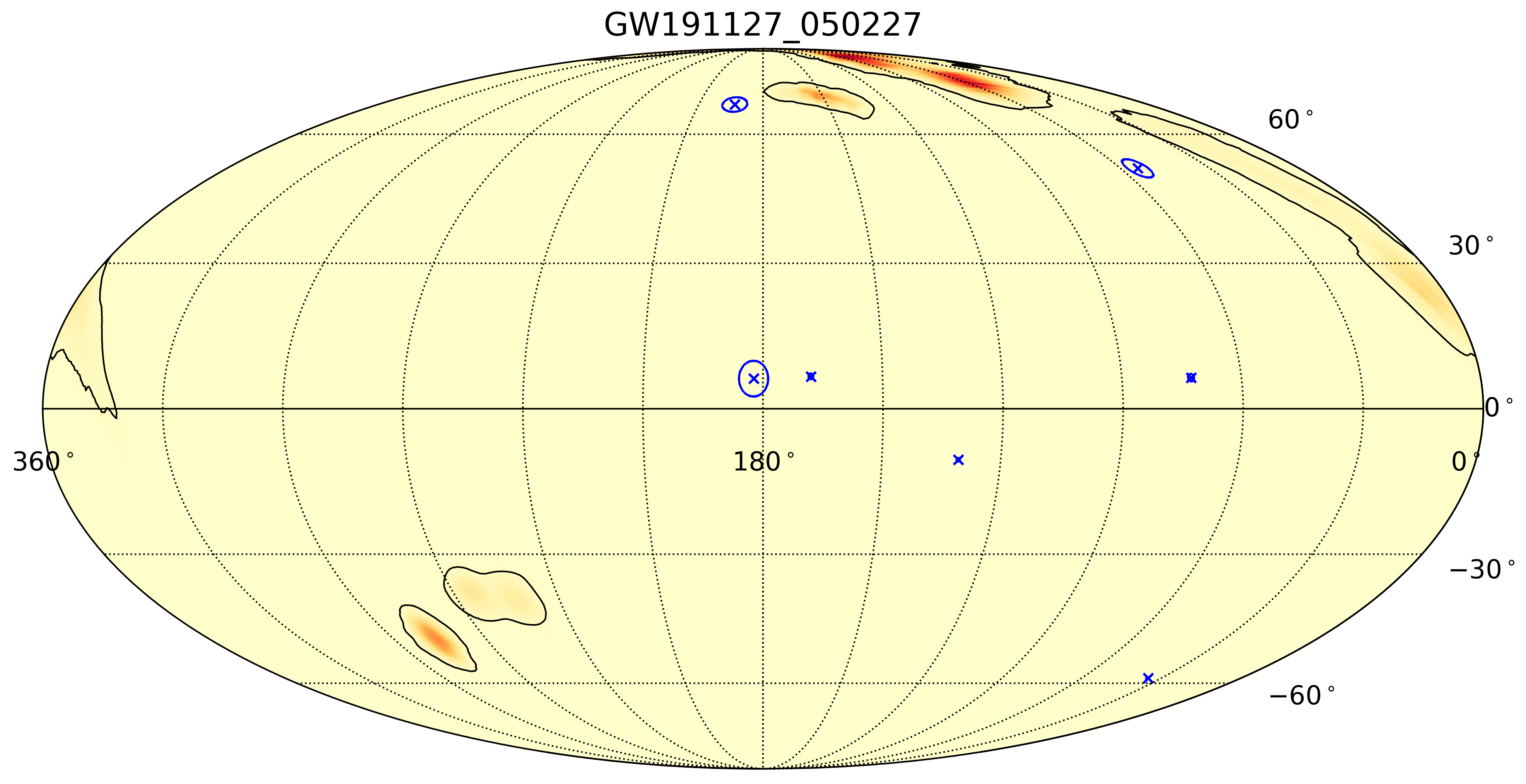}{0.31\textwidth}{(6)}}
\gridline{\fig{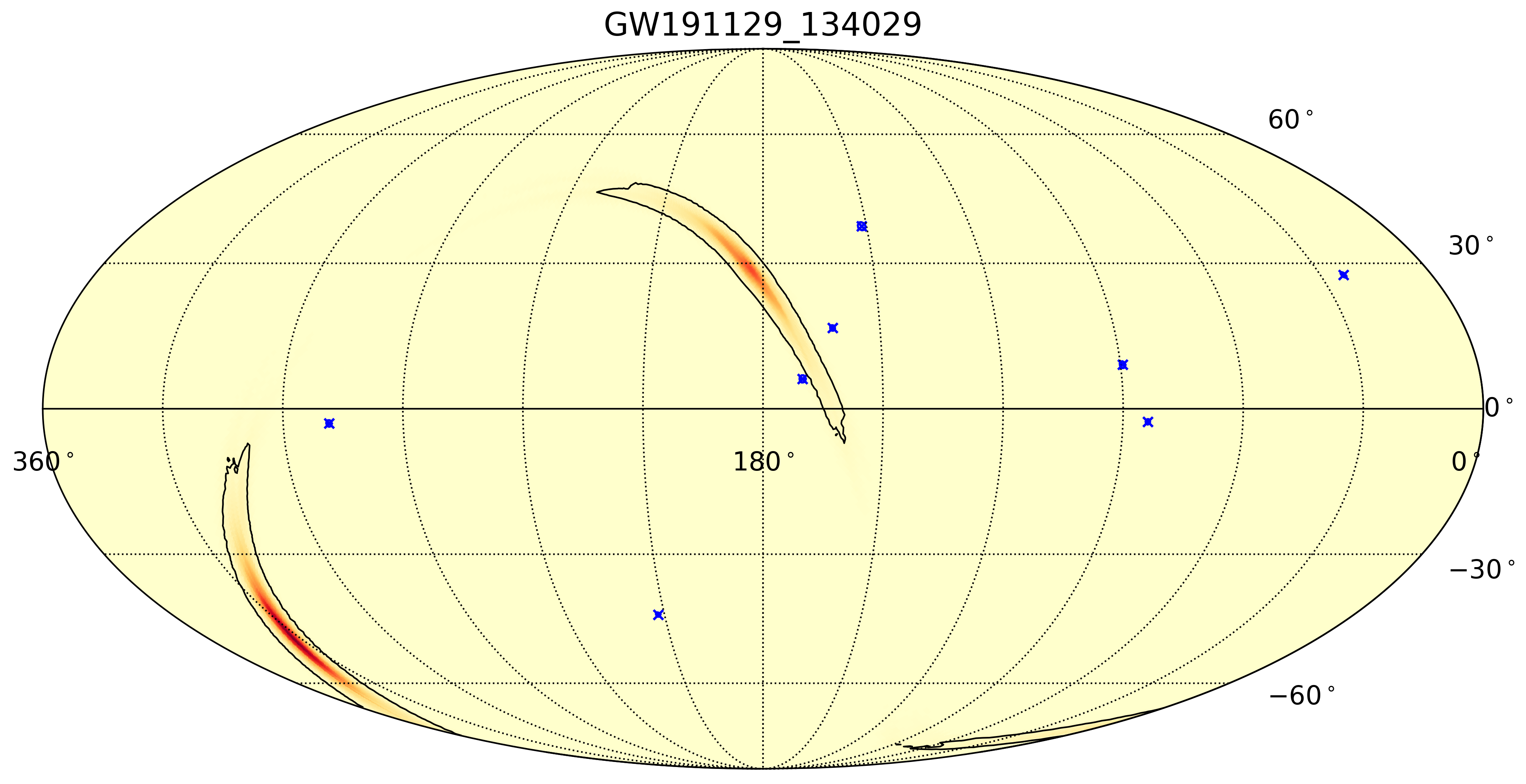}{0.31\textwidth}{(7)}
    \fig{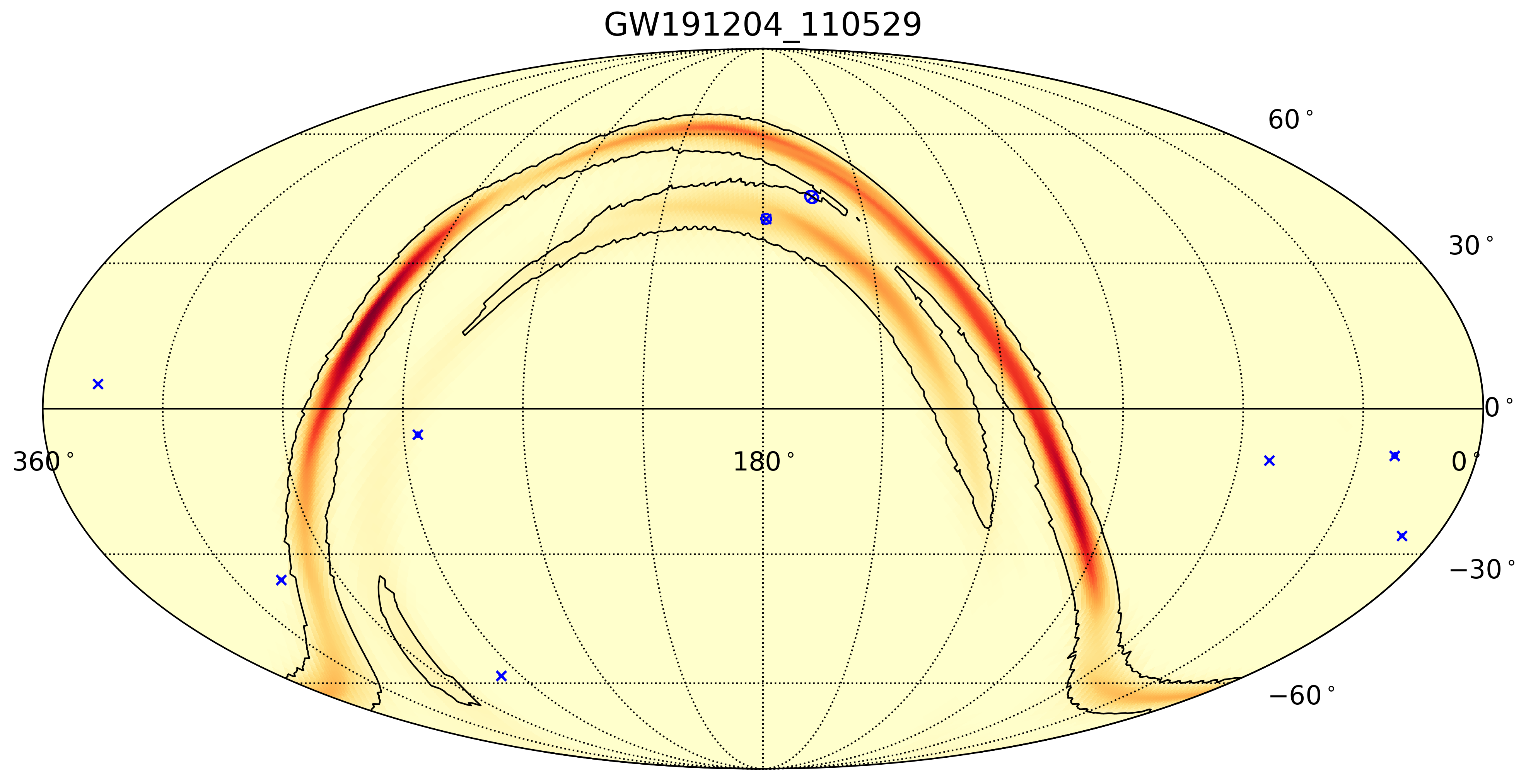}{0.31\textwidth}{(8)}
    \fig{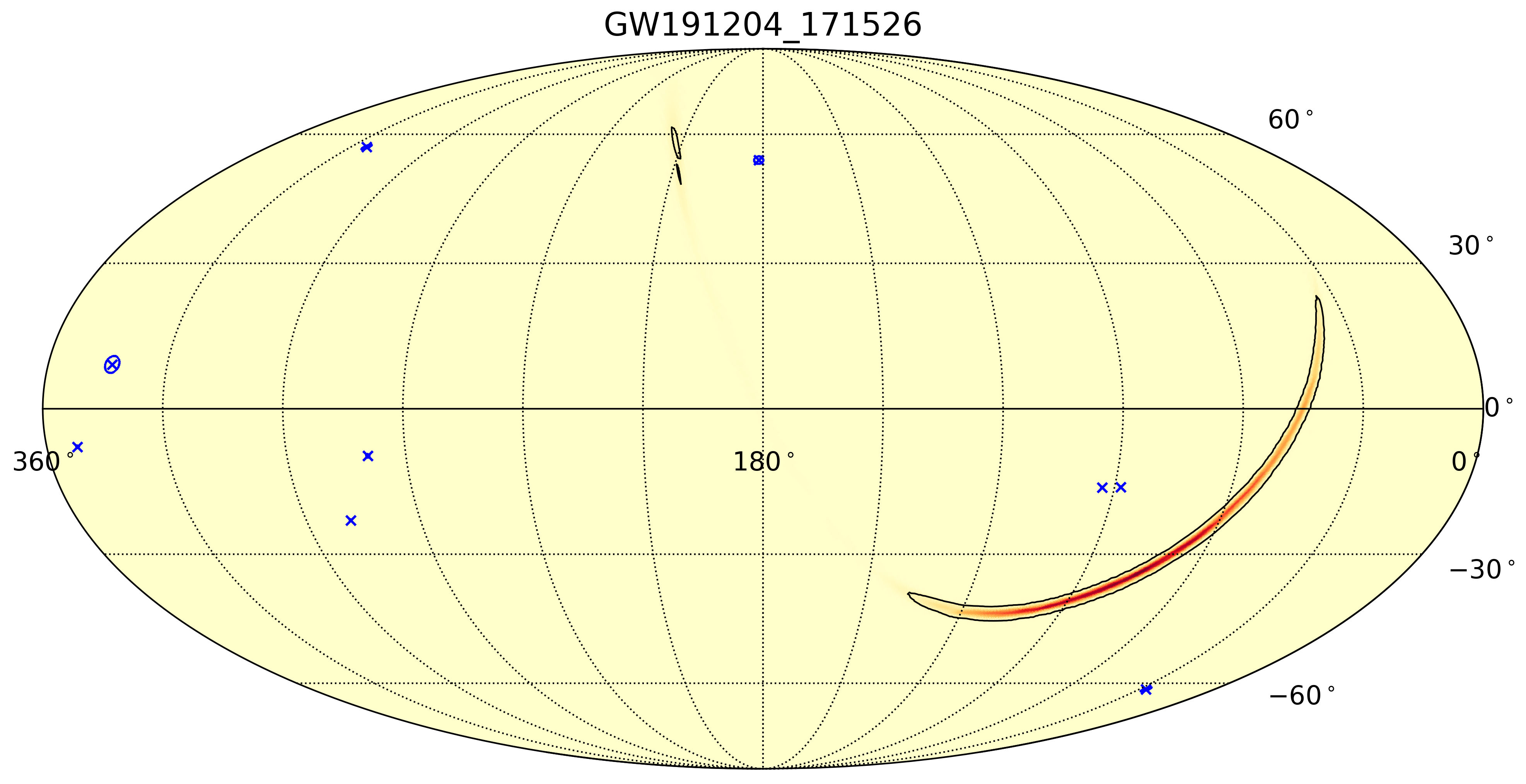}{0.31\textwidth}{(9)}}
\gridline{\fig{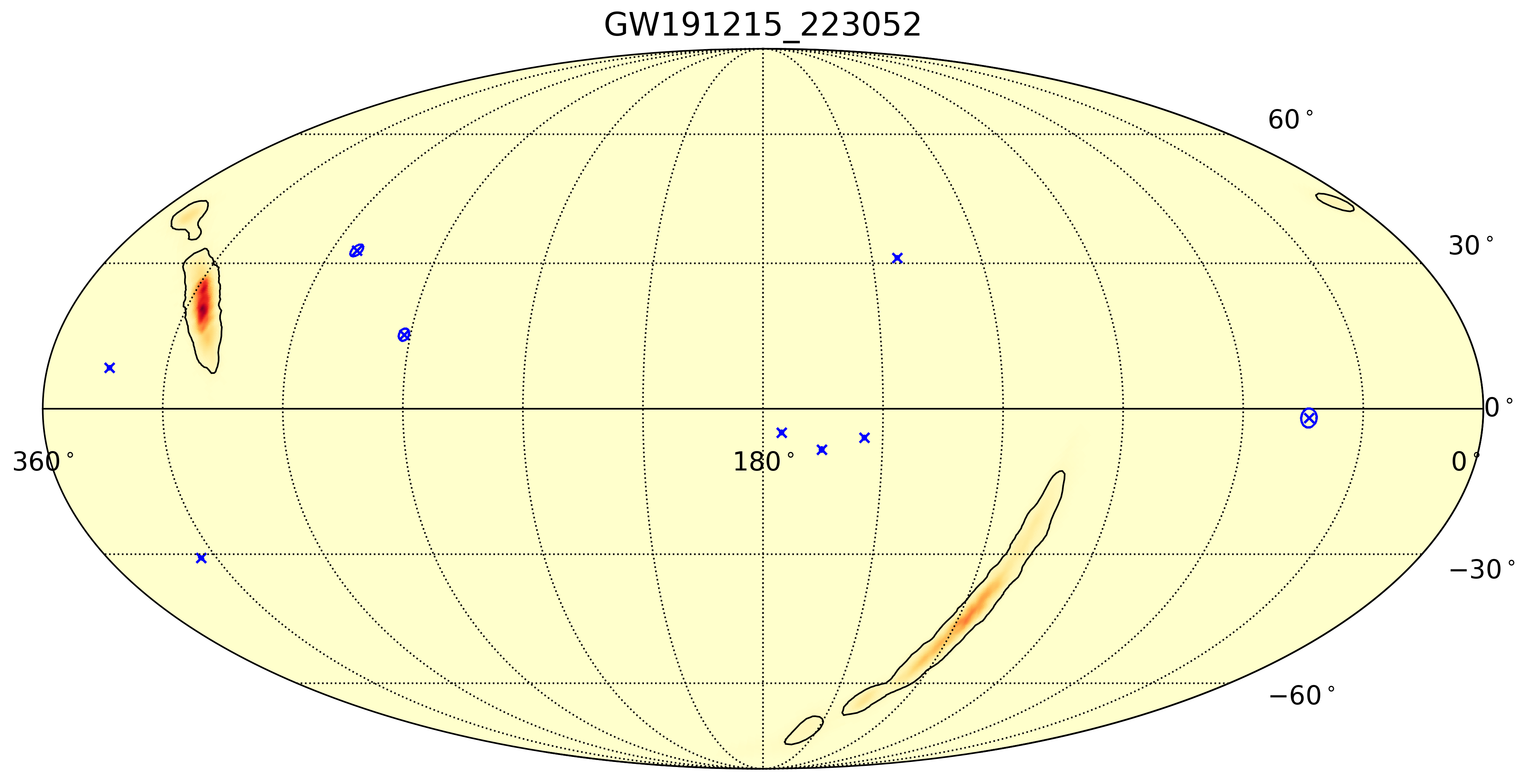}{0.31\textwidth}{(10)}
    \fig{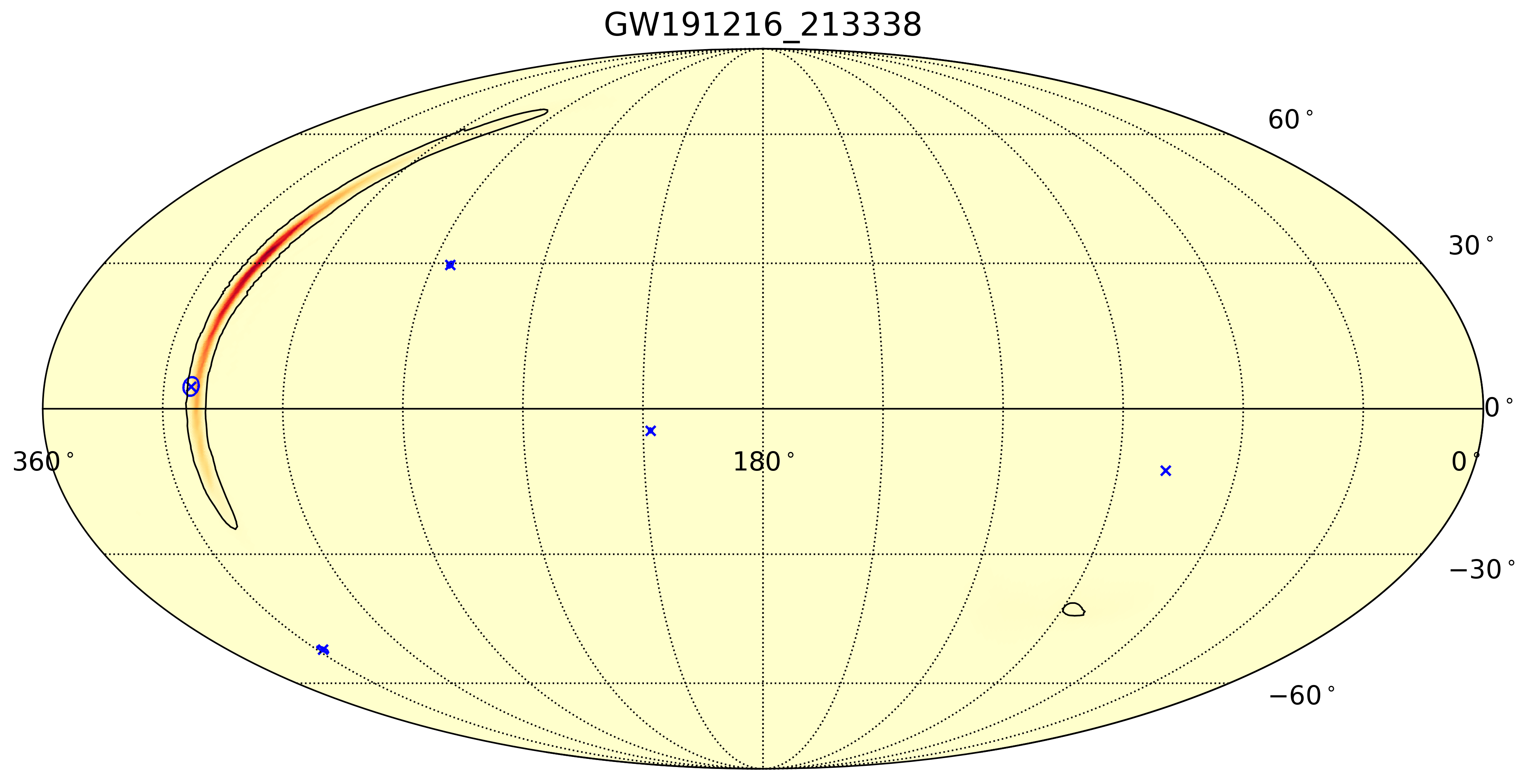}{0.31\textwidth}{(11)}
    \fig{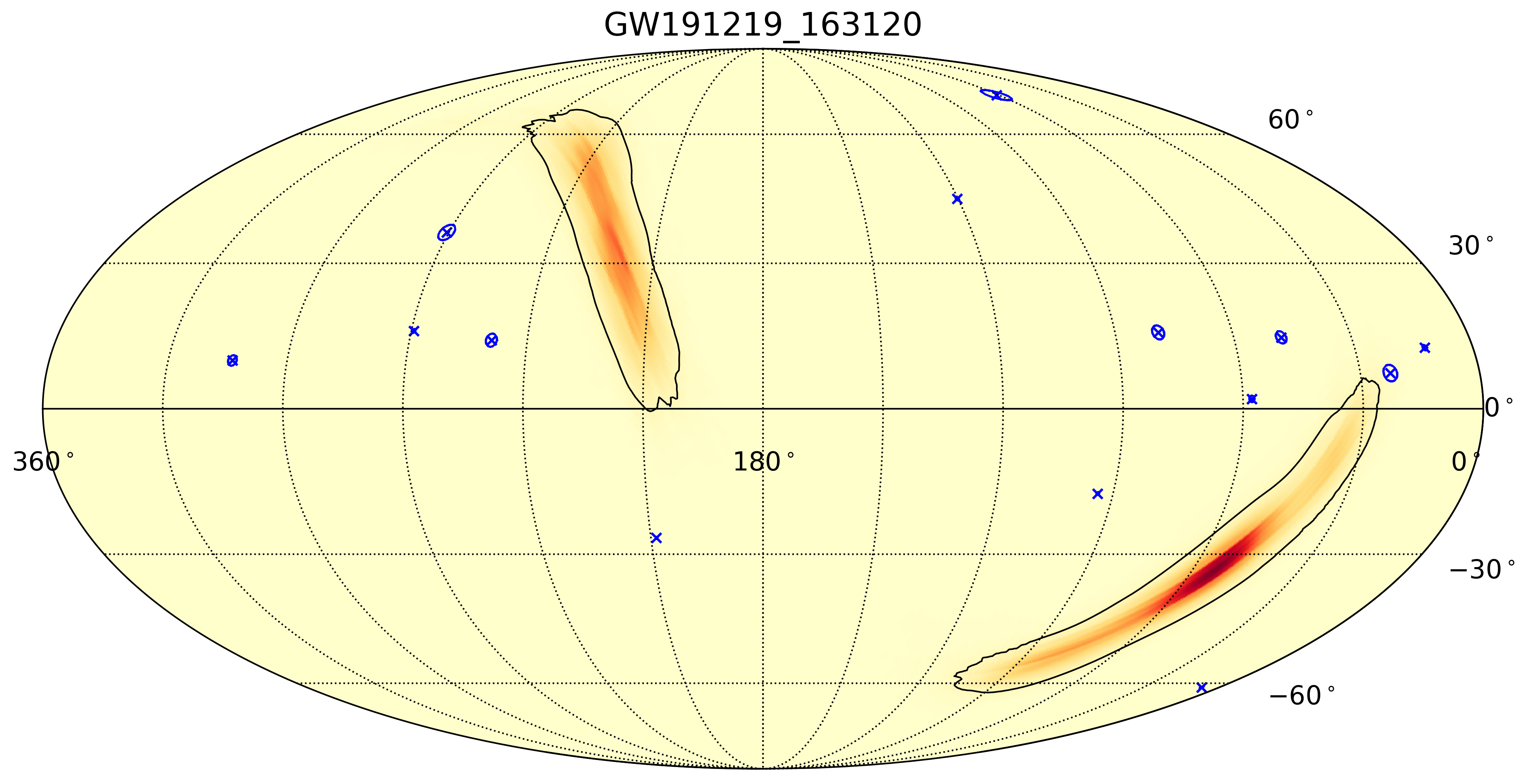}{0.31\textwidth}{(12)}}
\gridline{\fig{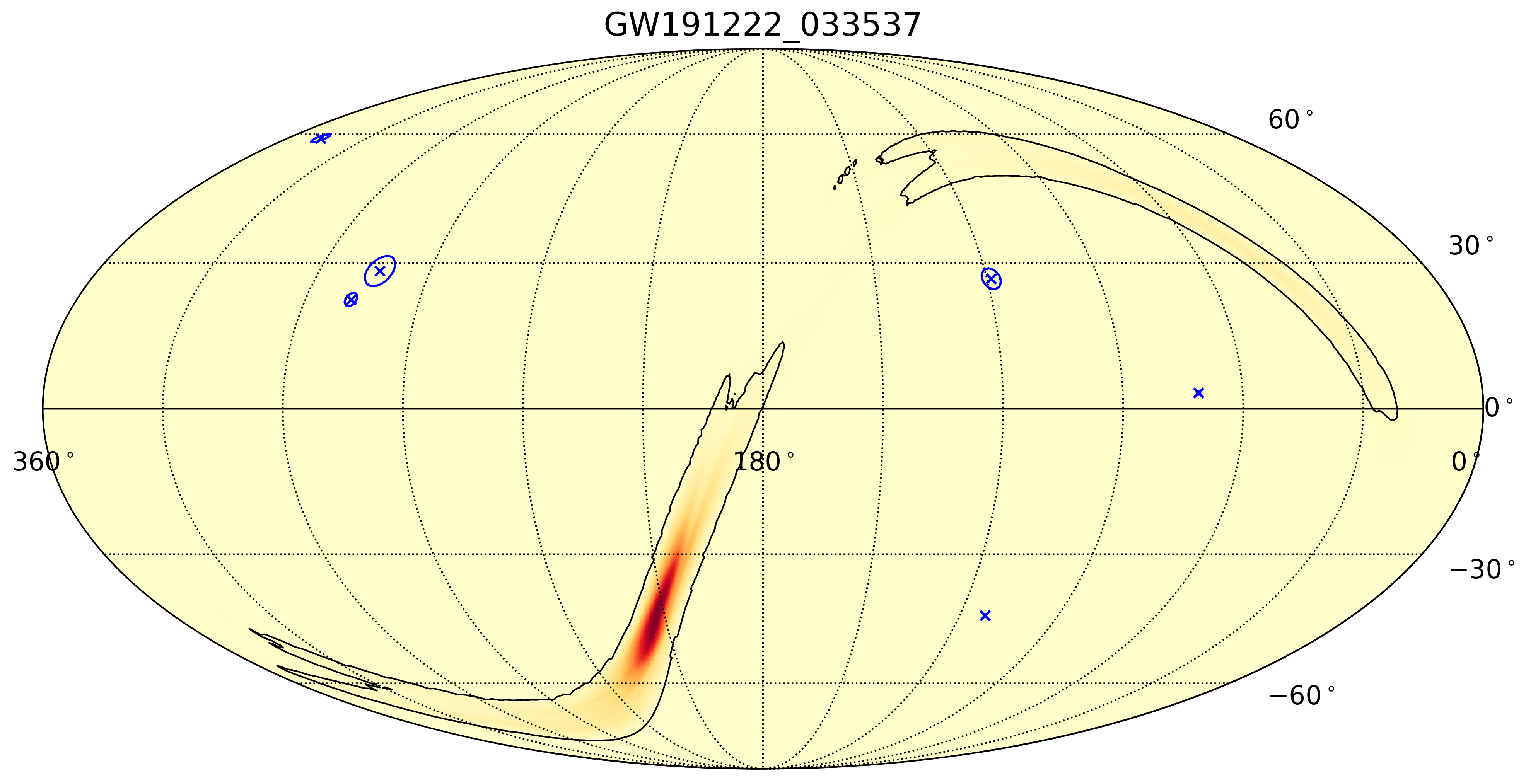}{0.31\textwidth}{(13)}
    \fig{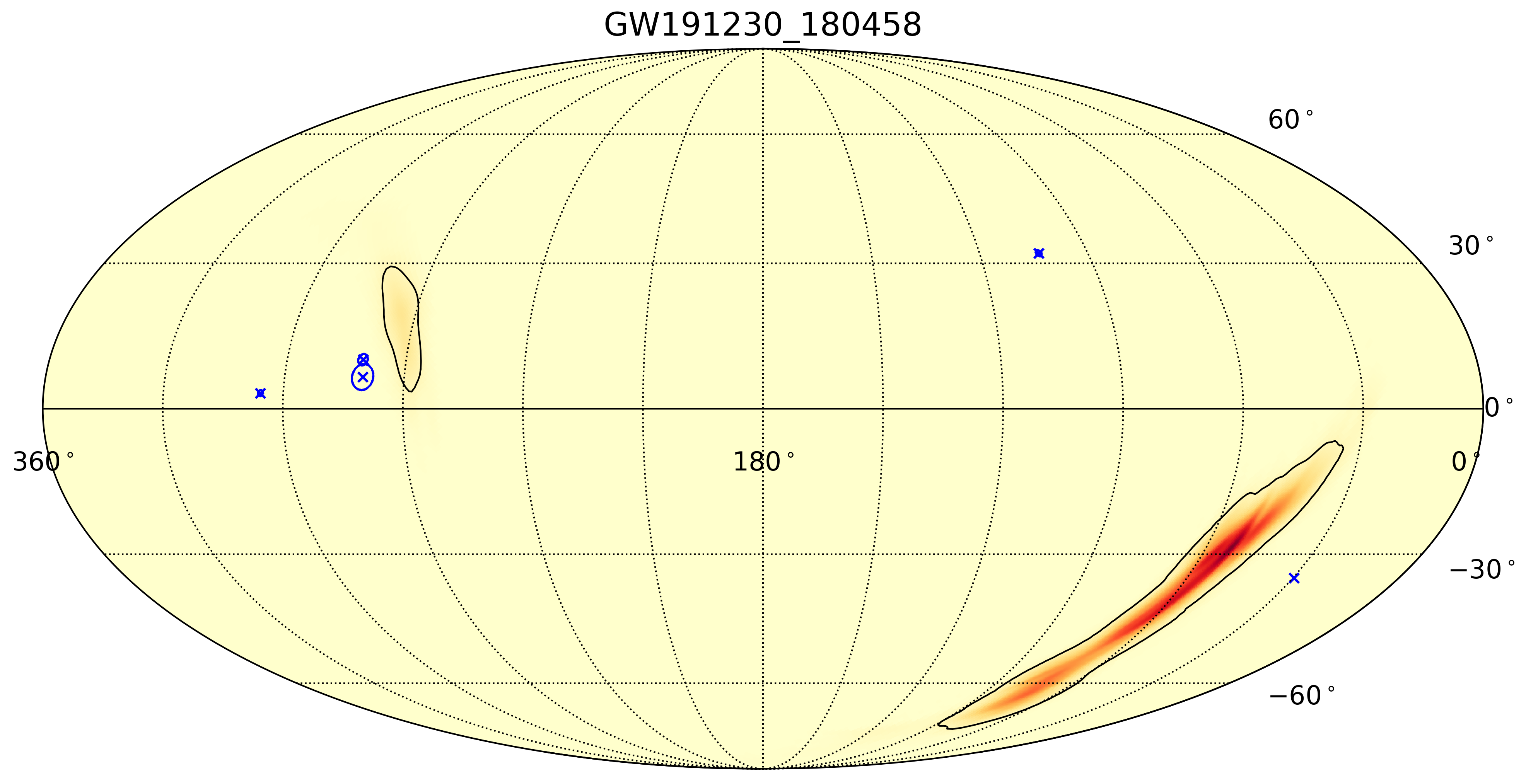}{0.31\textwidth}{(14)}
    \fig{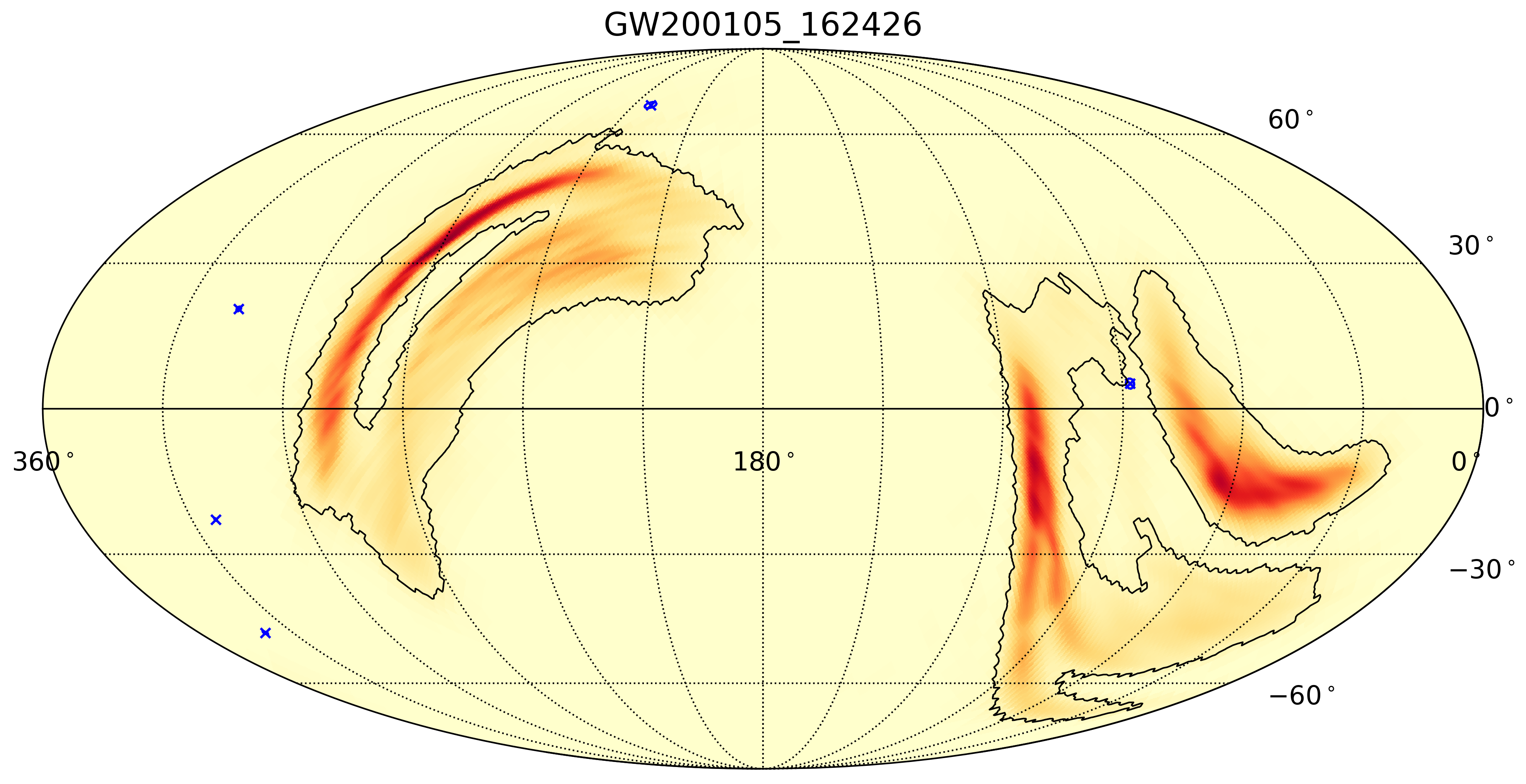}{0.31\textwidth}{(15)}}
\gridline{\fig{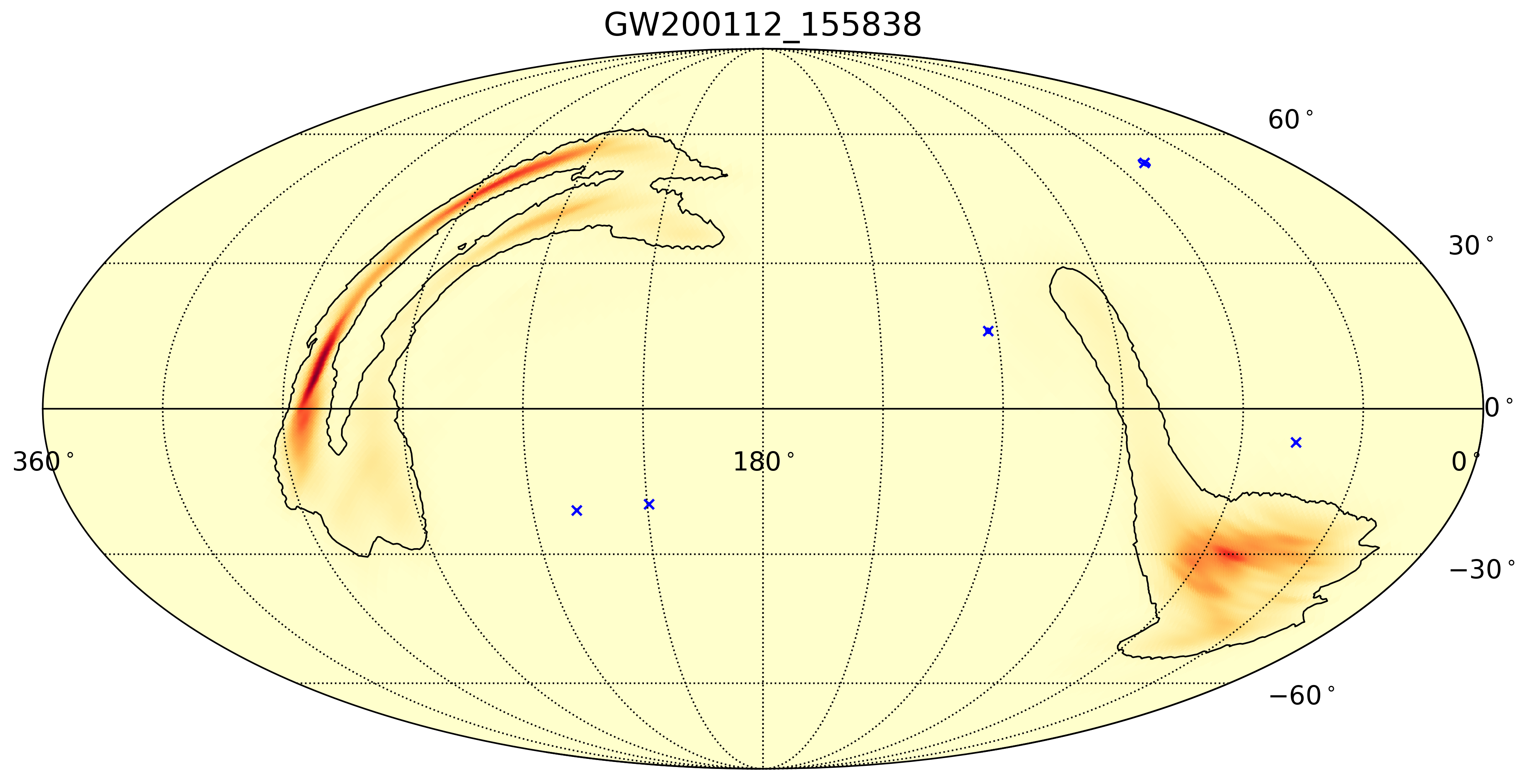}{0.31\textwidth}{(16)}
    \fig{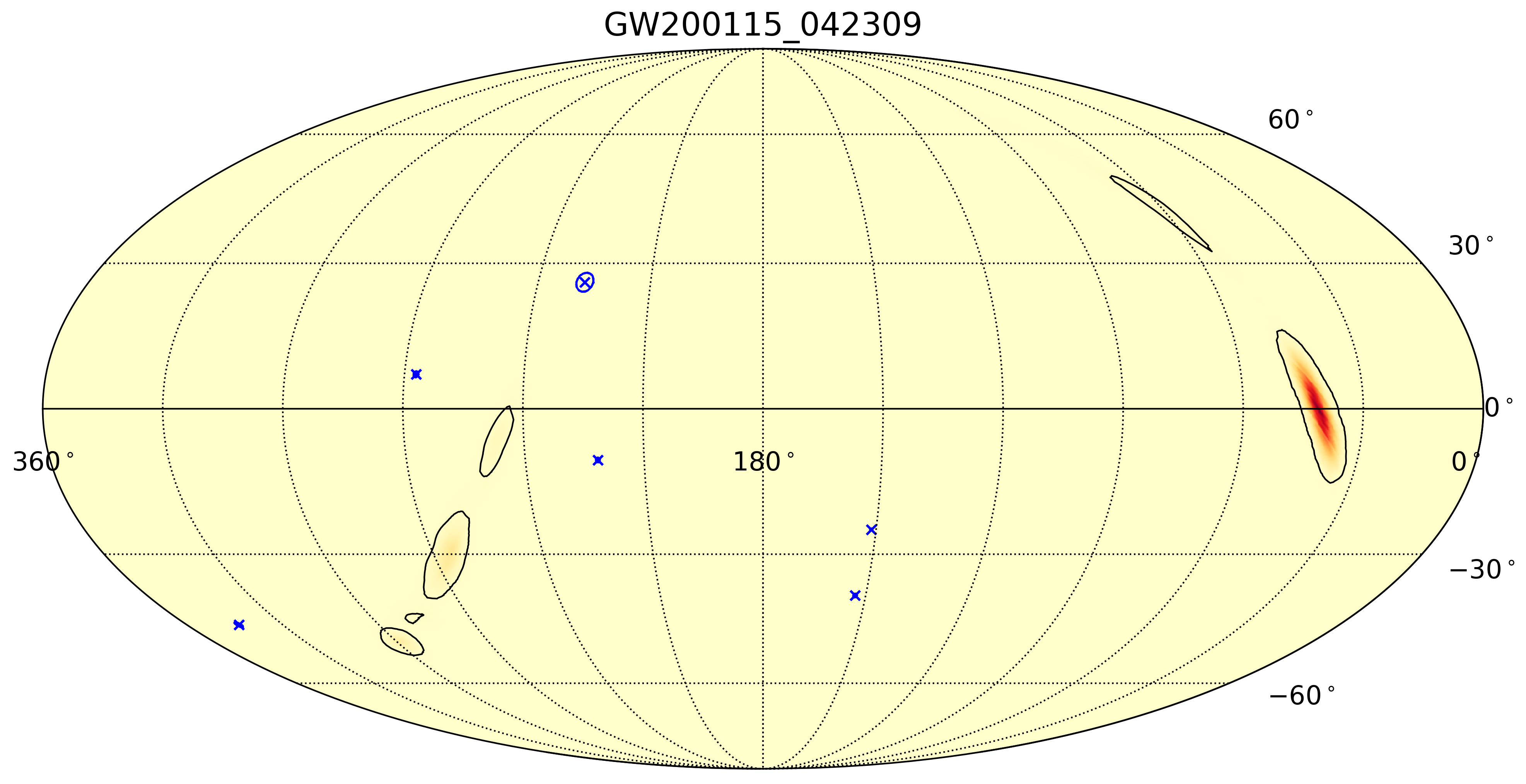}{0.31\textwidth}{(17)}
    \fig{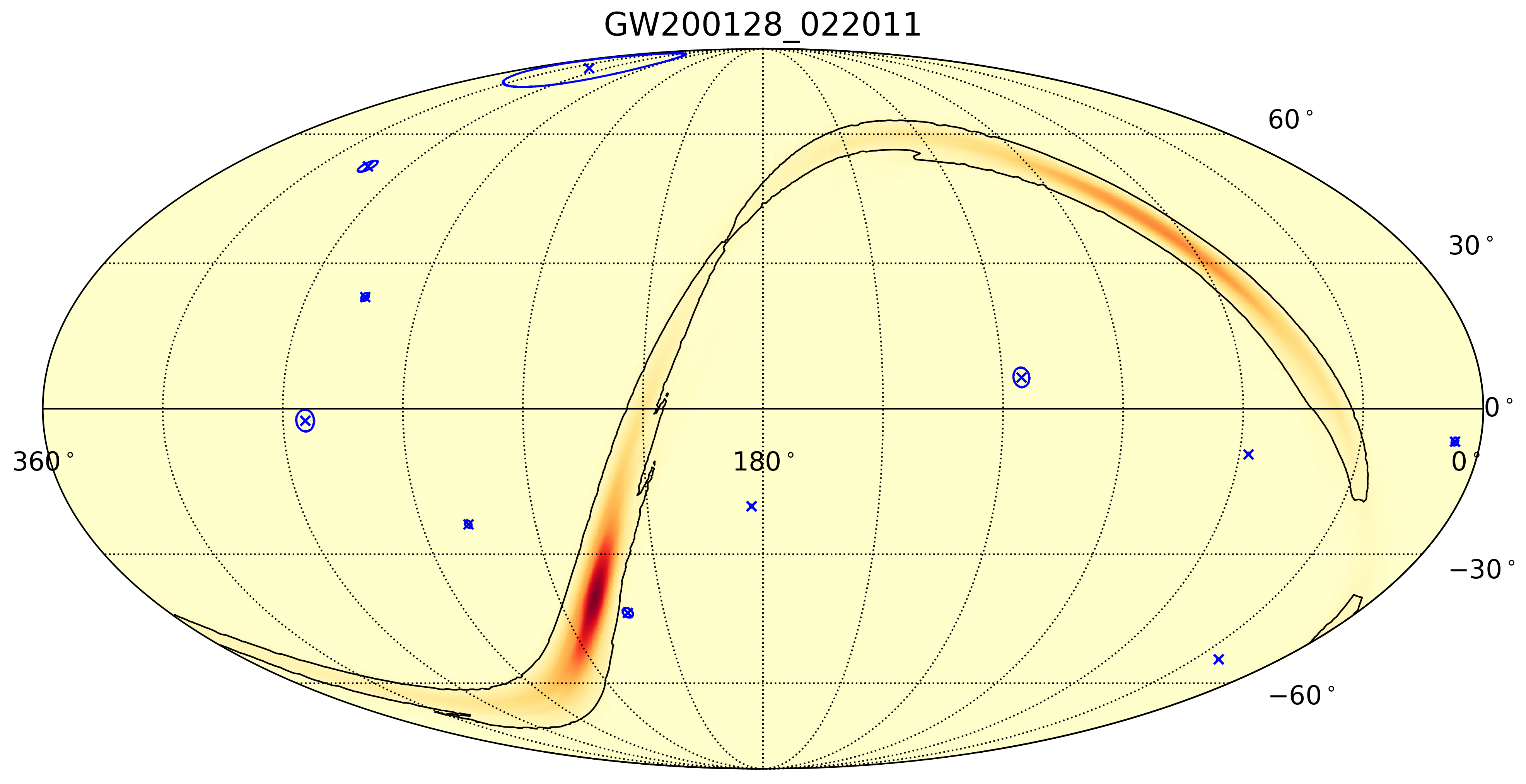}{0.31\textwidth}{(18)}}
\end{figure}

\setcounter{figure}{8}
\begin{figure} 
\gridline{\fig{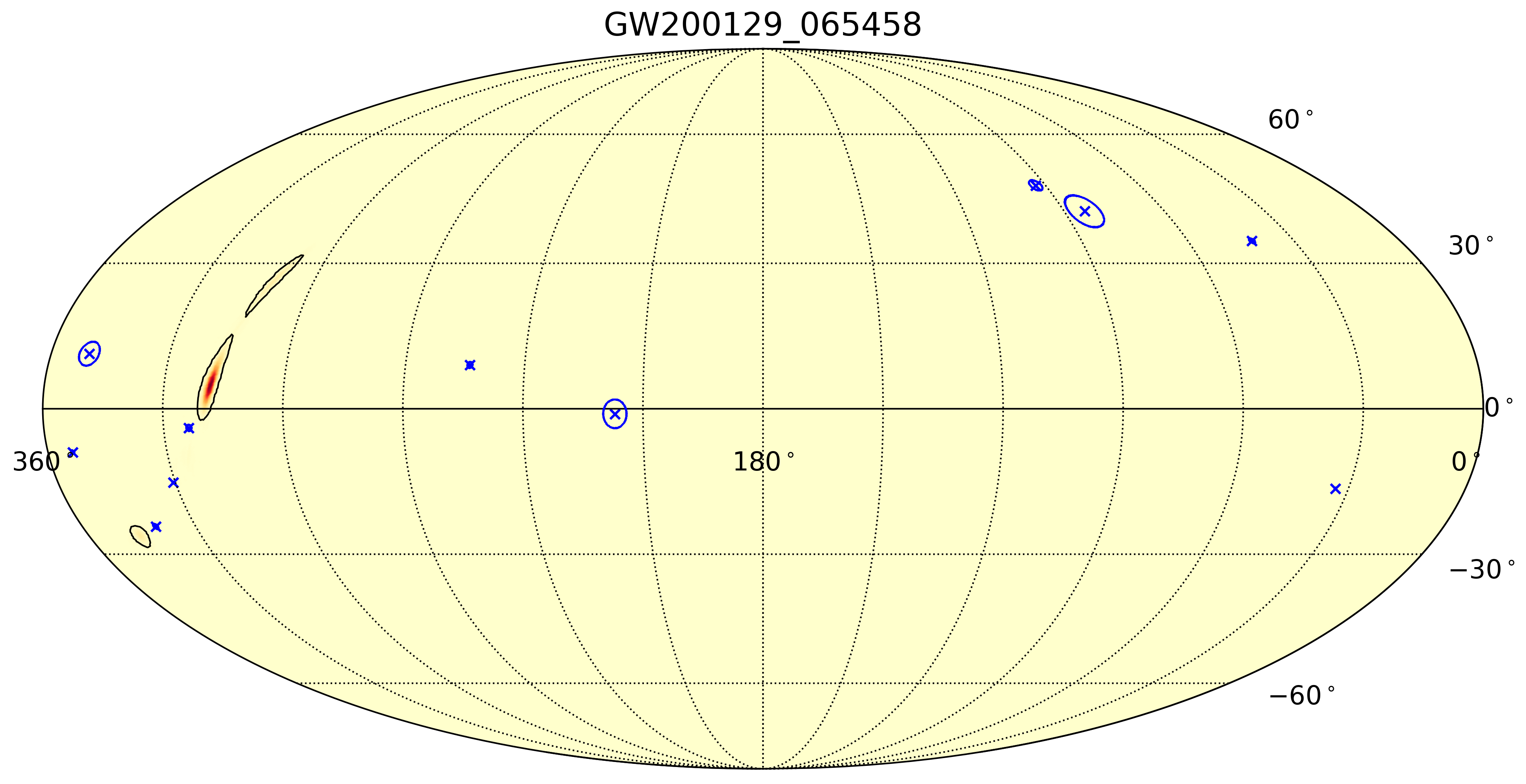}{0.31\textwidth}{(19)}
    \fig{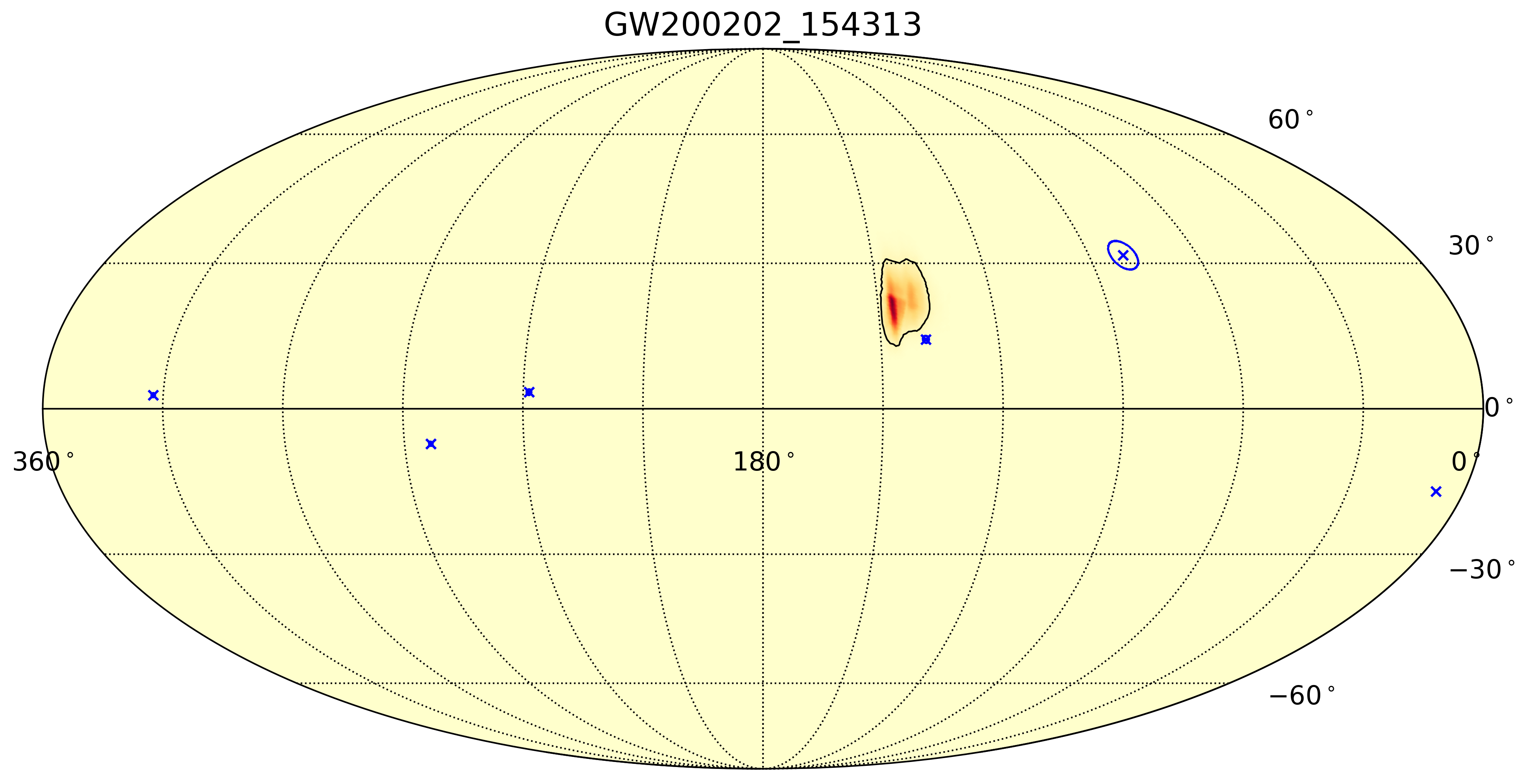}{0.31\textwidth}{(20)}
    \fig{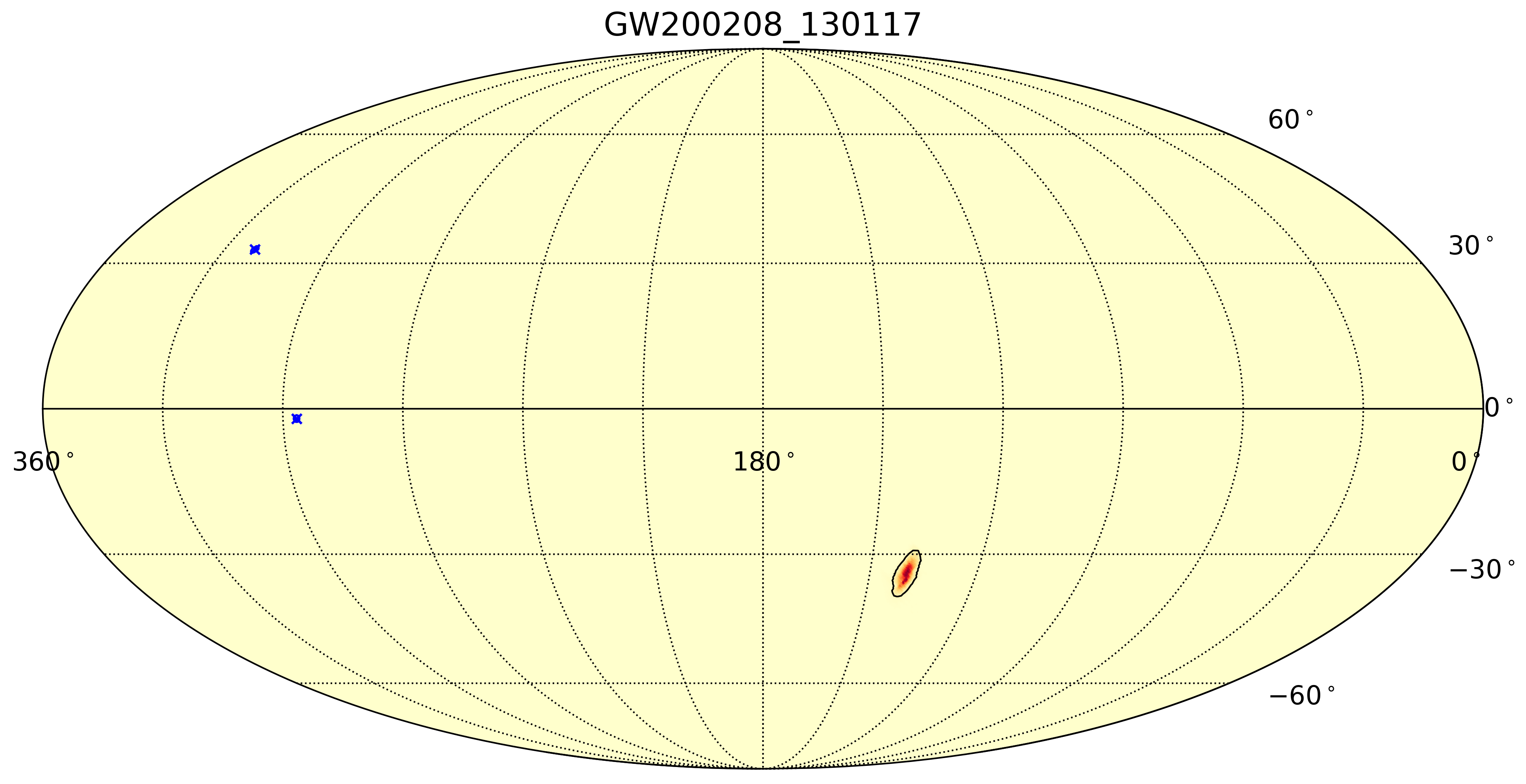}{0.31\textwidth}{(21)}}
\gridline{\fig{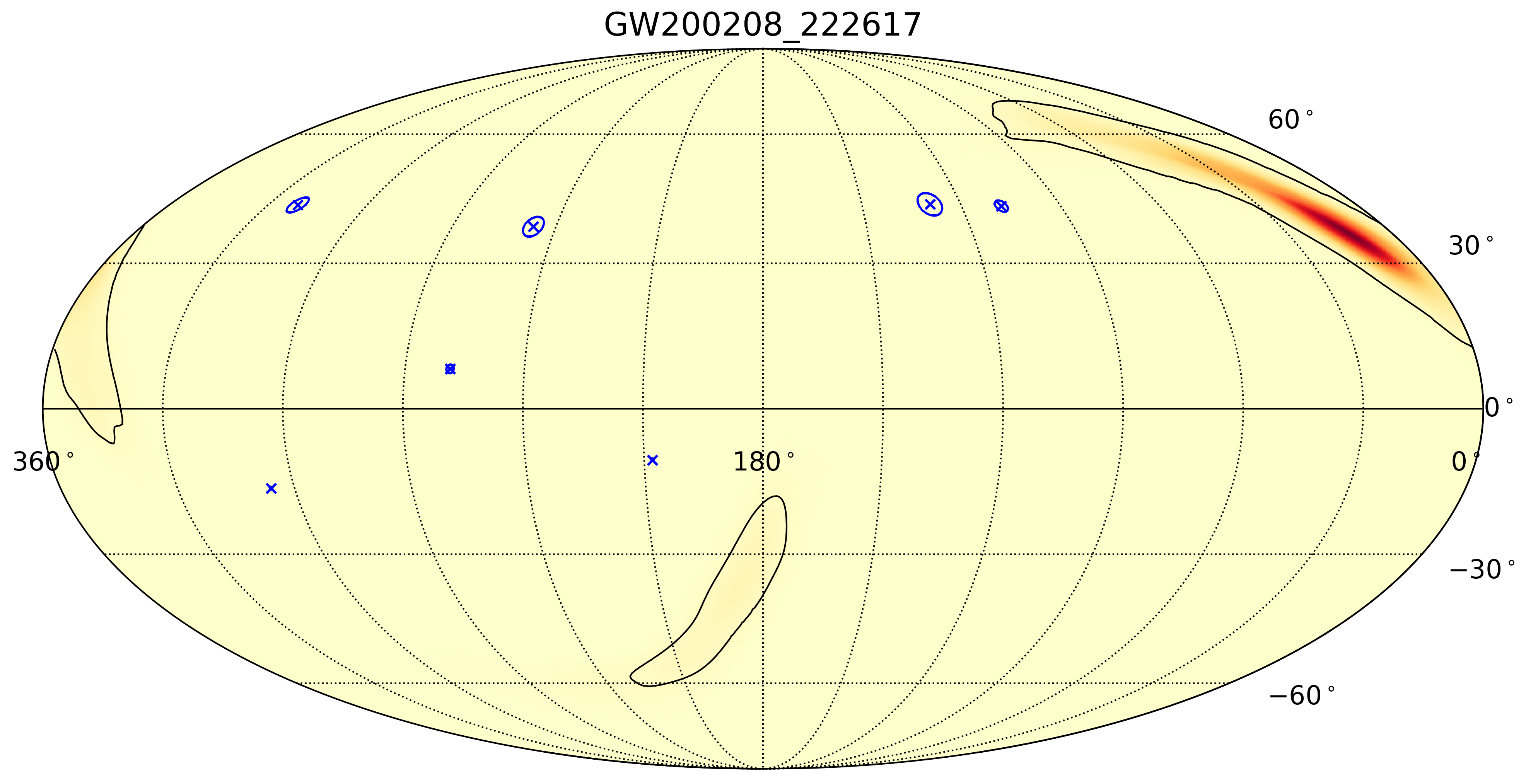}{0.31\textwidth}{(22)}
    \fig{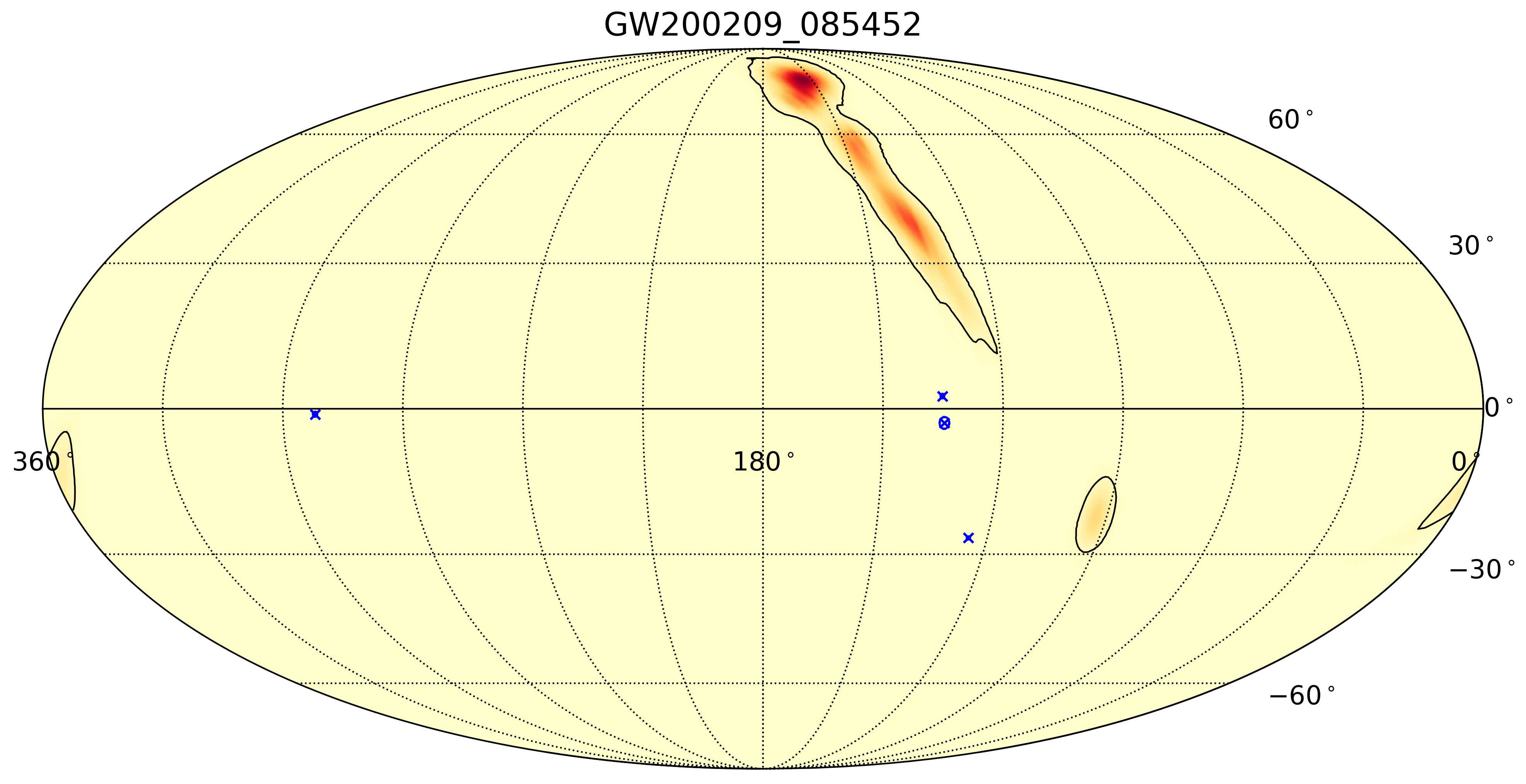}{0.31\textwidth}{(23)}
    \fig{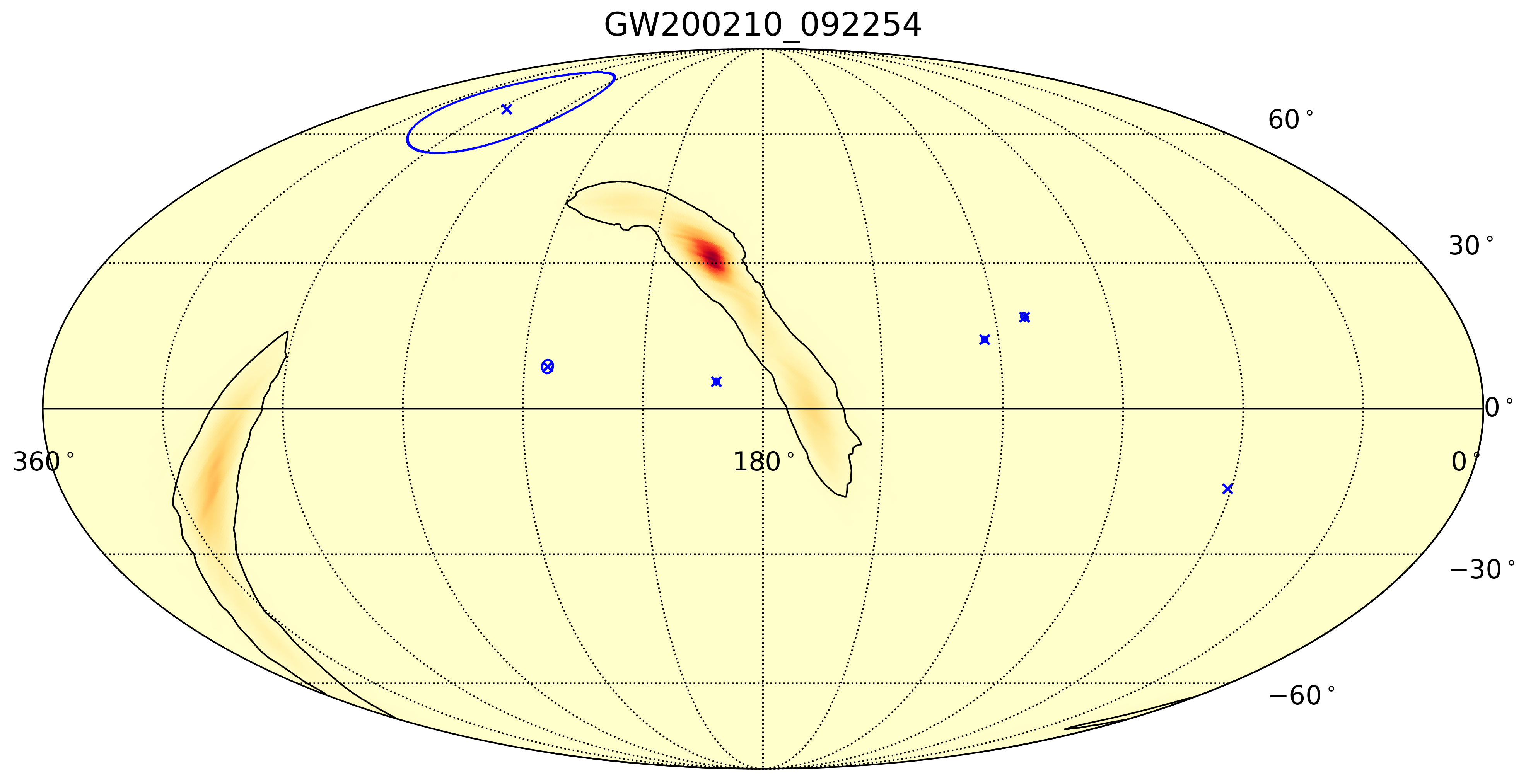}{0.31\textwidth}{(24)}}
\gridline{\fig{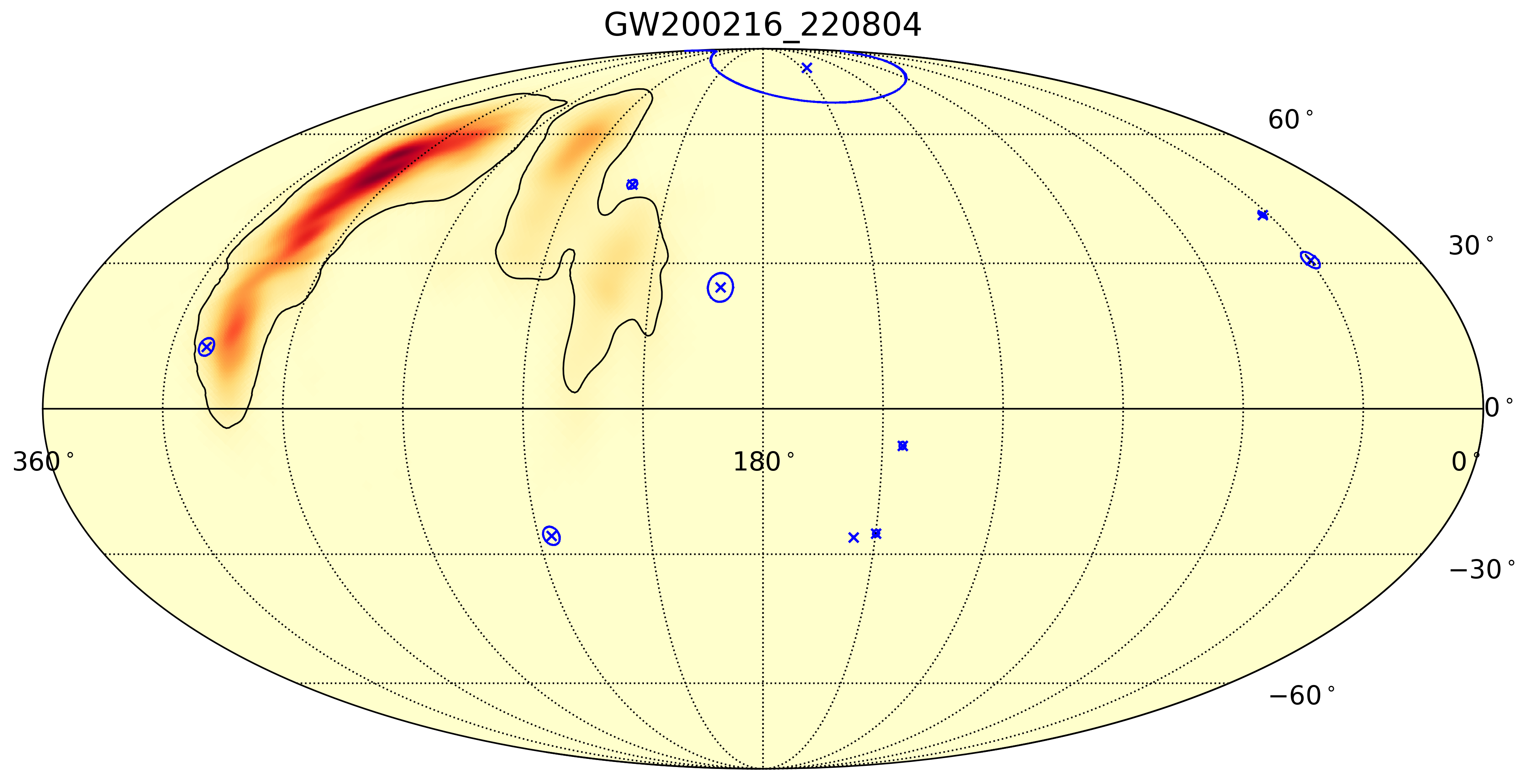}{0.31\textwidth}{(25)}
    \fig{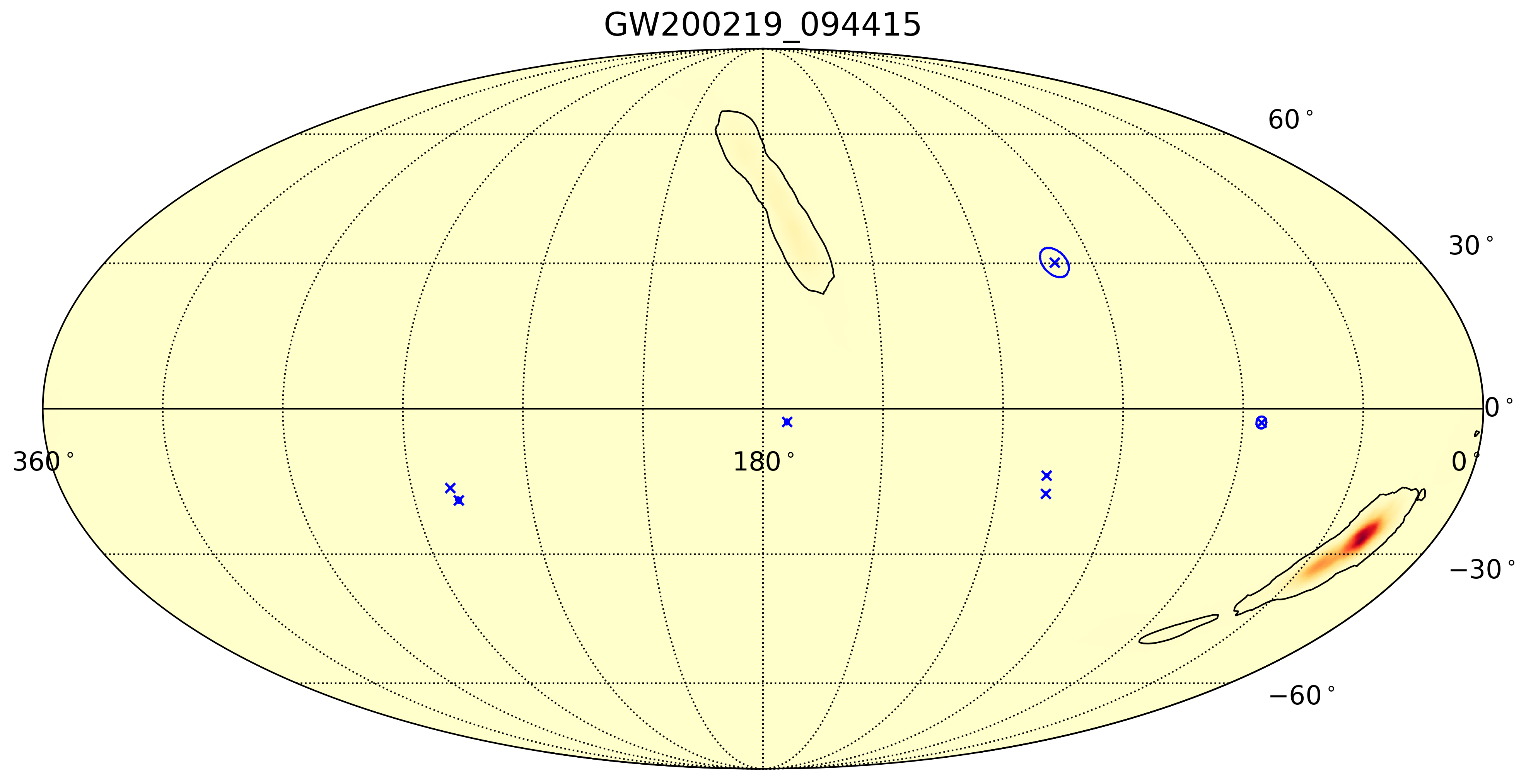}{0.31\textwidth}{(26)}
    \fig{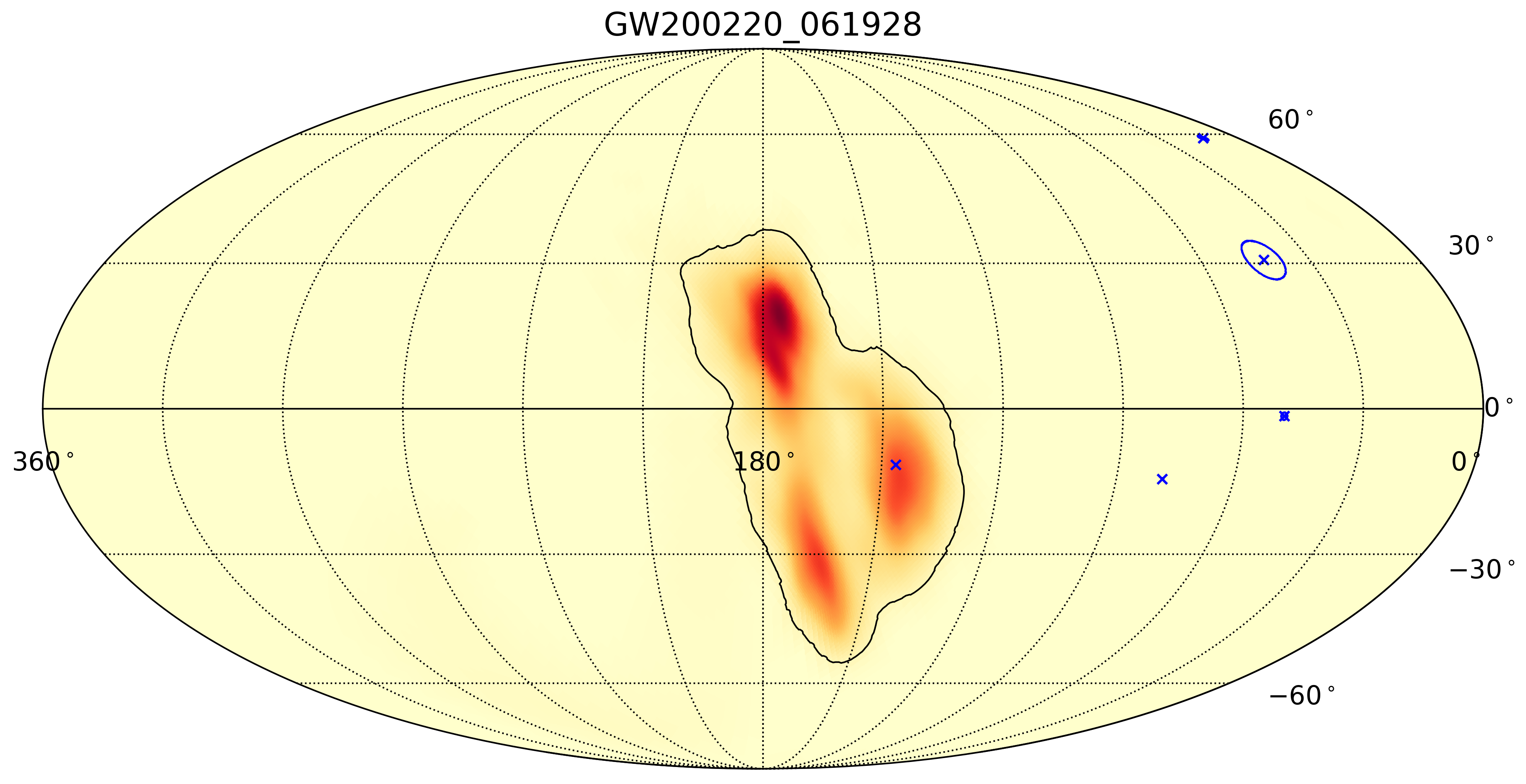}{0.31\textwidth}{(27)}}
\gridline{\fig{figures/skymaps/skymap_200216_220804.png}{0.31\textwidth}{(28)}
    \fig{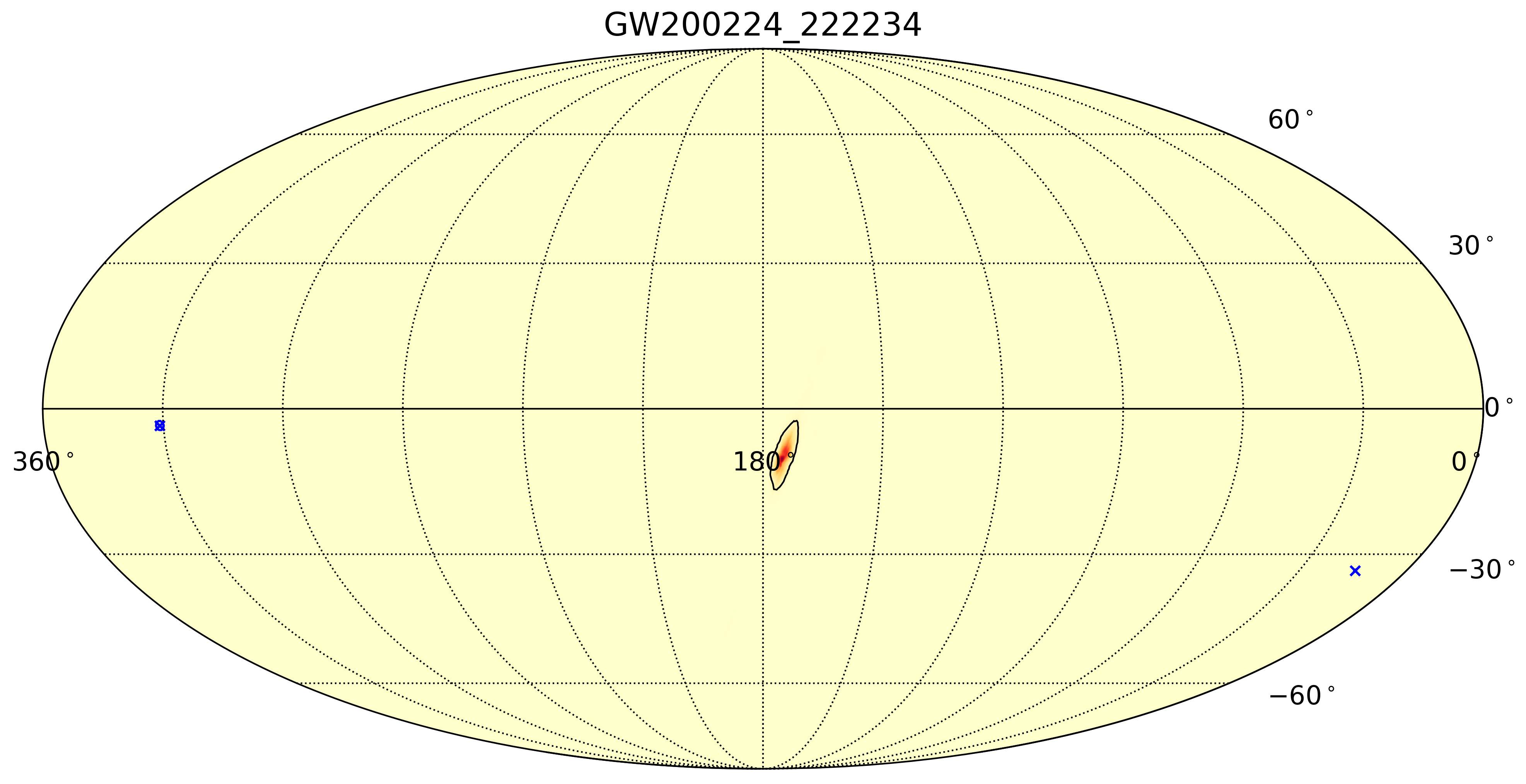}{0.31\textwidth}{(29)}
    \fig{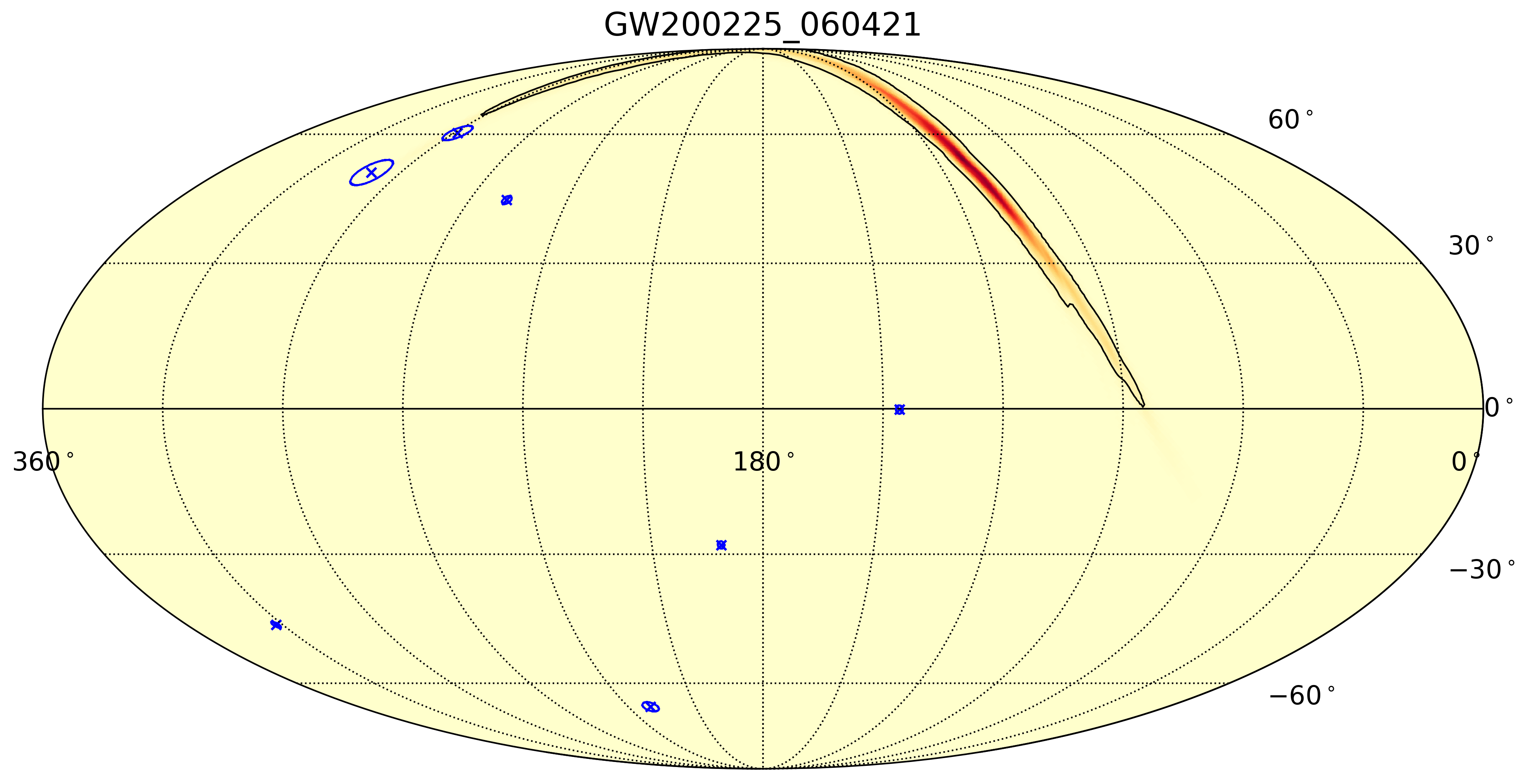}{0.31\textwidth}{(30)}}
\gridline{\fig{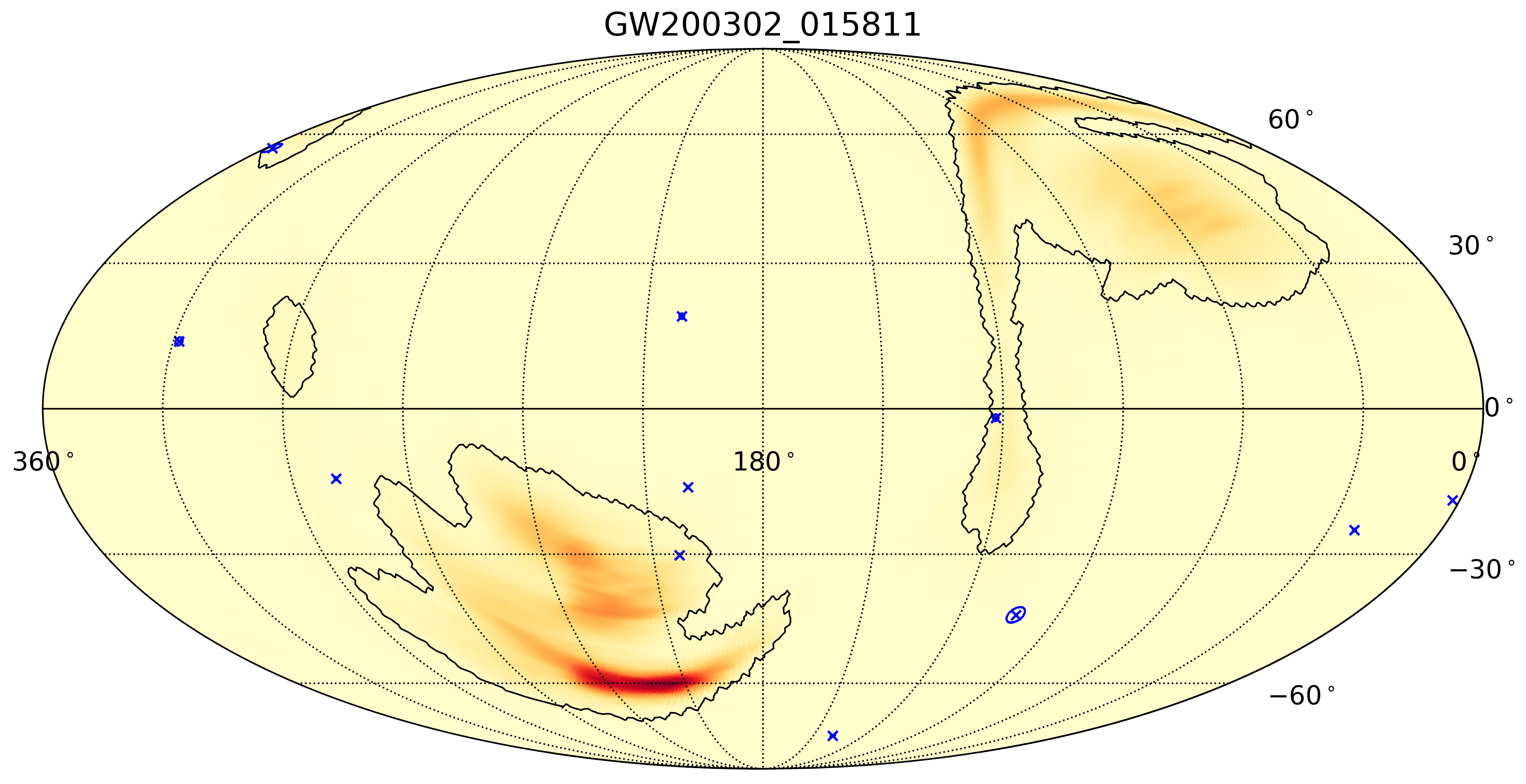}{0.31\textwidth}{(31)}
    \fig{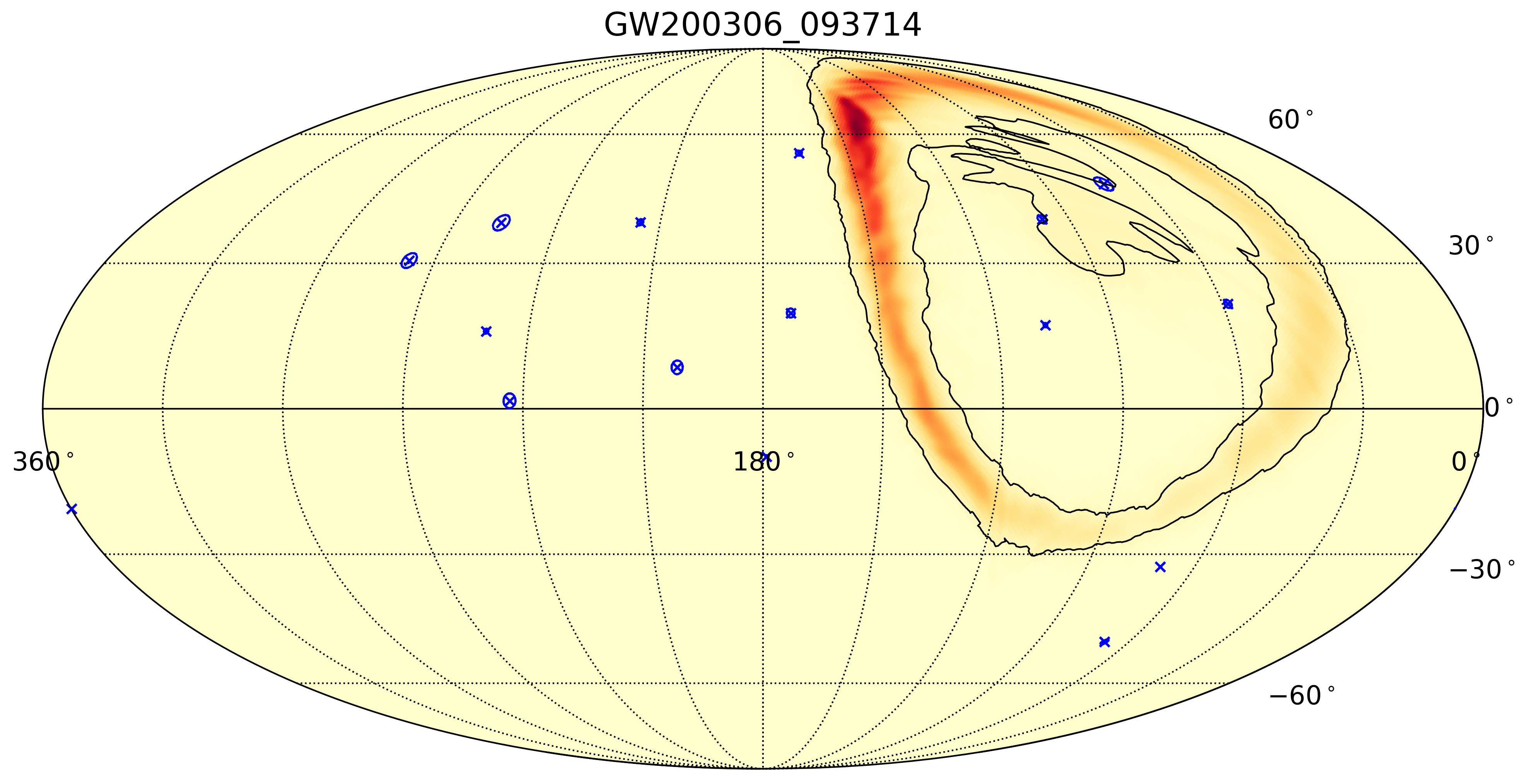}{0.31\textwidth}{(32)}
    \fig{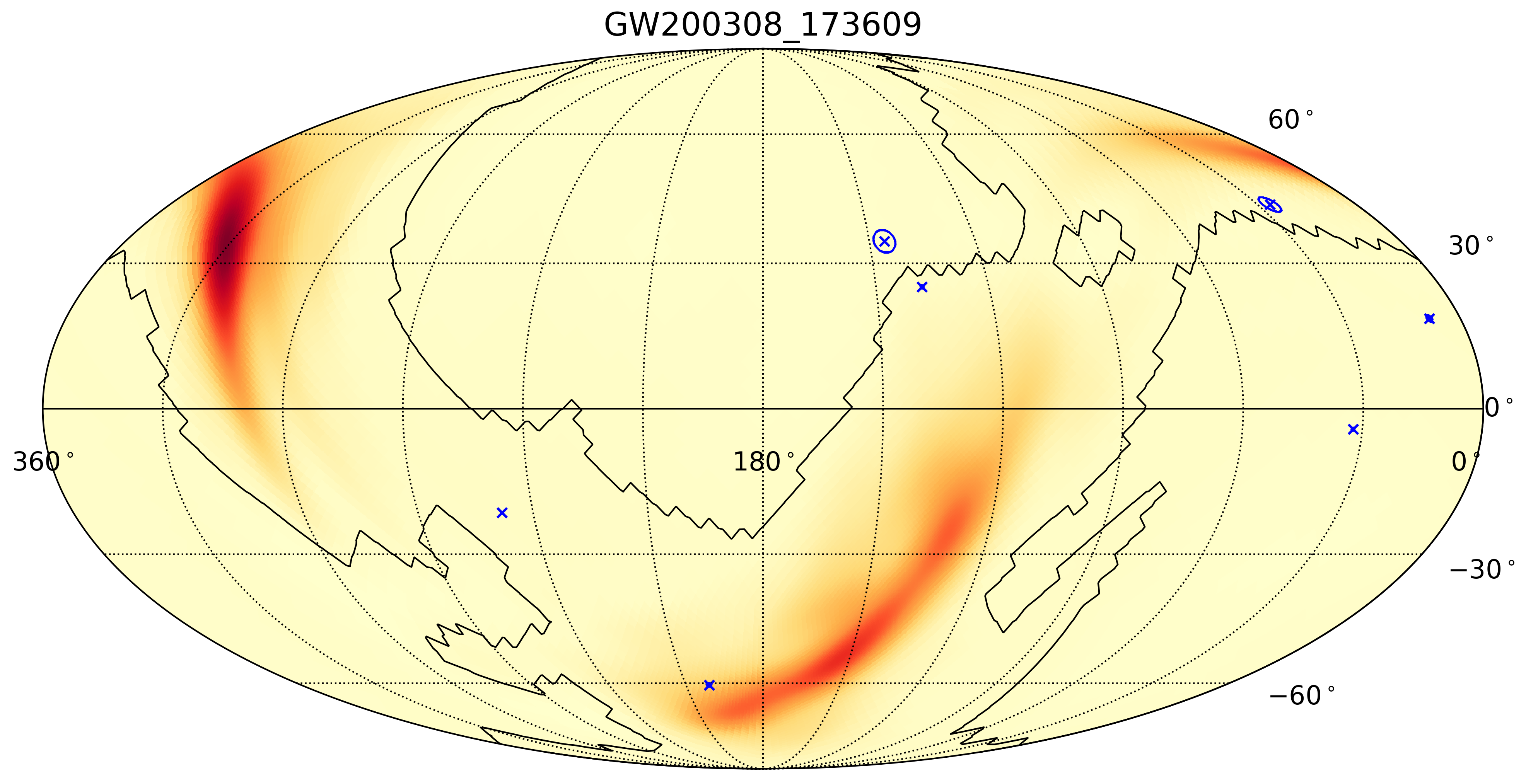}{0.31\textwidth}{(33)}}
\gridline{\fig{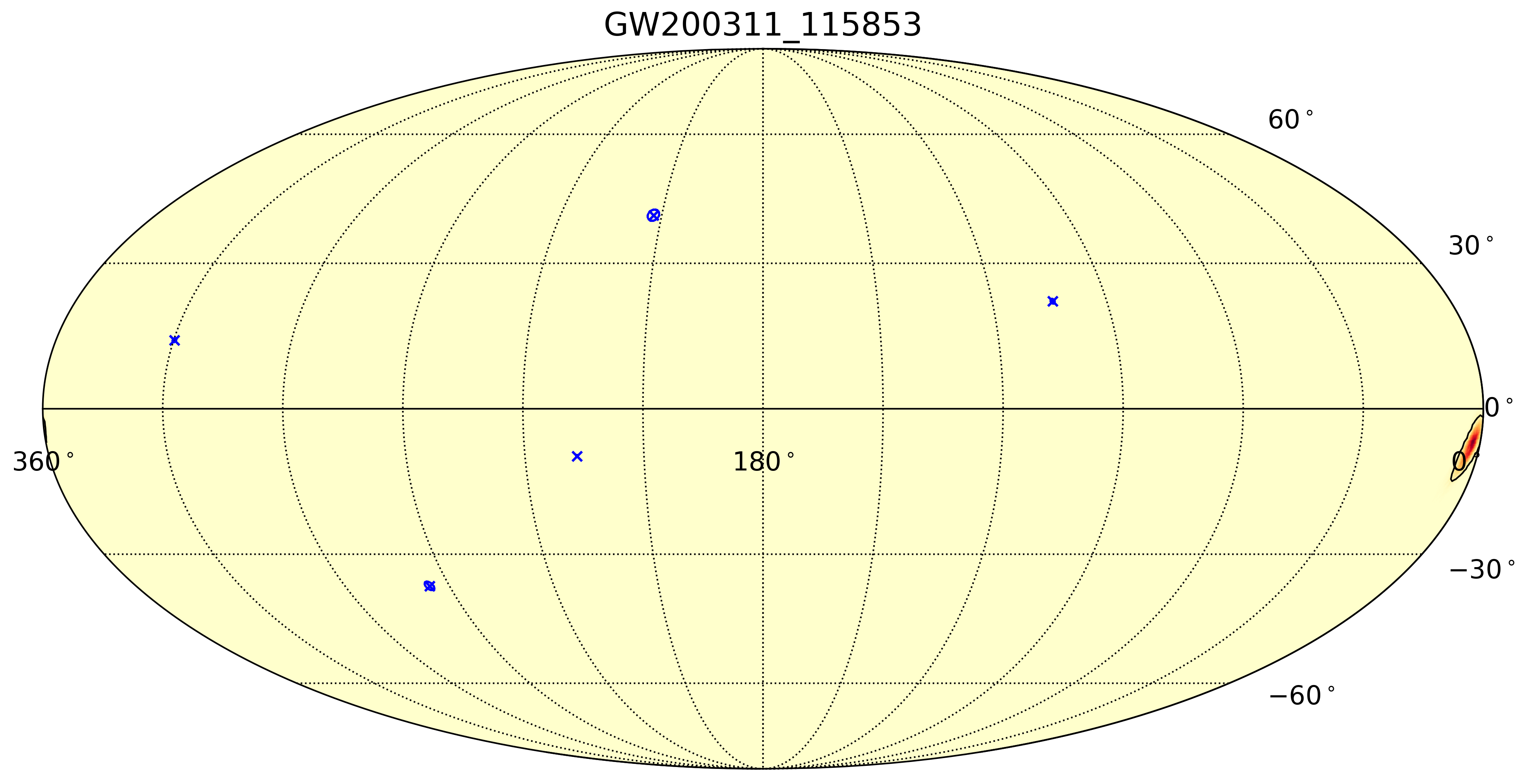}{0.31\textwidth}{(34)}
    \fig{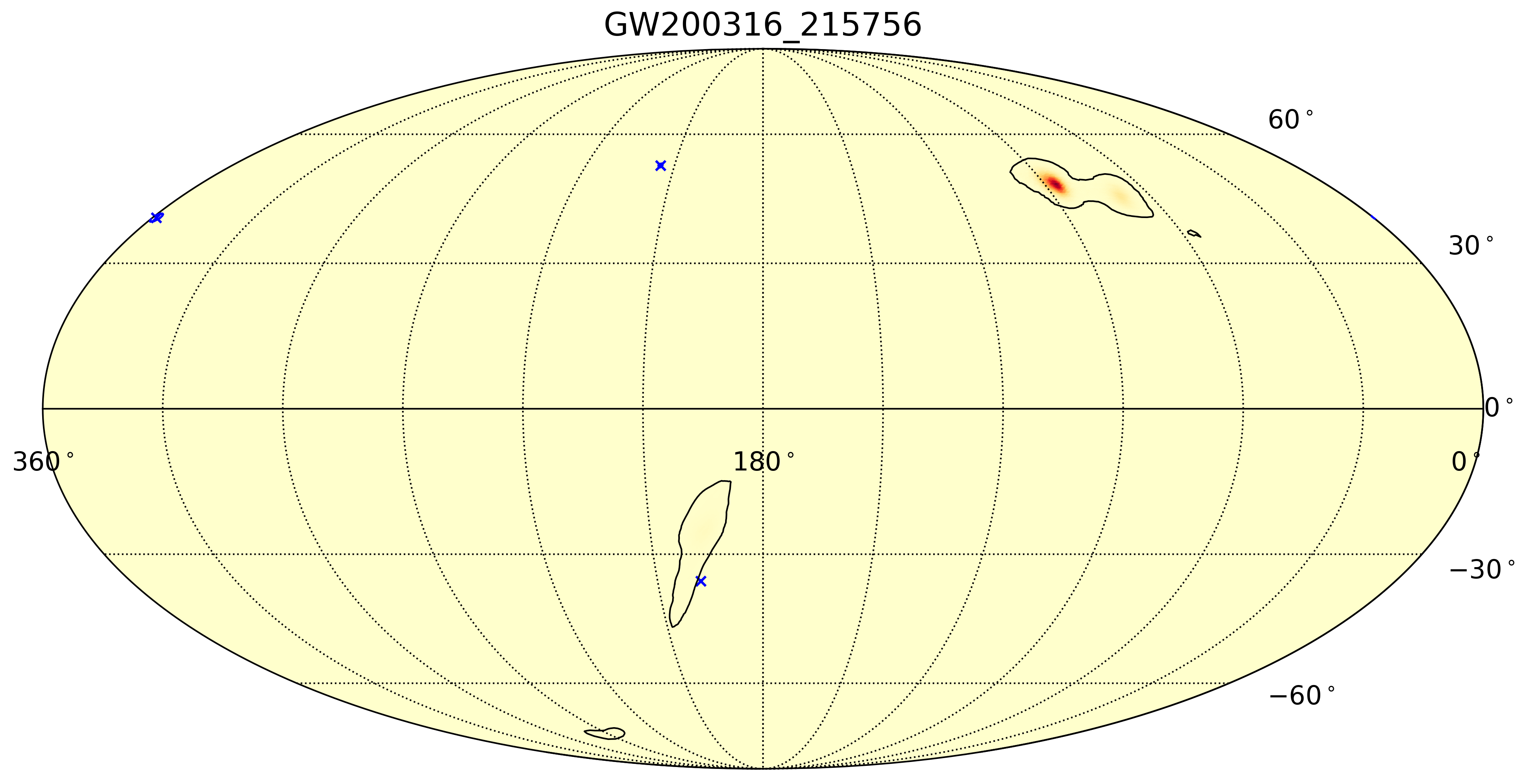}{0.31\textwidth}{(35)}
    \fig{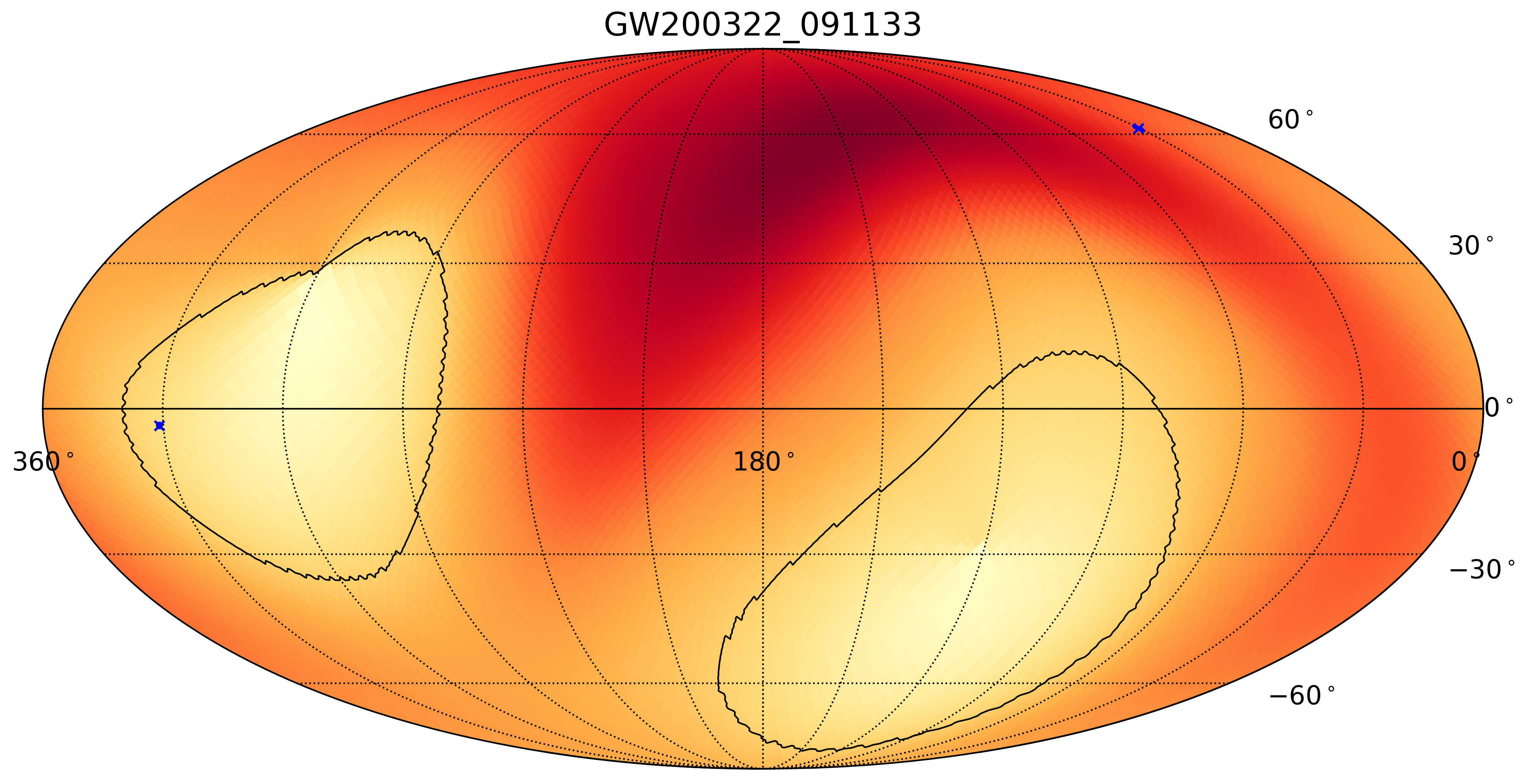}{0.31\textwidth}{(36)}}
    
\caption{Skymaps for the 1000~s follow-up of all events from the GWTC-3 \citep{LIGOScientific:2021djp} catalog.}
\label{fig:1000s_skymaps_gwtc3}
\end{figure}

\end{document}